\documentstyle[floats,prd,aps,epsf]{revtex}

\draft
\begin{document}

\newcommand{\beq}{\begin{equation}}
\newcommand{\eeq}{\end{equation}}
\newcommand{\beqn}{\begin{eqnarray}}
\newcommand{\eeqn}{\end{eqnarray}}
\newcommand{\pa}{\partial}
\newcommand{\vp}{\varphi}
\newcommand{\varep}{\varepsilon}
\def\zero{\hbox{$_{(0)}$}}
\def\bL{\hbox{$\,{\cal L}\!\!\!$--}}
\def\bI{\hbox{$\,I\!\!\!$--}}

\begin{center}
{\large\bf{Three-dimensional simulations of stellar core collapse in
full general relativity: Nonaxisymmetric dynamical instabilities
}}
~\\
~\\
Masaru Shibata and Yu-ichirou Sekiguchi\\
{\em Graduate School of Arts and Sciences,
University of Tokyo, Tokyo, 153-8902, Japan}\\
\end{center}

\begin{abstract}
We perform fully general relativistic simulations of 
rotating stellar core collapse in three spatial dimension. 
The hydrodynamic equations are solved 
using a high-resolution shock-capturing scheme. 
A parametric equation of state is adopted to model 
collapsing stellar cores and neutron stars following Dimmelmeier et al. 
The early stage of the collapse is followed by an axisymmetric code. 
When the stellar core becomes compact enough, we start a three-dimensional
simulation adding a bar-mode nonaxisymmetric density perturbation. 
The axisymmetric simulations are performed for a wide variety 
of initial conditions changing the rotational velocity profile, 
parameters of the equations of state, and the total mass.
It is clarified that the maximum density, the maximum value of
the compactness, and the maximum value of the ratio of the kinetic 
energy $T$ to the gravitational potential energy $W$ ($\beta \equiv T/W$) 
achieved during the stellar collapse and bounce depend sensitively on
the velocity profile and the total mass of the initial core,
and equations of state.
It is also found that for all the models with high 
degree of differential rotation, a funnel structure is formed 
around the rotational axis after the formation of neutron stars. 
For selected models in which the 
maximum value of $\beta$ is larger than $\sim 0.27$, three-dimensional
numerical simulations are performed. 
It is found that the bar-mode dynamical instability sets in
for the case that the following conditions
are satisfied: (i) the progenitor of the stellar core collapse should be 
rapidly rotating with the initial value of $0.01 \alt \beta \alt 0.02$, 
(ii) the degree of differential rotation for 
the velocity profile of the initial condition should be sufficiently high, 
and (iii) a depletion factor of pressure in an early stage of collapse
should be large enough to induce a significant contraction to form 
a compact stellar core for which an efficient spin-up can be
achieved surmounting the strong centrifugal force.
As a result of the onset of the bar-mode dynamical instabilities, 
the amplitude of gravitational waves can be by a factor of $\sim 10$ 
larger than that in the axisymmetric collapse. It is found that 
a dynamical instability with the $m=1$ mode is also induced for 
the dynamically unstable cases against the bar-mode, but the perturbation 
does not grow significantly and, hence, it does not 
contribute to an outstanding amplification of gravitational waves. 
No evidence for fragmentation of the protoneutron stars is found 
in the first a few $10$ msec after the bounce. 
\end{abstract}

\section{Introduction}

One of the most important issues of hydrodynamic simulations in
general relativity is to clarify stellar core collapse to a neutron star
or a black hole. The formation of neutron stars and black holes 
is among the most promising sources of gravitational waves. 
This fact has stimulated numerical simulations for the 
stellar core collapse \cite{Newton0,Newton,Newton1,Newton2,Newton3,Newton4,Newton45,Newton5,Newton6,HD,Siebel,SS2}. 
However, most of these works have been done in the Newtonian framework and 
in the assumption of axial symmetry. 
As demonstrated in \cite{HD,SS2}, general relativistic effects modify 
the dynamics of the collapse and the gravitational waveforms 
significantly in the formation of neutron stars.
Thus, the simulation should be performed
in the framework of general relativity. 
The assumption of axial symmetry is appropriate for
the case that the rotating stellar core is not rapidly rotating. 
However, for the sufficiently rapidly rotating cases, 
nonaxisymmetric instabilities may grow during the collapse and
the bounce \cite{Newton45}. As a result, the amplitude of 
gravitational waves may be increased significantly. 

To date, there has been no general relativistic work for 
the stellar core collapse in {\em three spatial dimensions}.
Three-dimensional simulations of the stellar core collapse
have been performed only in the 
framework of Newtonian gravity \cite{Newton2,Newton45}. 
Hydrodynamic simulations for gravitational collapse or for the onset of 
nonaxisymmetric instabilities of rotating 
{\em neutron stars} in full general relativity have been 
performed so far \cite{gr3d,SBS0,SBS,Duez,EURO}, but 
no simulation has been done for the rotating stellar core collapse to
a neutron star or a black hole. 
In this paper, we present the first numerical results 
of three-dimensional simulations for rapidly rotating 
stellar core collapse in full general relativity. 

Three-dimensional simulation is motivated by two major purposes. One is 
to clarify the criterion for the onset of nonaxisymmetric 
dynamical instabilities during the collapse, and the 
outcome after the onset of the instabilities. 
So far, a number of numerical simulations have illustrated that 
rapidly rotating stars {\em in isolation and in equilibrium}
are often subject to nonaxisymmetric 
dynamical instabilities not only in Newtonian theory 
\cite{TDM,DGTB,WT,TH,CT,PDD,Toman,New,brown,LL,SKE},
but also in post-Newtonian approximation \cite{SSBS}, 
and in general relativity \cite{SBS}. 
These simulations have shown that the dynamical bar-mode 
instabilities set in (i) when the ratio of the kinetic energy $T$ 
to the gravitational potential energy $W$ (hereafter $\beta \equiv T/W$) 
is larger than $\sim 0.27$ or (ii) when the rotating star is 
highly differentially rotating, even for $\beta \ll 0.27$ \cite{SKE}. 
As a result of the onset of the nonaxisymmetric instabilities, 
a bar and spiral arms are formed which can redistribute angular
momentum profile and change the density profile of the star. Also, 
a burst-type and subsequent quasiperiodic gravitational waves
with a high amplitude can be emitted in the case of rapidly
rotating neutron stars \cite{CT,SKE,SBS}. However, the numerical simulations 
have been performed mostly for isolated rotating stars in equilibrium.
To our knowledge, \cite{Newton45} is only one published paper in which 
the nonaxisymmetric dynamical instabilities 
{\em during stellar core collapse} have been investigated. In \cite{Newton45},
the authors performed Newtonian simulations for a few models and 
indicated that the dynamical instability sets in only for the case that
the value of $\beta$ exceeds much beyond 0.27. Such condition 
is satisfied only when the progenitor of the collapse
is rapidly and highly differentially rotating and the depletion of the
internal energy in an early stage of the collapse is large enough to produce
a very compact core for which a significant spin-up
can be achieved surmounting the strong centrifugal force \cite{Newton4}. 

Although the previous Newtonian work \cite{Newton45} indicated 
a criterion for the onset of nonaxisymmetric dynamical instabilities,
many unclear points still remain unsolved as follows.  First, 
in the Newtonian analysis \cite{Newton45}, 
the authors adopted a parametric equation of state, and performed simulations 
changing its own parameters. They found that for the onset of the
nonaxisymmetric instability during collapse, a soft equation of state
with $\Gamma_1=1.28$ and $\Gamma_2=2.5$ is necessary (see Sec. II B
for the definition of $\Gamma_1$ and $\Gamma_2$). 
Unfortunately, in the equations of state that they adopted, 
the maximum gravitational mass for cold spherical neutron stars 
in general relativity becomes $\approx 1.3M_{\odot}$ (see Table I),
which is too small to be adopted as a plausible equation of state
in general relativistic simulations since the maximum mass of neutron stars 
for a given equation of state should be larger than
$\approx 1.44M_{\odot}$ which is the precisely determined mass of a 
neutron star in PSRB1913+16 \cite{TW}.
A study with more plausible equations of state is required. 

Second, the authors in \cite{Newton45} focus little on the instabilities 
associated with $m=2$ bar mode, although it is
the fastest growing mode of the nonaxisymmetric dynamical
instabilities for equilibrium stars in most cases. 
(Here, $m$ denotes the azimuthal quantum number.) 
Thus, the criterion for the onset of the bar-mode
instabilities are not still clear. Also, they paid little attention to 
the bar-mode dynamical instabilities for highly differentially rotating
cases such as those recently reported in \cite{SKE}. 
This instability can set in even for a small value of
$\beta < 0.27$. This implies that for highly differentially rotating
initial conditions, attention should be also paid for small 
values of $\beta$. 

Third, in general relativity, the collapsed core can reach a more 
compact state than that simulated in the Newtonian theory due to 
the fact that self-gravity becomes stronger \cite{HD}. 
As a result, more efficient spin-up will be achieved. Therefore, 
the probability for the onset of nonaxisymmetric dynamical instabilities 
would be underestimated in the Newtonian simulation. 
This suggests that general relativistic 
analysis may be crucial for the study of nonaxisymmetric
dynamical instabilities. 

Finally, in \cite{Newton45}, 
the mass of the stellar core adopted is set to be in a narrow range 
between 1.5 and $1.7M_{\odot}$. According to the theory 
of stellar evolutions, in a very massive star of low metallicity
with the initial mass $50M_{\odot} \alt M \alt 100M_{\odot}$, 
the produced iron core may become 2--3$M_{\odot}$ \cite{Umeda,WHW,ET}.
This indicates that the mass of the core in nature may be in a 
wide range between $\sim 1M_{\odot}$ and $\sim 3M_{\odot}$. 
With the increase of the mass, the self-gravity becomes stronger, and hence, 
the collapsed stellar core can reach a more compact state
for which a spin-up may be enhanced effectively. 
Thus, the larger core mass may increase the probability for 
the onset of nonaxisymmetric dynamical instabilities. 

Motivated by the questions mentioned above, 
we perform general relativistic simulations choosing
rapidly and highly differentially rotating massive stars 
with plausible equations of state and with a wide mass range. 
Following \cite{Newton45}, we adopt a parametric equation of state. 
However, we choose sets of the parameters in which 
the maximum Arnowitt-Deser-Misner (ADM)
mass of cold spherical neutron star becomes
$\approx 1.6M_{\odot}$. With this setting, we choose the mass of the
stellar core in the range between $\sim 1.5$ and $\sim 3M_{\odot}$.
Furthermore, we pay particular attention to the bar-mode instabilities
which are likely to be the fastest growing mode. 

Another major role of three-dimensional simulations 
for the stellar core collapse is to determine the amplitude and 
the characteristic frequency of gravitational waves in the case that the
nonaxisymmetric dynamical instabilities set in. 
Axisymmetric numerical simulations have clarified 
that the amplitude of gravitational waves emitted in the
stellar core collapse is at most several $\times 10^{-23}$
at a distance of 10 Mpc (e.g., \cite{HD}), 
and the frequency is between 100 Hz and 1 kHz. 
Although the frequency is in the most sensitive band of
the laser interferometric gravitational wave detectors \cite{KIP}, 
the value of the amplitude is too small to be detected if the 
stellar core collapse occurs outside our local group of galaxies. 
In the three-dimensional process, on the other hand, the amplitude is
often by a factor of $\sim 10$ larger than that in the axisymmetric
phenomena because of the increase of the degree of asymmetry. 
Hence, if the nonaxisymmetric instabilities set in, 
the stellar core collapse may become a much stronger emitter of
gravitational waves than that considered so far. 

This paper is organized as follows. 
In Sec. II, we briefly review our formulation of 
general relativistic simulation, equations of state adopted
in this paper, and methods for extraction of gravitational waves. 
In Sec. III, initial conditions and computational setting
are described. In Sec. IV, numerical results of axisymmetric
simulations are presented paying attention to the value of
$\beta$ and to the profiles of the density and the angular velocity of
the outcomes. In Sec. V, numerical results of three-dimensional 
simulations are presented, clarifying the criterion for the onset of 
bar-mode dynamical instabilities. Gravitational waveforms
emitted in the growth of the bar-mode dynamical instabilities
are also presented. 
Sec. VI is devoted to a summary. Throughout this paper, we adopt the
geometrical units in which $G=c=1$ where $G$ and $c$ are the 
gravitational constant and speed of light, respectively. 

\section{Formulation}

\subsection{Summary of basic equations and implementations}

We perform hydrodynamic simulations in full general relativity 
using the same formulation as in~\cite{S2002,STU},
to which the reader may refer for details and basic 
equations. The fundamental variables for the hydrodynamics are 
$\rho$ : rest mass density, 
$\varep$ : specific internal energy, 
$P$ : pressure, 
$u^{\mu}$ : four velocity, and 
\beqn 
v^i ={dx^i \over dt}={u^i \over u^t},
\eeqn
where subscripts $i, j, k, \cdots$ denote $x, y$ and $z$, and
$\mu$ the spacetime components. 
As the fundamental variables to be evolved
in the numerical simulations, 
we define a weighted density $\rho_*$, 
a weighted four-velocity $\hat u_i$, and 
a specific energy density $\hat e$ as
\beqn
&& \rho_*=\rho w e^{6\phi}, \\
&& \hat u_i \equiv h u_i, \\
&&\hat e \equiv h w -{P \over \rho w}, 
\eeqn
where $e^{\phi}$ denotes the conformal factor, 
$w\equiv\alpha u^t$, and $h \equiv 1+\varepsilon+P/\rho$.
Here, $\hat e$ is computed from $T_{\mu\nu}n^{\mu}n^{\nu}/(\rho w)$
where $T_{\mu\nu}$ and $n^{\mu}$ denote the energy-momentum tensor
and a timelike unit normal vector. 
General relativistic hydrodynamic equations are solved using 
a high-resolution shock-capturing scheme \cite{Font,S2002}.
In axisymmetric and three-dimensional simulations, 
the cylindrical and Cartesian coordinates are used, respectively. 
The details of our hydrodynamic code are described in \cite{S2002}. 

For the following, we define 
the total baryon rest-mass, internal energy, and rotational
kinetic energy of the system as
\beqn
&&M_*=\int d^3 x \rho_* , \\
&&U  =\int d^3 x \rho_* \varep, \\
&&T  ={1 \over 2} \int d^3 x \rho_* \hat u_{\varphi} v^{\varphi}, 
\eeqn
where $M_*$ is the conserved quantity.
The definitions of $U$ and $T$ agree with those for
axisymmetric rotating stars in equilibrium \cite{ROT}.
In the axisymmetric case, the angular momentum $J$ is
a conserved quantity, and defined by 
\beqn
J =\int d^3 x \rho_* \hat u_{\varphi}. 
\eeqn
Note that in the nonaxisymmetric case, the equation of $J$ has
a different form (e.g., \cite{gr3d}). 

The fundamental variables for the geometry are 
$\alpha$: lapse function, $\beta^k$: shift vector, 
$\gamma_{ij}$: three-metric, 
$\gamma =e^{12\phi}={\rm det}(\gamma_{ij})$: trace of the three-metric, 
$\tilde \gamma_{ij}=e^{-4\phi}\gamma_{ij}$: conformal three-metric, 
and $K_{ij}$ : extrinsic~curvature. 
We evolve $\tilde \gamma_{ij}$, $\phi$, 
$\tilde A_{ij} \equiv e^{-4\phi}(K_{ij}-\gamma_{ij} K_k^{~k})$, 
and the trace of the extrinsic curvature $K_k^{~k}$ 
together with the three auxiliary variables 
$F_i\equiv \delta^{jk}\pa_{j} \tilde \gamma_{ik}$ 
with an unconstrained free evolution code as
in \cite{SN,gr3d,bina,S2002}.
The Einstein equations are solved in the Cartesian coordinates. 
In the axisymmetric case, the Cartoon method is used \cite{alcu,gw2d}.
In both cases, the equatorial reflection symmetry is assumed. 
The outer boundary conditions we adopt are the same as in 
the previous papers (e.g., \cite{gr3d,bina,S2002}).

As the slicing condition, we impose an ``approximate'' maximal slice
condition ($K_k^{~k} \approx 0$) which is 
the same as that adopted in previous papers (e.g., \cite{gr3d,SBS,bina}). 
As the spatial gauge condition, we adopt a hyperbolic gauge
condition \cite{S03,STU} in which we solve 
\beq
\pa_t \beta^k = \tilde \gamma^{kl} (F_l +\Delta t \pa_t F_l),
\label{dyn}
\eeq
where $\Delta t$ denotes a time step in numerical computation.

During numerical simulations, 
violations of the Hamiltonian constraint and conservation of
mass and angular momentum are monitored as code checks.
Numerical results for several test 
calculations, including stability and collapse of nonrotating
and rotating neutron stars, have been described in \cite{S2002}.
The axisymmetric code has been 
used for simulations of stellar core collapse to neutron stars and
black holes, producing numerically convergent results \cite{SS2}.
The three-dimensional code has been used particularly for 
simulations of merger of binary neutron stars \cite{bina,STU}.
In \cite{STU}, the details of the latest implementation are described, 
and we illustrate that accurate and convergent numerical results
on the outcomes after the merger as well as on gravitational waveforms
can be obtained with the present code. 

\subsection{Equations of state}

A parametric equation of state is adopted following M\"uller and
his collaborators \cite{Newton4,HD}. In this equation of state,
one assumes that the pressure consists of the sum of polytropic and
thermal parts as
\beq
P=P_{\rm P}+P_{\rm th}. \label{EOSII}
\eeq
The polytropic part is given by $P_{\rm P}=K_{\rm P}(\rho) \rho^{\Gamma(\rho)}$
where $K_{\rm P}$ and $\Gamma$ are not constants but functions of $\rho$.
This part corresponds to the cold (zero-temperature) part of the equation
of state. In this paper, we follow \cite{HD} for the choice of 
$K_{\rm P}(\rho)$ and $\Gamma(\rho)$: 
For the density smaller than the nuclear density which is defined as 
$\rho_{\rm nuc} \equiv 2\times 10^{14}~{\rm g/cm^3}$, 
$\Gamma=\Gamma_1(=$const) is set to be $\alt 4/3$, and 
for $\rho \geq \rho_{\rm nuc}$, $\Gamma=\Gamma_2(={\rm const}) \geq 2$.
Thus,
\beqn
P_{\rm P}=
\left\{
\begin{array}{ll}
K_1 \rho^{\Gamma_1}, & \rho \leq \rho_{\rm nuc}, \\
K_2 \rho^{\Gamma_2}, & \rho \geq \rho_{\rm nuc}, \\
\end{array}
\right.\label{P12EOS}
\eeqn
where $K_1$ and $K_2$ are constants. 
Since $P_{\rm P}$ should be continuous, the relation, 
$K_2=K_1\rho_{\rm nuc}^{\Gamma_1-\Gamma_2}$, is required. 
Following \cite{Newton4,HD}, the value of $K_1$ is fixed to be 
$5\times 10^{14}$ in the cgs unit.
With this choice, a realistic equation of state for 
$\rho < \rho_{\rm nuc}$, in which the degenerate pressure of electrons is 
dominant, is approximated. Since the specific internal energy 
should be continuous at $\rho=\rho_{\rm nuc}$, the polytropic specific 
internal energy $\varepsilon_{\rm P}$ is defined as 
\beqn
\varepsilon_{\rm P}=
\left\{
\begin{array}{ll}
\displaystyle
{K_1 \over \Gamma_1-1} \rho^{\Gamma_1-1}, & \rho \leq \rho_{\rm nuc}, \\
\displaystyle 
{K_2 \over \Gamma_2-1} \rho^{\Gamma_2-1}
+{(\Gamma_2-\Gamma_1)K_1 \rho_{\rm nuc}^{\Gamma_1-1}
\over (\Gamma_1-1)(\Gamma_2-1)},  & \rho \geq \rho_{\rm nuc}. \\
\end{array}
\right.
\eeqn
With this setting, a realistic equation of state for 
cold nuclear matter is mimicked for an appropriate choice of
$\Gamma_1$ and $\Gamma_2$. 

An advantage of the parametric equations of state is that
we can investigate the dependence of the dynamics of stellar collapse
on the equations of state systematically and very easily 
by changing the values of $\Gamma_1$ and $\Gamma_2$ appropriately.
A more realistic simulation with a realistic equation of state 
should be performed at the goal in this research field. However,
the equations of state for $\rho > \rho_{\rm nuc}$ are not still
well-known. Also, because of complexity of the microphysical processes, 
in simulations with such realistic equations of state,
it is often not easy to extract essential 
physical properties of stellar core collapse such as key
quantities that determine the maximum density in the collapse, 
the collapse time scale, the maximum value of $T/W$, the profiles 
of the density and angular velocity of formed protoneutron stars, 
$T/W$ of formed protoneutron stars, nonaxisymmetric dynamical 
stabilities, and amplitude of gravitational waves. 
Simulations with the parametric equations of state are 
helpful to systematically answer these questions. 

In this paper, we choose $(\Gamma_1, \Gamma_2)=(1.3, 2.5)$,
(1.32, 2.25), and (1.28, 2.75). In Table I, we list the
maximum mass and the corresponding density at the center
for the three sets of $\Gamma_1$ and $\Gamma_2$ with
$\rho_{\rm nuc}=2\times 10^{14}~{\rm g/cm^3}$. 
In all three cases, the maximum ADM mass becomes about
$1.6M_{\odot}$ which is a
reasonable value for neutron stars \cite{ST}. 
As a default, we set $\Gamma_1=1.3$ and $\Gamma_2=2.5$ in the following.
In a previous Newtonian three-dimensional simulation 
\cite{Newton45}, a different set as $\Gamma_1=1.28$ and $\Gamma_2=2.5$
is chosen, and the authors have found that only for such small value
of $\Gamma_1$, nonaxisymmetric dynamical instabilities are induced. 
The choice of this set is acceptable
in the Newtonian framework, but in general relativity it should not be 
adopted because with this choice, the maximum mass of a cold spherical
neutron star becomes about $1.3M_{\odot}$, which is too small
for the maximum mass. Such choice should be excluded 
in general relativistic simulations. 

The thermal part of the pressure $P_{\rm th}$ plays an important
role in the case
that shocks are generated. $P_{\rm th}$ is related to the thermal energy
density $\varepsilon_{\rm th}\equiv \varepsilon-\varepsilon_{\rm P}$ as 
\beq
P_{\rm th}=(\Gamma_{\rm th}-1)\rho \varepsilon_{\rm th}. 
\eeq
For simplicity, the value of $\Gamma_{\rm th}$, which determines
the strength of shocks, is chosen to be equal to $\Gamma_1(\approx 1.3)$.
Our previous numerical work \cite{SS2} showed that
the results depend very weakly on the value of $\Gamma_{\rm th}$
as far as it is in the range between $\sim 1.3$ and 5/3. 

For the simulation, first, 
equilibrium rotating stars with $\Gamma=4/3$ polytrope are given.
Then, the simulations are started with equations of state (\ref{EOSII}). 
Since the value of the adiabatic index is
slightly decreased from $\Gamma=4/3$ to $\Gamma_1(< 4/3)$,
the collapse is triggered. The equilibrium states are computed adopting
the polytropic equation of state 
\beq
P=K_0 \rho^{4/3},\label{EOS43}
\eeq
where $K_0$ is the adiabatic constant.
In this paper, $K_0$ is set to be $5\times 10^{14}$, $7 \times 10^{14}$,
and $8\times 10^{14}~{\rm cm^3/s^2/g^{1/3}}$.
The latter two are adopted to increase the
mass of the progenitor of stellar collapse: 
For the $\Gamma=4/3$ polytrope, the mass 
(both the baryon rest-mass and the ADM mass) of the stars is approximately 
written as $4.555(K_0/G)^{3/2}$ g, which
depends very weakly on the rotational velocity profile \cite{ST}
(cf. Table II). This implies that for
$K_{14} \equiv K_0/10^{14}~{\rm cm^3/s^2/g^{1/3}}=5$, 7, and 8,
the mass is about 1.5, 2.5, and $3M_{\odot}$,
respectively. Thus, for $K_{14} \geq 7$, 
the total mass of the system is much larger than the
maximum allowed mass of the cold spherical neutron stars chosen
in this paper $\approx 1.6M_{\odot}$. 

For $K_0 =K_{\rm deg} \approx 5 \times 10^{14}~{\rm cm^3/s^2/g^{1/3}}$
which is chosen in previous papers \cite{Newton4,HD,SS2}, 
a soft equation of state governed only by the electron degenerate pressure 
is approximated well \cite{ST,Bethe}. On the other hand, the 
radiation pressure is also approximated by the $\Gamma=4/3$ polytropic
equation of state. Thus, by choosing $K_0 > K_{\rm deg}$, we may
consider that the pressure is composed of the sum of
the electron degenerate pressure and the radiation pressure
with the ratio $K_{\rm deg}$ to $K_{\rm rad}\equiv K_0-K_{\rm deg}$ as
\beqn
P=K_{\rm deg} \rho^{4/3}+K_{\rm rad} \rho^{4/3}. 
\eeqn

In the simulation, $K_1$ is related to 
$K_{\rm deg}$ by $K_1=K_{\rm deg}\rho_0^{4/3-\Gamma_1}$
where we set $\rho_0=1~{\rm g/cm^3}$.
The specific internal energy is given by 
\beqn
\varep=3 K_0 \rho^{1/3},
\eeqn
and the pressure at the initial stage is written as 
\beqn
P=3(\Gamma_1-1)K_0 \rho^{4/3}, 
\eeqn
implying that for $\Gamma_1 < 4/3$, the pressure is depleted by
$(4-3\Gamma_1)=4$--16\% for $\Gamma_1=1.32$--1.28 
at the initial stage. 
Namely, in this setting, with the smaller value of $\Gamma_1$,
the pressure for a given value of $\rho < \rho_{\rm nuc}$ becomes
smaller, and also, 
the deletion factors of the pressure and the internal energy
at the initial condition are larger. As shown in Secs. IV and V,
the effect associated with the 
small change in $\Gamma_1$ significantly modifies the dynamics 
of the collapse and the stability against nonaxisymmetric
dynamical deformation. 

\begin{table}[tb]
\begin{center}
\begin{tabular}{|c|c|c|c|c|} \hline
$\Gamma_1$ & $\Gamma_2$ & $M_*(M_{\odot})$
& $M(M_{\odot})$ & $\rho_c({\rm g/cm^3})$ \\ \hline
1.3    & 2.5  & 1.810 & 1.600 &  2.87e15  \\ \hline
1.32   & 2.25 & 1.754 & 1.623 &  2.39e15  \\ \hline
1.28   & 2.75 & 1.869 & 1.597 &  3.18e15  \\ \hline \hline
1.28   & 2.5  & 1.486 & 1.298 &  4.42e15  \\ \hline
\end{tabular}
\caption{Maximum baryon rest-mass, ADM mass,
and the corresponding central density 
for spherical neutron stars of cold, parametric
equations of state (\ref{P12EOS}) for several choices 
of $\Gamma_1$ and $\Gamma_2$.
The first three sets of $\Gamma_1$ and $\Gamma_2$ are adopted in the paper.
The fourth set of the equation of state which is used in Refs. [6,7] 
is too soft and  the maximum ADM mass for cold, spherical neutron stars 
is too small to be adopted as a realistic parameter set. }
\end{center}
\vspace{-5mm}
\end{table}

\subsection{Wave extraction methods}

We extract gravitational waves using two methods.
One is a gauge-invariant wave extraction method in which we
perturbatively compute the Moncrief variables in a flat
spacetime background \cite{moncrief} as we have used
in our series of papers (e.g., \cite{gw3p2}). 
To compute them, first, we split $\gamma_{ij}$ 
into $\eta_{ij}+\sum_{lm} \zeta_{ij}^{lm}$
in the spherical polar coordinates, where 
$\eta_{ij}$ is the flat metric and $\zeta_{ij}^{lm}$ is given by 
\beqn
\zeta_{ij}^{lm}=&& \left(
\begin{array}{lll}
\displaystyle 
H_{2lm} Y_{lm} & h_{1lm} \pa_{\theta} Y_{lm} & h_{1lm}\pa_{\varphi}Y_{lm}\\
\ast & r^2(K_{lm}Y_{lm}+G_{lm}W_{lm}) & r^2G_{lm}X_{lm} \\
\ast & \ast & r^2\sin^2\theta(K_{lm}Y_{lm}-G_{lm}W_{lm}) \\
\end{array}
\right) \nonumber \\
&&+\left(\begin{array}{ccc}
0 &  -C_{lm} \pa_{\varphi} Y_{lm}/\sin\theta
& C_{lm} \pa_{\theta}Y_{lm}\sin\theta  \\
\ast & r^2D_{lm}X_{lm}/\sin\theta
             & -r^2 D_{lm}W_{lm}\sin\theta  \\
\ast & \ast & -r^2 D_{lm}X_{lm}\sin\theta \\
\end{array}
\right). 
\eeqn
Here, $\ast$ denotes the symmetric components. The quantities 
$H_{2lm}$, $h_{1lm}$, $K_{lm}$, $G_{lm}$, $C_{lm}$, and 
$D_{lm}$ are functions of $r$ and $t$, and are calculated 
by performing integrals over a two-sphere of a given coordinate radius 
[see \cite{SN} for details]. 
$Y_{lm}$ is the spherical harmonic function, and 
$W_{lm}$ and $X_{lm}$ are 
\beqn
W_{lm} \equiv \Bigl[ (\pa_{\theta})^2-\cot\theta \pa_{\theta}
-{1 \over \sin^2\theta} (\pa_{\varphi})^2 \Bigl] Y_{lm},
\hskip 5mm 
X_{lm} \equiv 2 \pa_{\varphi} \Bigl[ \pa_{\theta}-\cot\theta \Bigr] Y_{lm}. 
\eeqn
The gauge-invariant variables of even and odd parities are defined by 
\beqn
&&R_{lm}^{\rm E}(t,r) \equiv 
\sqrt{2(l-2)! \over (l+2)!}
\Bigl\{ 4k_{2lm}+l(l+1)k_{1lm} \Bigr\}, \\
&&R_{lm}^{\rm O}(t,r) \equiv \sqrt{2(l+2)! \over (l-2)!}
\biggl({C_{lm} \over r}+r \pa_r D_{lm}\biggr),
\eeqn
where
\beqn
k_{1lm}&& \equiv K_{lm}+l(l+1)G_{lm}+2r \pa_r G_{lm}-2{h_{1lm} \over r},\\
k_{2lm}&& \equiv {H_{2lm} \over 2} - {1 \over 2}{\pa \over \pa r}
\Bigl[r\{ K_{lm}+l(l+1)G_{lm} \} \Bigr].
\eeqn

%The cosine and sine components of the gauge-invariant 
%variables, which are real quantities, are also defined as 
%\beq
%R_{lm+}^{\rm E}={R_{lm}^{\rm E}+R_{l~-m}^{\rm E} \over \sqrt{2}}~~~
%{\rm and}~~~
%R_{lm-}^{\rm E}={R_{lm}^{\rm E}-R_{l~-m}^{\rm E} \over \sqrt{2}}~~~
%(m > 0). 
%\eeq

Using the gauge-invariant variables, the energy luminosity and the angular
momentum flux of gravitational waves can be defined by
\beqn
&&{dE \over dt}={r^2 \over 32\pi}\sum_{l,m}\Bigl[
|\pa_t R_{lm}^{\rm E}|^2+|\pa_t R_{lm}^{\rm O}|^2 \Bigr],
\label{dedt} \\
&&{dJ \over dt}={r^2 \over 32\pi}\sum_{l,m}\Bigl[
 |m(\pa_t R_{lm}^{\rm E}) R_{lm}^{\rm E} |
+|m(\pa_t R_{lm}^{\rm O}) R_{lm}^{\rm O} | \Bigr]. 
\label{dJdt} 
\eeqn
The total radiated energy and angular momentum are calculated by
\beq
\Delta E(t) = \int_0^t dt {dE \over dt}, \hskip 5mm
\Delta J(t) = \int_0^t dt {dJ \over dt}.
\eeq
In this paper, we pay attention only to even-parity modes
with $l=2$ which are the dominant modes. 

To search for the characteristic frequencies of gravitational waves,
the Fourier spectra are computed by 
\beq
\bar R_{lm}^{I}(f)=\int^{t_f}_{t_i} e^{2\pi i f t} R_{lm}^{I} dt,
\eeq
where $I$ denotes E and O.
In the analysis, $t_f$ is chosen as the time at which the simulation is
stopped. Before $t < r_{\rm obs}$ where $r_{\rm obs}$ denotes
a radius at which gravitational waves are extracted, no waves 
propagate to $r_{\rm obs}$, so that we choose $t_i \approx r_{\rm obs}$. 

Using the Fourier spectrum, the energy power spectrum is written as
\beq
{dE \over df}={\pi \over 4}r^2
\sum_{l\geq 2, m\geq 0}f^2
(|\bar R_{lm}^{\rm E}(f)|^2+|\bar R_{lm}^{\rm O}(f)|^2)~~~(f > 0), 
\label{dedf}
\eeq
where for $m\not=0$, we define 
\beq
\bar R_{lm}^{I}
\equiv \sqrt{|\bar R_{l m}^{I}(f)|^2 + |\bar R_{l -m}^I(f)|^2}~~(m>0). 
\eeq
Note that in deriving Eq. (\ref{dedf}), we use the relation 
$|\bar R_{lm}^{I}(-f)|=|\bar R_{lm}^{I}(f)|$. 

Computation of gravitational waves is also carried out 
in terms of a quadrupole formula which is described in \cite{SS1,SS2}.
As shown in \cite{SS1}, a kind of quadrupole formula can provide 
an approximate gravitational waveforms from oscillating compact stars. 
The quadrupole formula is in particular 
useful when the amplitude of gravitational waves is
smaller than the numerical noise because in such case, it is difficult to 
extract gravitational waves from the metric in the wave zone. 

In quadrupole formulas, we compute gravitational waves from 
\beq
h_{ij}=\biggl[P_{i}^{~k} P_{j}^{~l}-{1 \over 2}P_{ij}P^{kl}\biggr]
\biggl({2 \over r}{d^2\bI_{kl} \over dt^2}\biggr),\label{quadf}
\eeq
where $\bI_{ij}$ and $P_i^{~j}=\delta_{ij}-n_i n_j$ ($n_i=x^i/r$)
denote a tracefree quadrupole moment and a projection tensor. 

In fully general relativistic and dynamical spacetimes, 
there is no unique definition for the quadrupole moment $I_{ij}$.
Following \cite{SS1,SS2}, we choose the formula as 
\beq
I_{ij} = \int \rho_* x^i x^j d^3x. 
\eeq
Then, using the continuity equation, 
we can compute the first time derivative as 
\beq
\dot I_{ij} = \int \rho_* (v^i x^j +x^i v^j)d^3x.
\eeq
To compute $\ddot I_{ij}$, we 
carried out the finite differencing of the numerical result 
for $\dot I_{ij}$.

In this paper, we focus only on $l=2$ mass quadrupole modes. 
Then, the gravitational waveforms are described by 
\beqn
&&h_+=
{1 \over r} \biggl[ {\ddot I_{xx}-\ddot I_{yy} \over 2}
(1+\cos^2\theta)\cos(2\varphi)
+\ddot I_{xy}(1+\cos^2 \theta) \sin(2\varphi)
+\biggl( \ddot I_{zz}-{\ddot I_{xx}+\ddot I_{yy} \over 2} \biggr)
\sin^2\theta \biggr],\\
&&h_{\times}=
{2 \over r} \biggl[ -{\ddot I_{xx}-\ddot I_{yy} \over 2}
\cos\theta \sin(2\varphi)
+\ddot I_{xy} \cos\theta \cos(2\varphi) \biggr], 
\eeqn
in the quadrupole formula, and
\beqn
&&h_+=
{1 \over r}\biggl[ \sqrt{{5 \over 64\pi}}
\{ R_{22+}(1+\cos^2\theta)\cos(2\varphi)
+  R_{22-}(1+\cos^2\theta)\sin(2\varphi) \}
+ \sqrt{{15 \over 64\pi}}R_{20} \sin^2\theta \biggr],\label{eq34} \\
&&h_{\times}={2 \over r} \sqrt{{5 \over 64\pi}}
\Bigl[ -R_{22+} \cos\theta \sin(2\varphi)
+R_{22-} \cos\theta \cos(2\varphi)\Bigr], \label{eq35}
\eeqn
in the gauge-invariant wave extraction where
\beqn
&& R_{22\pm}={R_{22}^{\rm E} \pm R_{2~-2}^{\rm E} \over \sqrt{2}}r,\\ 
&& R_{20}=R_{20}^{\rm E}r. 
\eeqn
For derivation of $h_+$ and $h_{\times}$, 
we assume that the wave part of the
spatial metric in the wave zone is written as 
\beqn
dl^2=dr^2+r^2[(1+h_+)d\theta^2+\sin^2\theta(1-h_+)d\varphi^2
+2 \sin\theta h_{\times} d\theta d\varphi], 
\eeqn
and set $I_{xz}=I_{yz}=0$ and $R_{2~\pm 1}^{\rm E}=0$ 
since we assume the reflection symmetry 
with respect to the equatorial plane. 

In the following, we present
\beqn
&& A_+=\ddot I_{xx}-\ddot I_{yy},\\
&& A_{\times}=2 \ddot I_{xy},\\
&& A_0={2\ddot I_{zz}-\ddot I_{xx}-\ddot I_{yy} \over 2}, 
\eeqn
in the quadrupole formula, and as the corresponding variables,
\beqn
&& R_+=\sqrt{{5 \over 16\pi}}R_{22+},\\
&& R_{\times}=\sqrt{{5 \over 16\pi}}R_{22-},\\
&& R_0=\sqrt{{15 \over 64\pi}}R_{20},
\eeqn
in the gauge-invariant wave extraction method.
These provide the amplitude of a given mode
measured by an observer located in the most 
optimistic direction. 

\section{Setting}

\subsection{Initial conditions for axisymmetric simulation}

\begin{table}[tb]
\begin{center}
\begin{tabular}{|c|c|c|c|c|c|c|c|c|c|c|c|} \hline
Model & $A$ & $K_0$ & $M_* $ & $\beta_{\rm init}$ & $J/M^2$ & $P_0$ & $R_e$
& $\alpha_{\rm Min}$ & $\rho_{\rm Max}$ & $\beta_{\rm max}$ & Outcome \\ \hline
M5a1 &
$\infty$& 5e14&1.503 & 0.00891&1.235 &1.53  &2.27e3  &0.76 &6.6e14 &0.11
& NS \\\hline \hline
M5c1 &
0.1  & 5e14 & 1.545 & 0.0177 &1.201 &0.127 & 1.48e3 &0.79 &3.2e14 &0.28
& O-A \\\hline 
M5c2 &
0.1  & 5e14 & 1.521 & 0.0124 &1.028 &0.155 & 1.50e3 &0.74 &5.0e14 &0.28
& O-A $\rightarrow$ NS \\\hline 
M5c3 &
0.1  & 5e14 & 1.496 & 0.00730&0.784 &0.212 & 1.53e3 &0.71 &6.6e14 &0.21
& NS \\\hline \hline
%%%%%%%%%%%%%%%%%%%%%%
M7a1 &
$\infty$  & 7e14 & 2.476 & 0.00886 &1.045 &1.53 &2.68e3& 0.57& 1.1e15&0.10 
& NS \\\hline
M7a2 &
$\infty$  & 7e14 & 2.458 & 0.00649 &0.888 &1.77 &2.16e3& 0.56& 1.1e15&0.081
& NS \\\hline
M7a3 &
$\infty$  & 7e14 & 2.449 & 0.00526 &0.792 &1.97 &2.06e3 & --- & --- & --- 
& BH \\\hline 
M7a4 &
$\infty$  & 7e14 & 2.438 & 0.00367 &0.663 &2.33 &1.98e3 & --- & --- & --- 
& BH \\\hline \hline
M7b1 &
0.25  & 7e14 & 2.579 & 0.0218 & 1.423 & 0.370 &1.88e3 & 0.82 & 7.9e13 &0.26
& O-B \\\hline
M7b2 &
0.25  & 7e14 & 2.545 & 0.0177 & 1.283 & 0.411 &1.87e3 & 0.73 & 2.6e14 &0.29
& O-A \\\hline
M7b3 &
0.25  & 7e14 & 2.514 & 0.0138 & 1.134 & 0.466 &1.87e3 & 0.65 & 4.8e14 &0.29
& NS \\\hline
M7b4 &
0.25  & 7e14 & 2.495 & 0.0113 & 1.027 & 0.515 &1.86e3 & 0.62 & 6.4e14 &0.26
& NS   \\\hline
M7b5 &
0.25  & 7e14 & 2.451 & 0.00543& 0.712 & 0.744 &1.85e3 & 0.55 & 1.1e15 &0.16
& NS \\\hline
M7b6 &
0.25  & 7e14 & 2.434 & 0.00321& 0.547 & 0.969 &1.85e3 & 0.52 & 1.4e15 &0.10 
& NS \\\hline\hline
M7c1 &
0.1 & 7e14 & 2.579 & 0.0219 & 1.126 & 0.111 &1.72e3 & 0.70 & 3.8e14 &0.31
& O-A \\\hline
M7c2 &
0.1 & 7e14 & 2.544 & 0.0177 & 1.018 & 0.127 &1.75e3 & 0.61 & 6.0e14 &0.33
& O-A \\\hline
M7c3 &
0.1 & 7e14 & 2.505 & 0.0127 & 0.871 & 0.155 &1.78e3 & 0.51 & 9.2e14 &0.30
& NS \\\hline
M7c4 &
0.1 & 7e14 & 2.505 & 0.00994 & 0.773 & 0.179 &1.79e3 &0.46 & 1.2e15 &0.27 
& NS \\\hline
M7c5 &
0.1 & 7e14 & 2.464 & 0.00728& 0.664 & 0.213 &1.81e3 & 0.42 & 1.4e15 &0.22
& NS \\\hline
M7c6 &
0.1 & 7e14 & 2.439 & 0.00392& 0.489 & 0.296 &1.83e3 & 0.42 & 1.4e15 &0.14 
& NS \\\hline
M7c7 &
0.1 & 7e14 & 2.427 & 0.00232& 0.377 & 0.389 &1.83e3 & 0.44 & 1.7e15 &0.088
& NS \\\hline\hline
M8a1 &
$\infty$ & 8e14 & 3.016 & 0.00884& 0.978 & 1.54 & 2.86e3 & --- & --- & ---
& BH \\\hline\hline
M8c1 &
0.1 & 8e14 & 3.187 & 0.0263 & 1.151 & 0.0984 &1.82e3 & 0.70 & 3.0e14 &0.31 
& O-A \\\hline
M8c2 &
0.1 & 8e14 & 3.141 & 0.0219 & 1.055 & 0.111 & 1.84e3 & 0.59 & 5.4e14 &0.34 
& O-A \\\hline
M8c3 &
0.1 & 8e14 & 3.099 & 0.0176 & 0.953 & 0.128 &1.87e3 & 0.47 & 8.4e14 &0.35 
& O-A \\\hline
M8c4 &
0.1 & 8e14 & 3.052 & 0.0127 & 0.815 & 0.156 &1.90e3 & 0.29 & 1.5e15 &0.30
& NS  \\\hline
M8c5 &
0.1 & 8e14 & 3.010 & 0.00814& 0.657 & 0.200 &1.93e3 &  --- & --- & ---
& BH \\\hline
\end{tabular}
\caption{Quantities for selected set of the initial conditions
and the results of collapse are listed.
$K_0$, $M_*$, $P_0(\equiv 2\pi/\Omega_a)$, $R_e$, and $\rho_{\rm Max}$ are
shown in units of ${\rm cm^3/s^2/g^{1/3}}$, $M_{\odot}$, sec,
km, and ${\rm g/cm^3}$, respectively.
Here, $\rho_{\rm Max}$ and $\alpha_{\rm Min}$ are the
maximum and minimum achieved during the whole evolution. 
$\beta_{\rm init}$ and $\beta_{\rm max}$ denote the initial value of
$T/W$ and the maximum value of $T/(T+U)$ achieved during the collapse. 
The baryon rest-mass $M_*$ is nearly equal to the ADM mass $M$ 
for all the models. In the last column, the outcomes for 
$\Gamma_1=1.3$, $\Gamma_2=2.5$, and
$\rho_{\rm nuc}=2\times 10^{14}~{\rm g/cm^3}$ are shown. 
Here, NS, O-A, and O-B 
denote that the outcomes are neutron star, oscillating star with 
the maximum density larger than $\rho_{\rm nuc}$, and 
oscillating star of subnuclear density, respectively.
Note that for $K_0=8 \times 10^{14}$ (cgs) and $A \rightarrow \infty$, 
any star collapses to a black hole and that
for $K_0 \leq 6 \times 10^{14}$ (cgs),
any star does not collapse to a black hole.}
\end{center}
\vspace{-5mm}
\end{table}

Rotating stellar cores in equilibrium with the $\Gamma=4/3$ polytropic 
equation of state [Eq. (\ref{EOS43})] are prepared 
for the initial conditions. Following \cite{HD,SS2}, the maximum density
is chosen as $\rho_{\rm max} = 10^{10}~{\rm g/cm^3}$
irrespective of the velocity profile and the value of $K_0$. 

The velocity profiles of equilibrium rotating stellar cores are given 
according to a popular relation \cite{KEH,Ster,HD} 
\beq
u^t u_{\varphi} = \varpi_d^2( \Omega_a - \Omega ), 
\eeq 
where $\Omega=v^{\varphi}$ denotes the
angular velocity, $\Omega_a$ is that 
along the rotational axis, and $\varpi_d$ is a constant. 
In the Newtonian limit, the rotational profile is written as 
\beq
\Omega = \Omega_a{\varpi_d^2 \over \varpi^2 + \varpi_d^2}. \label{omenew}
\eeq
Thus, $\varpi_d$ indicates the steepness of differential rotation.
Since the compactness of the initial data adopted in this paper is not
so large with $M/R \sim 10^{-3}$, where $R$ denotes a stellar radius, that
general relativistic effects are weak. As a result, 
the profile of the rotational angular velocity is approximately
written by Eq. (\ref{omenew}). 
In the following, we adopt rigidly rotating models in which
$\varpi_d \rightarrow \infty$, 
and differentially rotating models 
with $A \equiv \varpi_d/R_e=0.25$ and 0.1, where $R_e$ is the
coordinate radius at the equatorial surface.
The ratio of the angular velocity at the equatorial surface to $\Omega_a$
is $\approx 1/17$ and $1/101$ for $A=0.25$ and 0.1, indicating that
Eq. (\ref{omenew}) is approximately satisfied. 
We pay particular attention to the case with high degrees of
differential rotation in this paper, since in the collapse, 
a large value of $\beta$ can be achieved only for such cases.
Indeed, a study for presupernova evolution of
rotating massive stars \cite{HLW} indicates that the velocity
profile of the iron core just before the onset of collapse
may be differentially rotating. 

As introduced in Sec. I, the ratio ($\beta=T/W$) of the
rotational kinetic energy $T$ to the gravitational potential energy
$W$ is often referred in the following. In general relativity, 
$W$ is defined by \cite{ROT}
\beqn
W= M_* + U + T - M. 
\eeqn
Here, $W$ is defined to be positive.
For stable rotating stars in equilibrium with $\Gamma=4/3$, $M_*$ is
nearly equal to $M$. Thus,
\beqn
W \approx U + T, ~~{\rm and}~~\beta \approx {T \over U+T}. 
\eeqn
Even in the dynamical evolution, 
$M_*$ is the conserved quantity and $M$ is approximately
conserved in the case that luminosity of gravitational waves is small.
Thus, if other components of the energy such as the kinetic energy 
associated with radial velocity are small, 
the above approximate relation for $\beta$
in terms of $U$ and $T$ may be used even in the dynamical spacetime
[see also discussion around Eq. (\ref{beta1})].

%%In the following, the relation in Eq. (\ref{beta}) is used for
%%approximately computing $\beta$ even in the collapse. 

In Table II, several quantities for the models adopted 
in the present numerical computation are summarized.
In the first column, we describe the name of each model. 
We refer to the models
with $(K_{14}, A)=(5, \infty)$, (5, 0.1),
$(7, \infty)$, (7, 0.25), (7, 0.1), $(8, \infty)$, 
and (8, 0.1) as M5a, M5c, M7a, M7b, M7c, M8a, and M8c, respectively.

For the rigidly rotating case with $A \rightarrow \infty$, 
the magnitude of the angular velocity has to be smaller than the
Kepler angular velocity at the equatorial surface.
This restricts the maximum value of $\beta$ to be
smaller than $\approx 0.0089$ for all the values of $K_0$. This implies 
that the angular velocity for models M5a1, M7a1, and M8a1 is approximately 
maximum among the rigidly rotating cases for a given value of $K_0$. 
The final outcome of M8a1 is a black hole. This implies that 
any star with $(K_{14}, A)=(8, \infty)$  
collapses to a black hole because the
mass is too large and the angular momentum is too small to halt
the collapse. On the other hand, for $K_{14} \leq 6$, 
any star does not collapse to a black hole since the mass is not
large enough. The detail on the criterion of the formation
of black hole is also described in the companion paper \cite{SS3}.

\subsection{Method of axisymmetric simulation}

During the collapse, the maximum density increases from
$10^{10}~{\rm g/cm^3}$ to $\sim 10^{15}~{\rm g/cm^3}$ in the neutron
star formation and to $\gg 10^{15}~{\rm g/cm^3}$ in the black hole
formation.
This implies that the characteristic length scale of the system varies by
a factor of $\agt 100$. In the early phase of the collapse which 
proceeds in a nearly homologous manner, we may follow the collapse 
with a relatively small number of grid points by moving the outer
boundary inward while decreasing the grid spacing, 
without increasing the grid number by a large factor. 
As the collapse proceeds, the central region shrinks more rapidly than 
the outer region does and, hence, a better grid resolution is 
necessary to accurately follow such a rapid collapse in the central region. 
On the other hand, the location of the outer boundaries
should not be changed by a large factor to avoid discarding the matter
in the outer envelopes. 

To compute such a collapse accurately while saving the CPU time efficiently,
a regridding technique as described and used in \cite{SS,SS2} is adopted. 
The regridding is carried out whenever the characteristic radius
of the collapsing star decreases by a factor of a few. At each regridding, 
the grid spacing is decreased by a factor of 2. 
All the quantities in the new grid are calculated
using the cubic interpolation. 
To avoid discarding the matter in the outer region, we also increase the 
grid number at the regridding, keeping a rule that 
the discarded baryon rest-mass has to be less than 1\% of the total. 

Specifically, $N$ and $L$ in the present work 
are chosen using a relativistic gravitational potential
defined by $\Phi_c \equiv 1 -\alpha_c~ (\Phi_c>0)$, which  
is $\sim 0.01$ at $t=0$. Here, $\alpha_c$ denotes the
central value of the lapse function. 
Since $\Phi_c$ is approximately proportional to $M/R$, 
$\Phi_c^{-1}$ can be used as a measure of the characteristic 
length scale for the regridding. Typically, 
the value of $N$ is chosen monitoring the magnitude
of $\Phi_c$ in the following manner; 
for $\Phi_c \leq 0.04$, we set $N=420$; 
for $0.04 \leq \Phi_c \leq 0.1$, we set $N=700$; 
for $0.1 \leq \Phi_c \leq 0.2$, we set $N=1200$; and 
for $0.2 \leq \Phi_c $, we set $N=1800$, and keep 
this number until the termination of the simulations.
Note that at $t=0$, the equatorial radius is covered by 400 grid points
in this case. With this setting, 
the total discarded fraction of the baryon rest-mass
which is located outside new regridded domains is $\alt 1\%$.
The grid spacing in $N=1800$ is $\sim 0.6$ km for differentially
rotating initial condition and $\sim 0.6$--0.8 km for
rigidly rotating cases. 
A previous work \cite{SS2} illustrates that with these grid resolutions 
a convergent result is obtained for most cases. 

Nevertheless, we should be very careful in judging black hole formation 
since the criterion for the black hole formation near a threshold 
depends sensitively on the 
strength of shocks that are formed when the density around the central
region exceeds $\rho_{\rm nuc}$. The shocks in
numerical simulations in general become stronger with improving the grid
resolutions. This implies that a black hole may be spuriously formed
in a coarse grid resolution in which the strength of the
shocks is underestimated.
To avoid such misjudging, in the case that a black hole is likely
to be formed,
we perform simulations with a finer grid resolution as follows; 
for $\Phi_c \leq 0.04$, we set $N=620$; 
for $0.04 \leq \Phi_c \leq 0.1$, we set $N=1020$; 
for $0.1 \leq \Phi_c \leq 0.2$, we set $N=1700$; and 
for $0.2 \leq \Phi_c $, we set $N=2500$. 
Note that at $t=0$, the equatorial radius is covered by 600 grid points
in this case. If we find a convergent result on the black hole formation 
in both resolutions, we judge that the black hole is formed. 

Simulations for each model with the typical grid resolution 
are performed for 40,000--50,000 time steps. 
The required CPU time for one model is about 20 hours using 4
processors of FACOM VPP5000 at the data processing center
of National Astronomical Observatory of Japan, and
about 10 hours using 8 processors of NEC SX6 at the
data processing center of ISAS in JAXA. 

\subsection{Method of three-dimensional simulation}

Since computational resources are restricted and
we cannot take the grid number per one direction as large as
that in the axisymmetric case, it is not a good idea to perform 
three-dimensional simulations from the initial conditions 
with $\rho_{\rm max} =10^{10}~{\rm g/cm^3}$. 
To save computational time, we always follow 
the early stage of the collapse using the axisymmetric code. 
After the collapse proceeds sufficiently, we 
change to the three-dimensional code.
In preparing the initial conditions of three-dimensional computations, 
numerical results of the axisymmetric simulations are used. 
In the early stage of collapse 
at which the value of $\beta$ of the collapsing star is not still
very large ($\alt 0.2$), 
nonaxisymmetric dynamical instabilities will not be induced. 
For highly differentially rotating cases, nonaxisymmetric instabilities 
could be induced even with a low value of $\beta$ \cite{SKE}.
However in such cases, the growth time of the nonaxisymmetric instabilities
would be much longer than the collapse time scale \cite{SKE}. Therefore, 
the method that we adopt is appropriate. 

Specifically, the initial condition for the three-dimensional
simulations is prepared when the central value of the lapse function becomes
$\alpha_c = 0.8$ in the axisymmetric simulations.
(For some case in which the minimum value of $\alpha_c$ is slightly 
larger than 0.8, we choose the value as 0.85.) 
Since our major purpose in 
the three-dimensional simulations is to investigate
the nonaxisymmetric dynamical stability of the collapsing star, 
we add a nonaxisymmetric density perturbation to 
the axisymmetric state. Associated with this change, 
the metric should be also perturbed, but we do not know how to do. 
For this reason, we adopt a very simple method for
setting the initial conditions as follows. 

First, we note that for $\alpha_c \geq 0.8$, the magnitude of 
$\tilde \gamma_{ij}-\delta_{ij}$ is very small ($\ll 0.01$)
for all the components, and hence,
the spatial hypersurface is approximately conformally flat. 
Also, the trace of the extrinsic curvature is nearly equal to zero
because of our choice of the slicing condition. 
Thus, in setting the initial conditions of the 
three-dimensional simulations, we assume that the
three-hypersurface is conformally flat and $K_k^{~k}=0$ for simplicity.
Then, we determine the conformal factor and the trace-free
extrinsic curvature using the constraint equations. 
In this case, for a solution of the constraint equations, 
we only need to extract $\rho_*$, $\hat e$, and $\hat u_i$ 
from the numerical results of the axisymmetric simulations in 
the following method. 

In the first step, we solve the momentum constraint equation using the
York's procedure \cite{york}. Setting 
\beqn
\hat A_{ij}\equiv \tilde A_{ij} \psi^6 = \pa_i W_j + \pa_j W_i
- {2 \over 3}\delta_{ij} \pa_k W_k, 
\eeqn
where $W_k$ is a three-vector, the momentum constraint is written as
\beqn
\Delta_{\rm flat} W_i + {1 \over 3}\pa_i \pa_j W_j = 8\pi \rho_* \hat u_i, 
\eeqn
where $\Delta_{\rm flat}$ is the flat Laplacian. Since 
$\rho_*$ and $\hat u_i$ are given, this equation is solved 
by a standard procedure (e.g., \cite{gr3d}) to give the tracefree
part of the extrinsic curvature. 

In the next step, the Hamiltonian constraint equation is solved. In the
conformally flat spatial hypersurface, 
the equation for the conformal factor $\psi$ is written as 
\beqn
\Delta_{\rm flat} \psi =-2\pi \rho_* \hat e \psi^{-1}
-{\psi^{-7} \over 8} \hat A_{ij} \hat A^{ij}. 
\eeqn
Since $\rho_*$ and $\hat e$ were given, and $\hat A_{ij}$ was 
already computed in the first step, $\psi$ is also computed
by a standard procedure often used in the initial value problem. 

The simulations were performed using a fixed uniform grid 
and assuming reflection symmetry with respect to the equatorial plane. 
The typical grid size is $(2N+1, 2N+1, N+1)$ for $(x, y, z)$, and
we adopt $N=156$, 188, and 220. 
The grid covers the region $-L \leq x \leq L$, $-L \leq y \leq L$, and 
$0 \leq z \leq L$ 
where $L$ is the location of the outer boundaries along each axis.
For a given model, we take the identical value of 
$L$ irrespective of the value of $N$. 
The grid spacing $\Delta x=L/N$ is chosen to be larger than that
adopted in the corresponding axisymmetric simulation because of
restricted computational resources for the three-dimensional case.
In the case of $N=156$, we choose the grid spacing
twice as large as that of the corresponding axisymmetric simulations.
The typical computation is performed with $N=188$, and 
to check the convergence, the value of $N$ is varied.
For models in which a bar-mode instability sets in,
simulations are performed with $N=220$. 

The value of $L$ is much smaller than that of the axisymmetric
simulation. This implies that we discard the matter located in the
outer region of the collapsing core. Specifically, we discarded the
matter outside a sphere of radius $\approx r_0$ by the rule 
\beqn
\rho_*({\rm new}) =\rho_*({\rm axisymmetric})
\times {1 \over e^{(r-r_0)/\delta r}+1}, \label{eq54}
\eeqn
where $r_0 \approx 0.95L$ and $\delta r =2 \Delta x$. 
In this method, the fraction of the discarded baryon mass located 
for $r > r_0$ is about 10--20\% (compare Tables II and IV). 

In this paper, we focus primarily on the dynamical stability
against $m=2$ bar-mode deformation, 
since it is expected to be the fastest growing mode.
Specifically, we superimpose a density perturbation in the form 
\beqn
\rho_*=\rho_*({\rm new})\biggl(1+0.4{x^2-y^2 \over L^2}\biggr). 
\label{barmode}
\eeqn
To check that the bar-mode perturbation grows for dynamically unstable
models even when the initial configuration is nearly axisymmetric,
we also performed simulations without adding nonaxisymmetric
perturbation besides random numerical noises 
for selected unstable models. We found that in such cases,
the bar-mode perturbation indeed grows 
although it takes more computational time to follow the growth. 

In the case that the equations of state are very soft, 
the degree of the differential rotation is very high, and
the value of $\beta$ is large enough ($\agt 0.14$) for 
a collapsed star, $m=1$ modes may grow faster than the $m=2$ mode
\cite{M1,M11}. In the formation of neutron stars
in which $\rho_{\rm max} > \rho_{\rm nuc}$, 
the equation of state is stiff, and hence, the $m=1$ mode may not
be very important. On the other hand, 
in the formation of oscillating stars, 
equations of state can be soft for $\rho_{\rm max} < \rho_{\rm nuc}$.
However, the value of $\beta$ in such phase of 
subnuclear density are not very large. 
Thus, it is expected that even if the $m=1$ mode becomes unstable, 
the perturbation may not grow as significantly 
as found in \cite{M1,M11}. 
Hence, we do not pay particular attention to this mode in this paper. 
Since nonaxisymmetric numerical noises are randomly 
included at $t=0$, in some models, the $m=1$ mode 
grows as found in Sec. V. However, the amplitude of the
perturbation is indeed not as large as that for $m=2$. 

Since we assume the conformal flatness in spite of the 
fact that the conformal three-metric is slightly different from zero 
in reality, a small systematic error is introduced in setting
the initial data. Moreover, we discard the matter located in the outer 
region of the collapsing core according to Eq. (\ref{eq54}).
This could also introduce a 
systematic error. To confirm that the magnitude of such error induced by
these approximate treatments is small, 
we compare the results in the three-dimensional simulations with those in
the axisymmetric ones. We have found that 
the results agree well each other and 
the systematic error is not very large. 
This will be illustrated in Sec. V  (cf. Fig. 13). 

Simulations for each model with the grid size (441, 441, 221) ($N=220$) 
were performed for about 15,000 time steps. 
The required CPU time for computing one model is about 30 hours using 32 
processors of FACOM VPP 5000 at the data processing center 
of National Astronomical Observatory of Japan.

\section{Numerical results of axisymmetric simulations}

\begin{figure}[htb]
\vspace*{-4mm}
\begin{center}
\epsfxsize=3.in
\leavevmode
(a)\epsffile{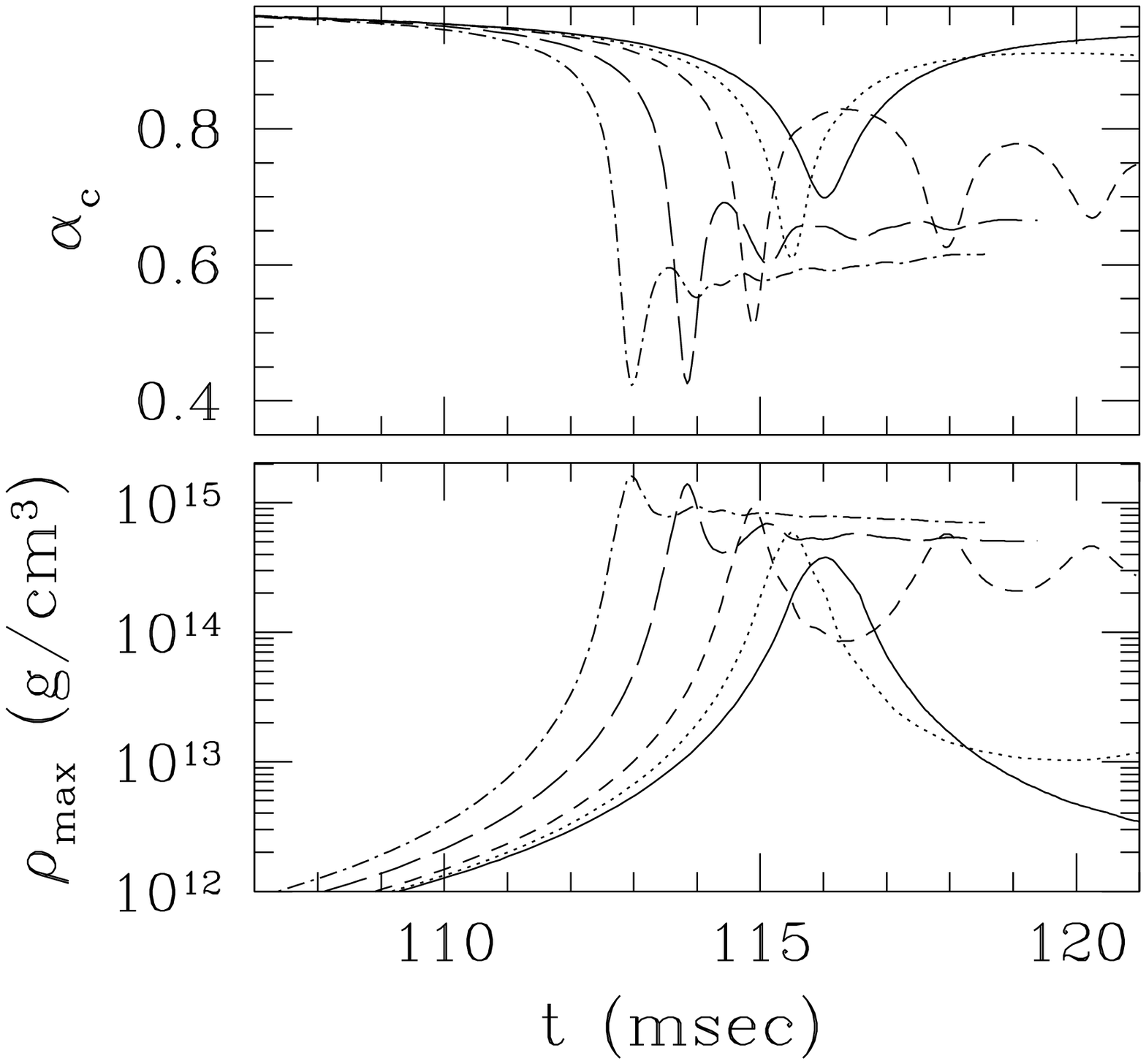}
\epsfxsize=3.in
\leavevmode
~~~(b)\epsffile{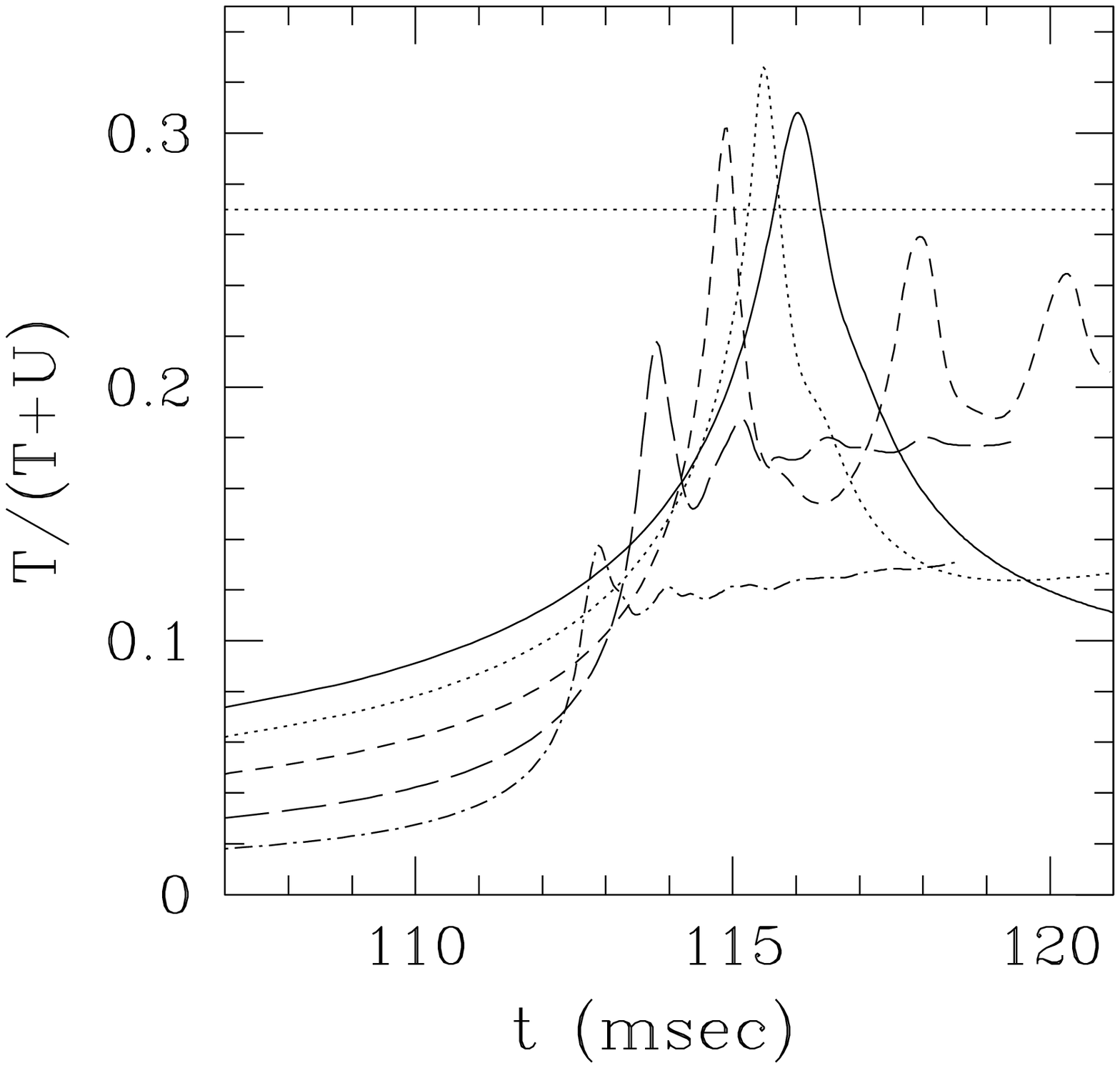}
%\vspace*{-4mm}
\caption{Evolution of (a) $\alpha_c$ and $\rho_{\rm max}$
and (b) $\beta=T/(T+U)$ for models M7c1 (solid curve),
M7c2 (dotted curve), M7c3 (dashed curve), M7c5 (long-dashed curve),
and M7c6 (dotted-dashed curve). 
The dotted horizontal line denotes $\beta=0.27$. 
\label{FIG1}}
\end{center}
\end{figure}

\begin{figure}[htb]
\vspace*{-4mm}
\begin{center}
\epsfxsize=3.in
\leavevmode
(a)\epsffile{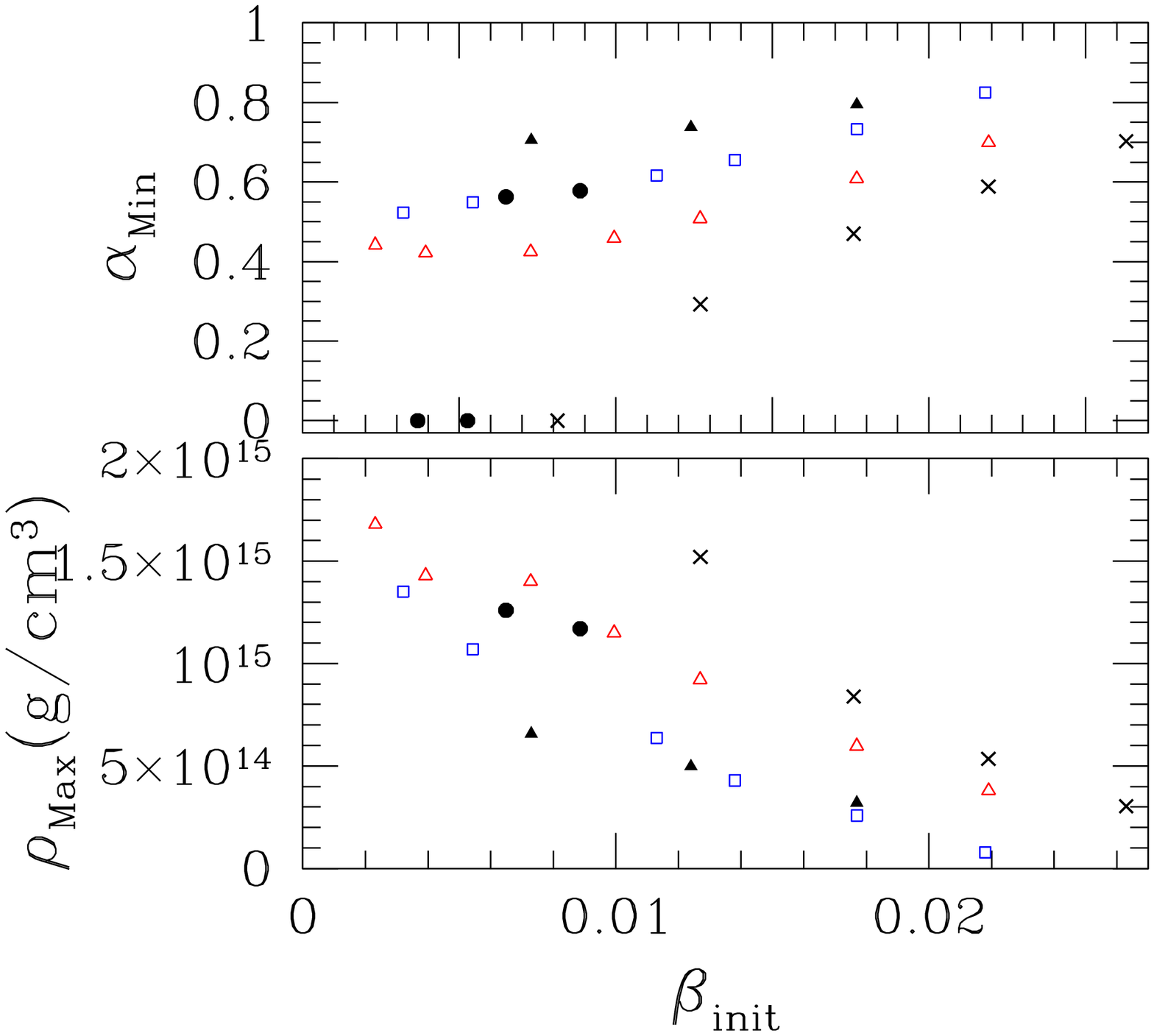}
\epsfxsize=3.in
\leavevmode
~~~(b)\epsffile{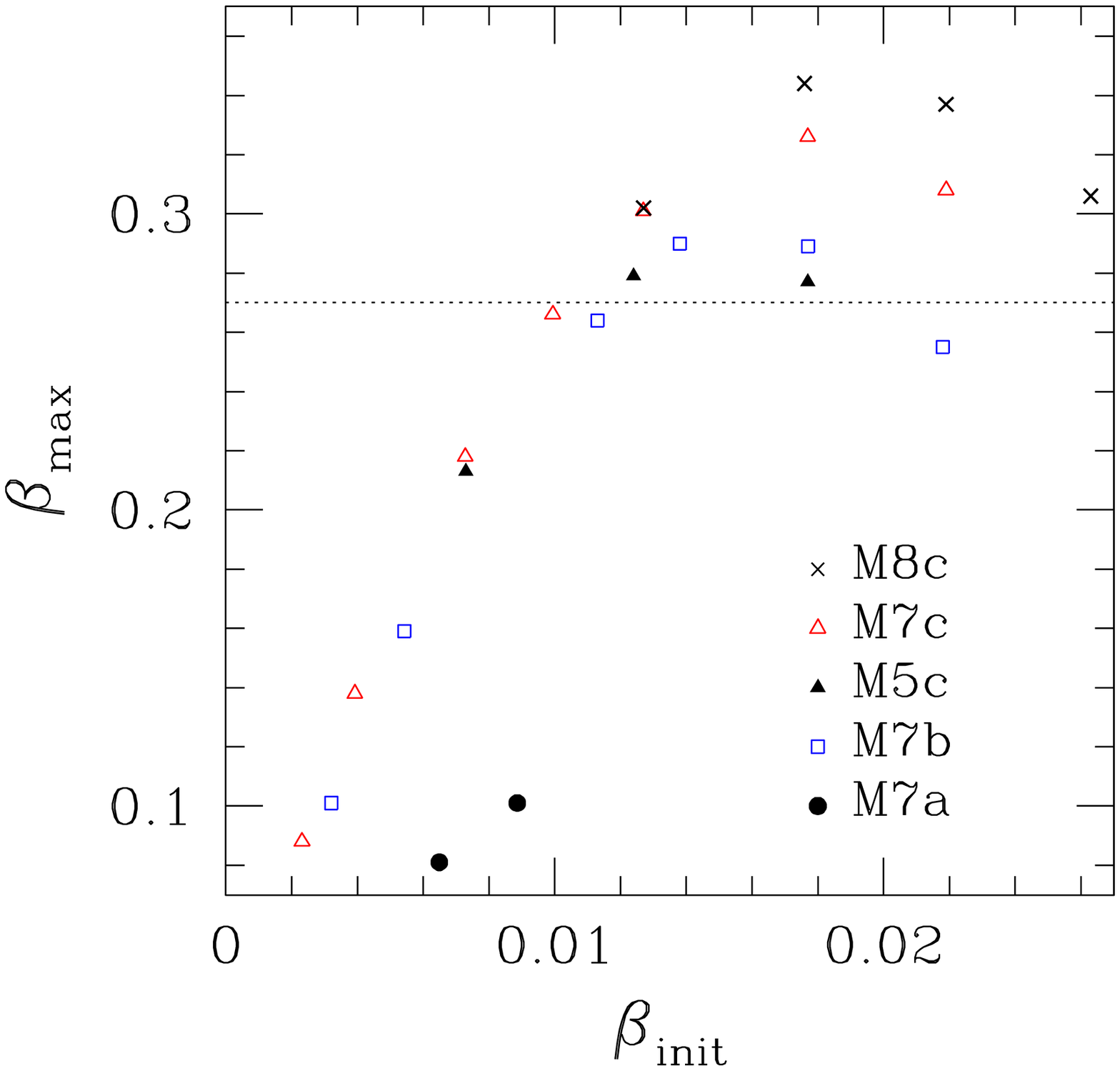}
%\vspace*{-4mm}
\caption{(a) $\alpha_{\rm Min}$ and $\rho_{\rm Max}$ 
and (b) $\beta_{\rm max}$ for various values of 
$\beta_{\rm init}$. In both panels, the solid triangle, 
the solid circles, open squares, 
open triangles, and crosses denote models M5c, M7a, M7b, M7c, and M8c,
respectively. 
The dotted line in panel (b) denotes $\beta_{\rm max}=0.27$. 
\label{FIG2}}
\end{center}
\end{figure}

\subsection{Outcomes}

In the last column of Table II, we summarize the outcomes of 
stellar core collapse in the axisymmetric simulations
for $\Gamma_1=1.3$ and $\Gamma_2=2.5$. 
They are divided into three types: Black hole, neutron star, 
and oscillating star for which the maximum density
inside the star is not always larger than $\rho_{\rm nuc}$.
For given values of $K_0 (\geq 7\times 10^{14}~{\rm cgs})$ and $A$,
a black hole is formed when the initial value of $\beta$
(hereafter $\beta_{\rm init}$)
is smaller than critical values that depend on $A$.
As described in Sec. III A, $\beta$ in the collapse is defined by
\beqn
\beta \equiv {T \over T+U}. \label{beta1}
\eeqn
In the dynamical spacetime with $M_* \approx M$ for $\Gamma=4/3$,
$W$ would be approximately written as 
\beqn
W \approx U+T+T_{\rm other},
\eeqn
where $T_{\rm other}$ denotes a partial kinetic energy obtained 
by subtracting the rotational kinetic energy from the total.
Thus, $T/W$ should be approximated by $T/(U+T+T_{\rm other})$, but 
we do not know how to appropriately define $T_{\rm other}$.
Fortunately, it would be much smaller than $T$ at the initial state, 
at the maximum compression at which the spin of the collapsing
star becomes maximum, and in a late phase when the outcome relaxes to a
quasistationary state. 
This implies that using the definition of (\ref{beta1}), 
the maximum value and a final relaxed value
of $\beta$ will be computed approximately. 
In other phases, $\beta$ computed by Eq. (\ref{beta1})
gives an overestimated value.  

In Fig. 1, we show the evolution of 
the central value of the lapse function ($\alpha_{c}$), 
the maximum value of the density ($\rho_{\rm max}$), and $\beta$
for models M7c1, M7c2, M7c3, M7c5, and M7c6. In the following,
we denote the maximum density and minimum value of the lapse 
achieved in the whole evolution as 
$\rho_{\rm Max}$ and $\alpha_{\rm Min}$, respectively.
On the other hand, the maximum density at a given time is denoted by 
$\rho_{\rm max}$.

Figure 1 shows that for most cases, the value of $\rho_{\rm Max}$ 
becomes larger than $\rho_{\rm nuc}$. 
However, with the increase of $\beta_{\rm init}$, 
it decreases significantly. Also, for
several cases, $\rho_{\rm max}$ 
drops below $\rho_{\rm nuc}$ soon after it reaches the maximum. 
Such oscillating stars for which the values of $\rho_{\rm max}$ oscillate 
between $\rho_{1}(>\rho_{\rm nuc})$ and $\rho_{2}(<\rho_{\rm nuc})$ 
are referred to type O-A in Table II. On the other hand, 
if $\beta_{\rm init}$ is not very large and neither is the 
maximum value of $\beta$ (hereafter $\beta_{\rm max}$), 
a neutron star or a black hole is formed. 
Here, formation of a neutron star implies that $\rho_{\rm max}$ 
achieved after the stellar collapse is always 
larger than $\rho_{\rm nuc}$. Formation of a black hole 
implies that we confirm the formation of apparent horizon. 

In Figs. 2(a) and (b), we show $\alpha_{\rm Min}$, $\rho_{\rm Max}$,
and $\beta_{\rm max}$ for various values of $\beta_{\rm init}$. 
For $\beta_{\rm init} \agt 0.02$ with $K_0 = 7\times 10^{14}$ cgs, 
$\rho_{\rm Max}$ is smaller than $\rho_{\rm nuc}$,
and the resulting star is quasiradially oscillating 
with the subnuclear density. Such stars are referred to as O-B in Table II. 

Figure 1 and Table II show that initial high degrees of differential 
rotation with $A=0.25$ and 0.1 
have an effect for preventing black hole formation.
(Compare the parameters among the models with $K_0=7\times 10^{14}$ cgs.) 
For the rigidly rotating cases, the stars with
$\beta_{\rm init} \alt 0.005$ 
collapse to a black hole. On the other hand, 
the stars with $\beta_{\rm init} \sim 0.003$ do not collapse to a black hole 
but form a neutron star for $A=0.1$ and 0.25. 
This is simply because 
the stars with such high degrees of differential rotation
have a large centrifugal force near the rotational axis, and
hence, even in the case that the global value $\beta_{\rm init}$ is
small, the effective local value of the centrifugal force would be 
large enough to prevent cores from collapsing to a black hole.

\subsection{Evolution of $\beta$ for $\Gamma_1=1.3$
and $\Gamma_2=2.5$}

\begin{figure}[htb]
\vspace*{-5mm}
\begin{center}
\epsfxsize=2.8in
\leavevmode
(a)\epsffile{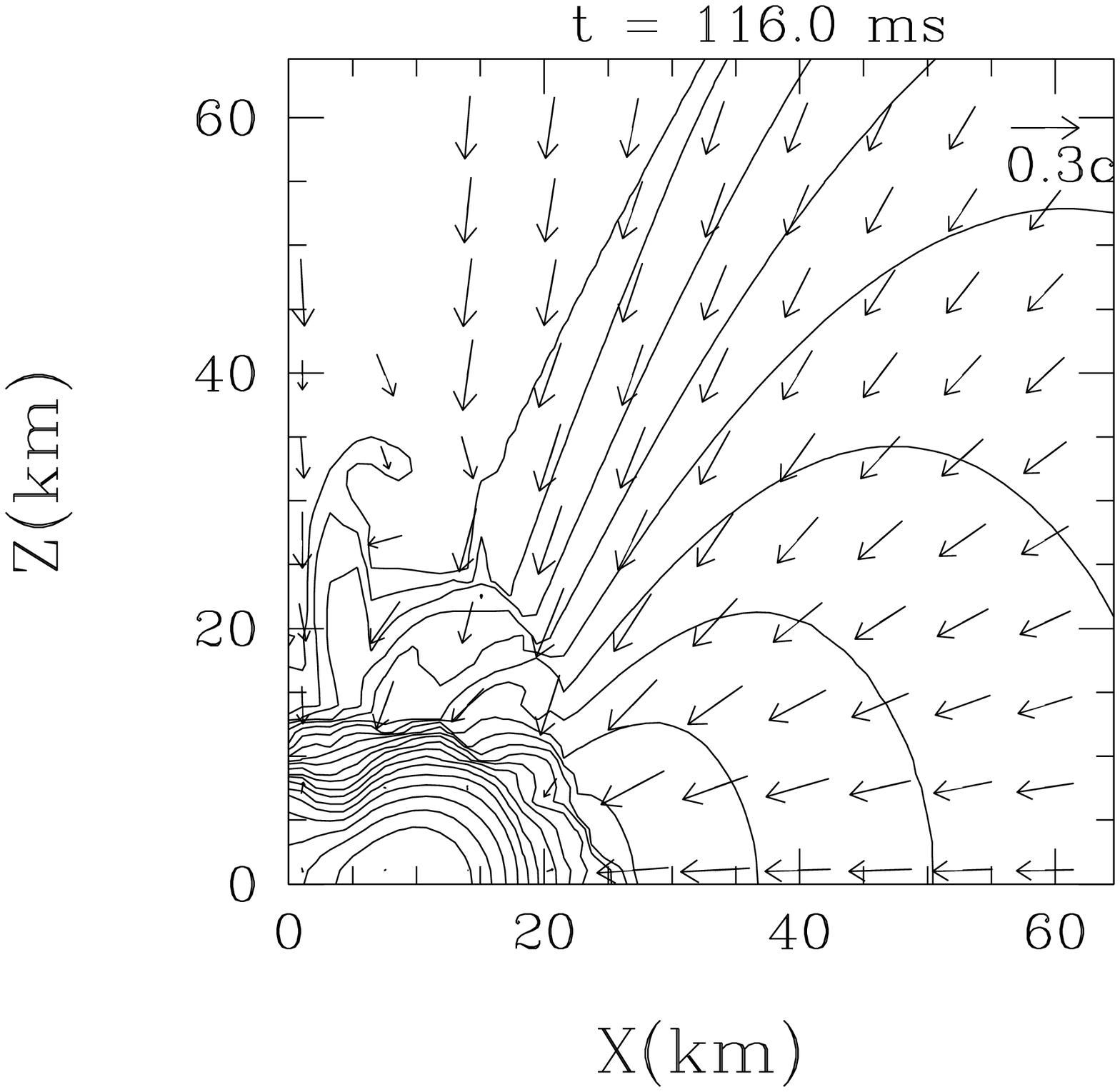}
\epsfxsize=2.8in
\leavevmode
(b)\epsffile{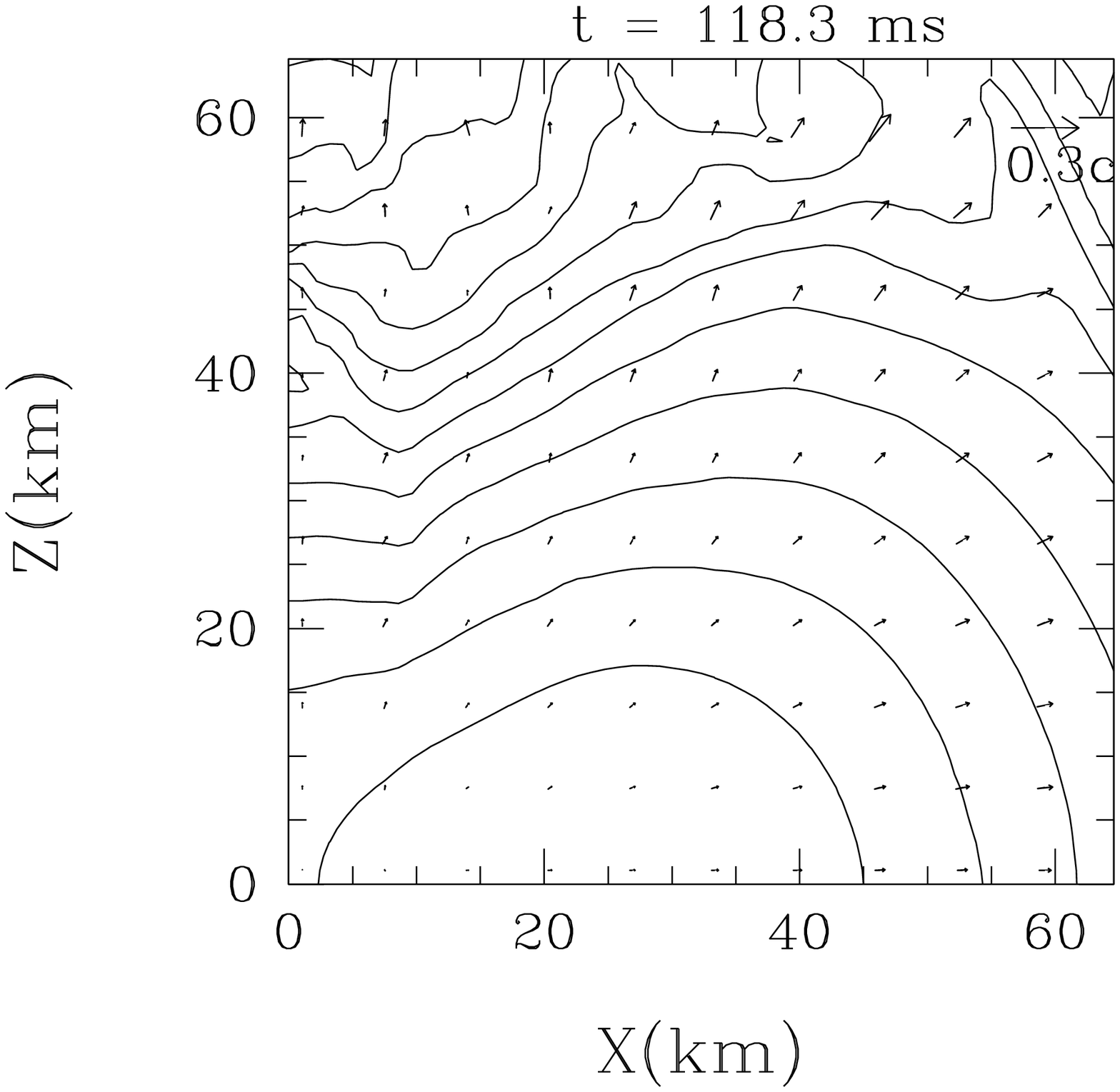}\\
\vspace*{-4mm}
\epsfxsize=2.8in
\leavevmode
(c)\epsffile{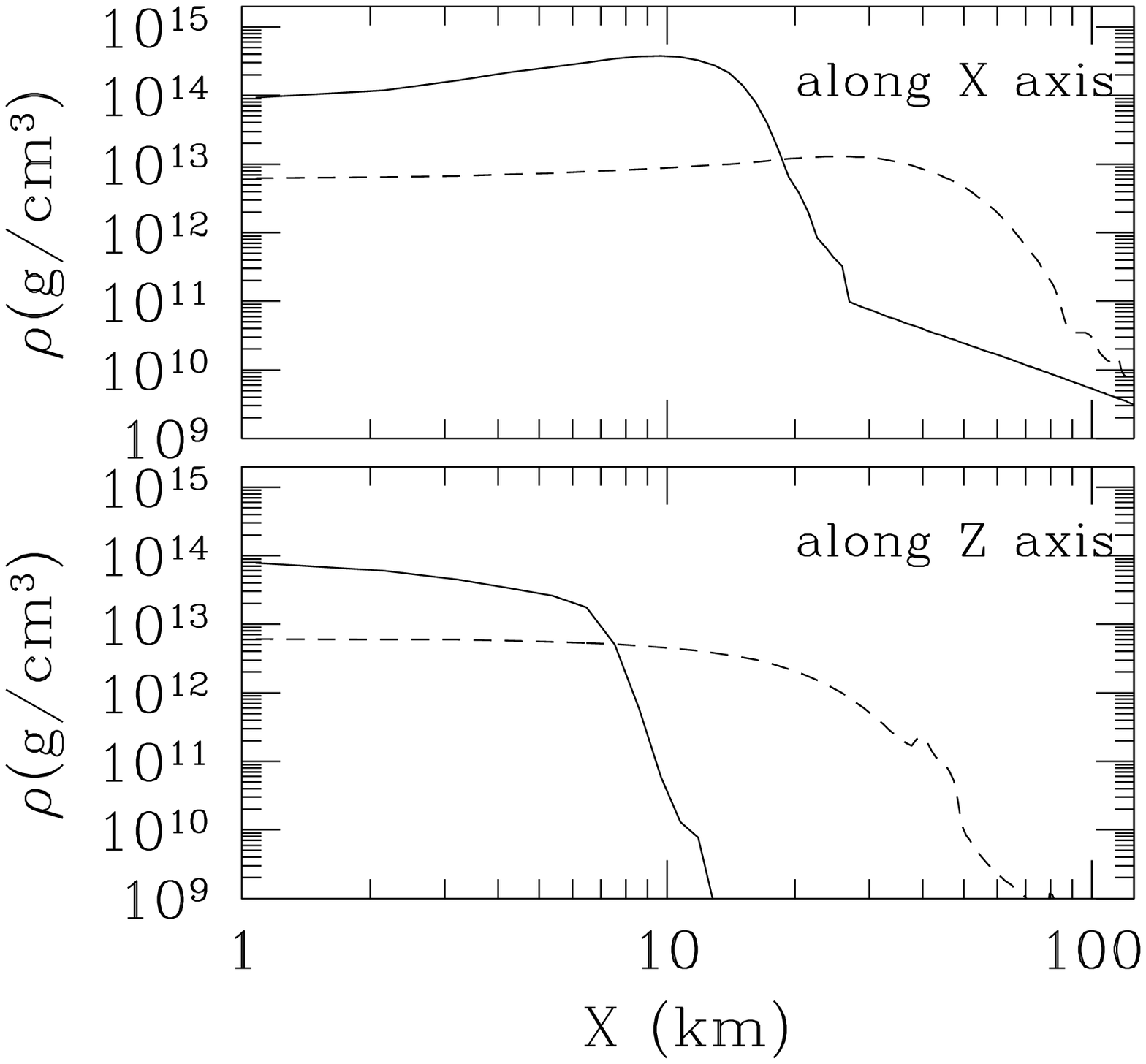}
\epsfxsize=2.8in
\leavevmode
(d)\epsffile{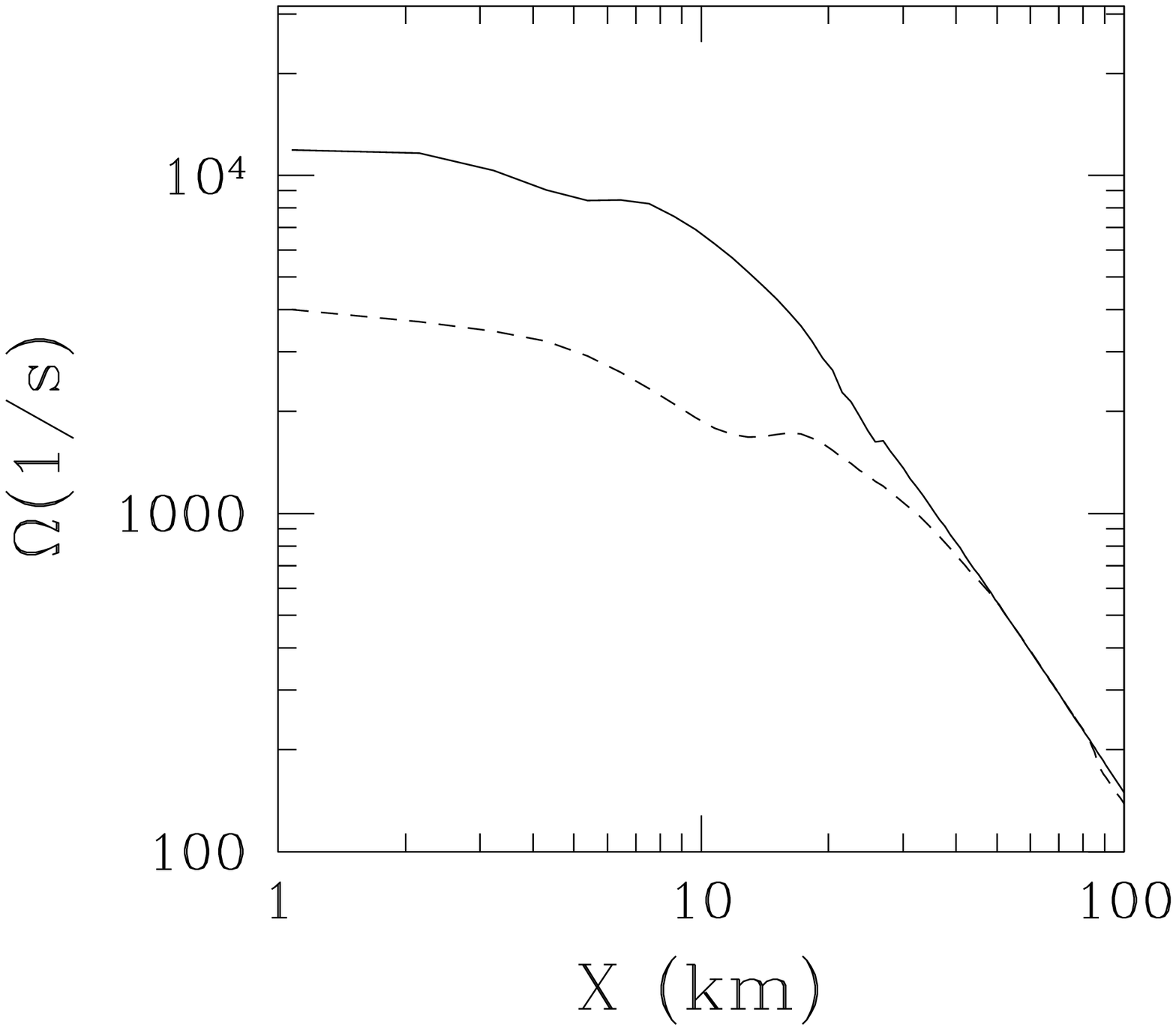}
%\vspace*{-4mm}
\caption{(a) and (b): 
The density contour curves for $\rho$ for model M7c1
at $t=116.0$ and 118.3 msec. 
The solid contour curves are drawn for $\rho/\rho_{\rm max}=e^{-0.3j}$ 
for $j=1, 2, 3,\cdots,20$ where $\rho_{\rm max}$ denotes the maximum
of $\rho$ at the given times, which are found from Fig. 1(a). 
Vectors indicate the local velocity field $(v^x,v^z)$, and the scale 
is shown in the upper right-hand corner.
(c) Density profiles along the $x$ and $z$ axes
at $t=116.0$ (solid curve) and 118.3 msec (dashed curve). 
(d) The same as (c) but for angular velocity
along the radial coordinate in the equatorial plane. 
\label{FIG3}}
\end{center}
\end{figure}

\begin{figure}[htb]
\vspace*{-5mm}
\begin{center}
\epsfxsize=2.8in
\leavevmode
(a)\epsffile{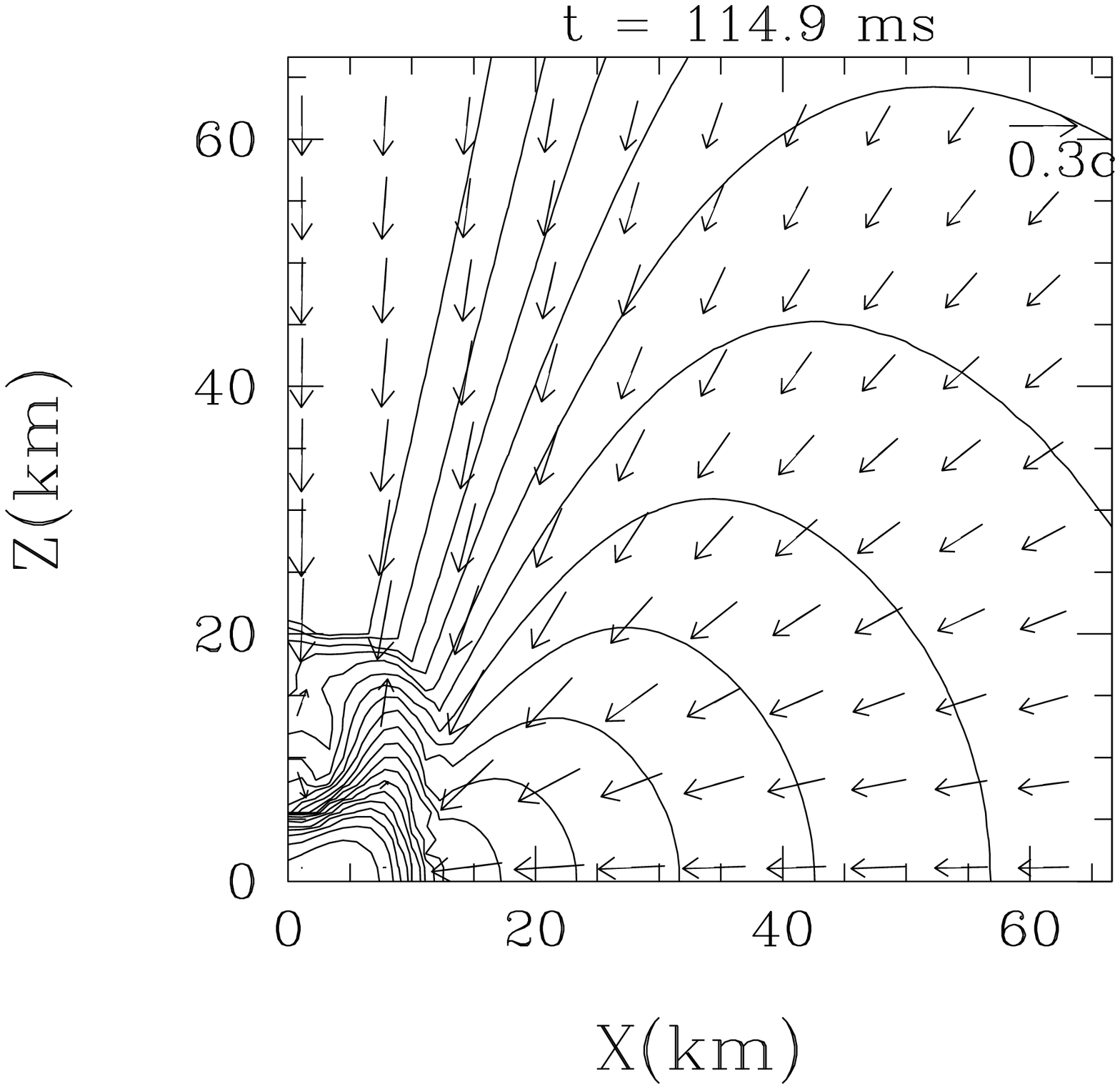}
\epsfxsize=2.8in
\leavevmode
(b)\epsffile{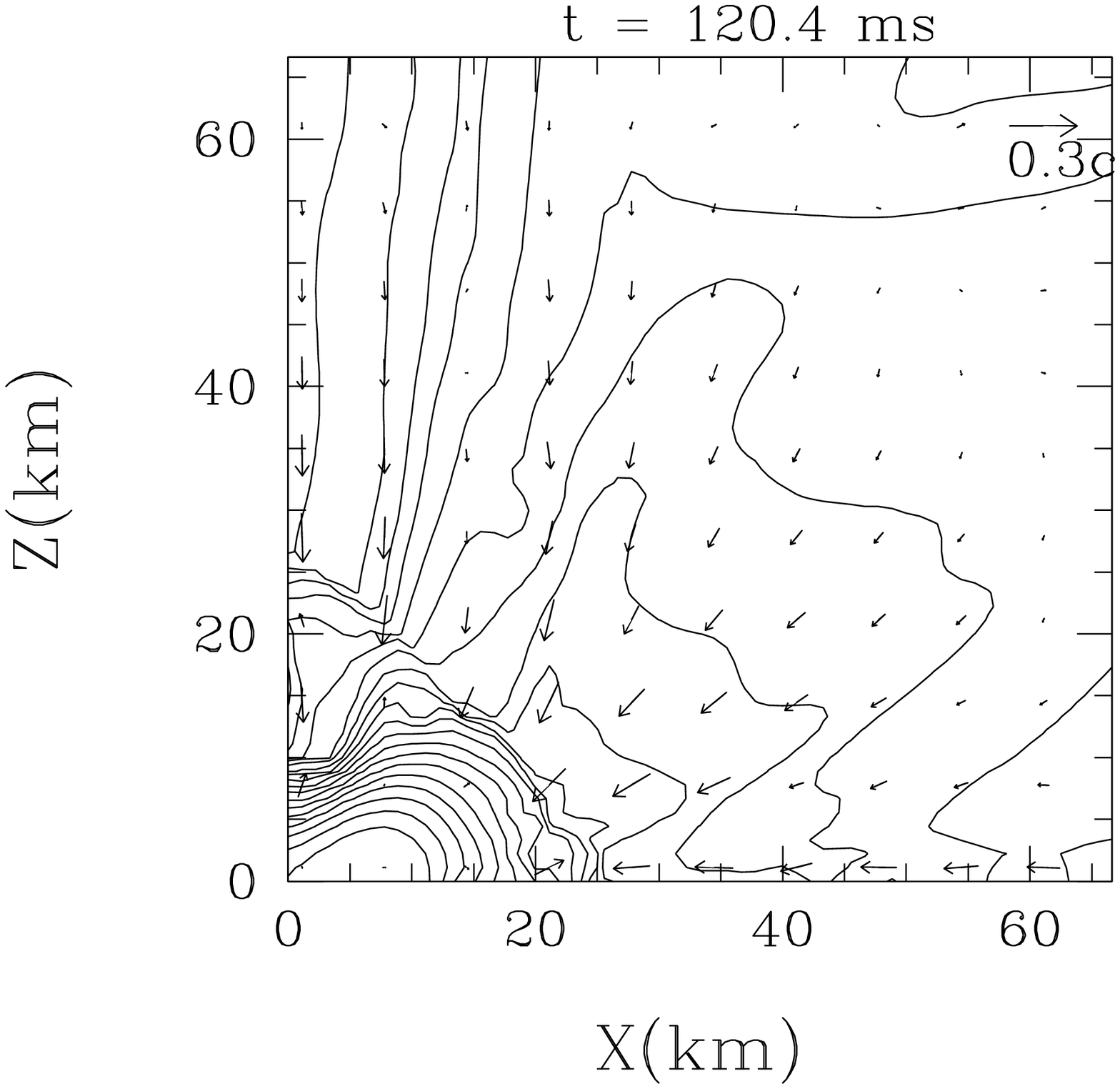}\\
\vspace*{-4mm}
\epsfxsize=2.8in
\leavevmode
(c)\epsffile{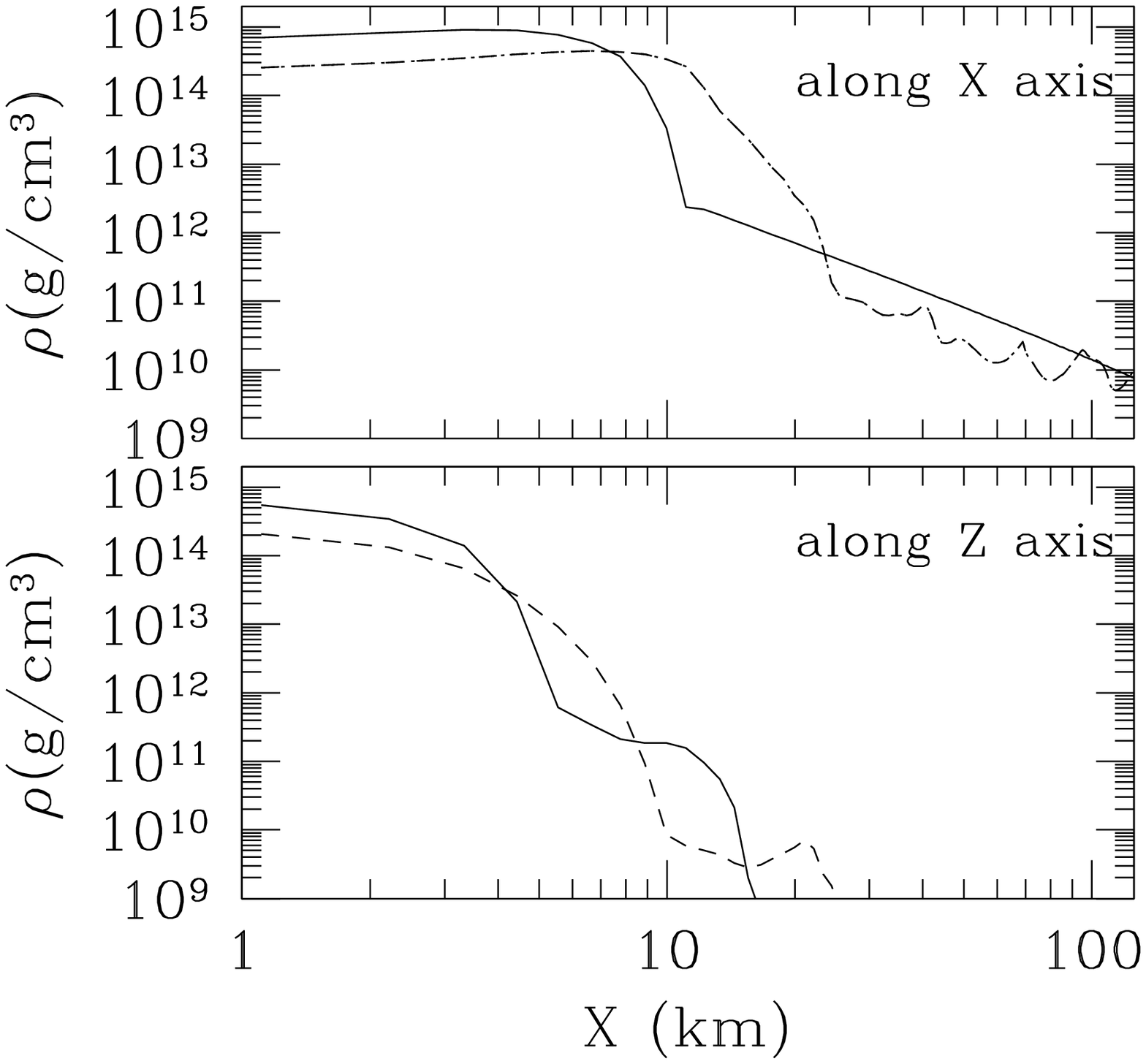}
\epsfxsize=2.8in
\leavevmode
(d)\epsffile{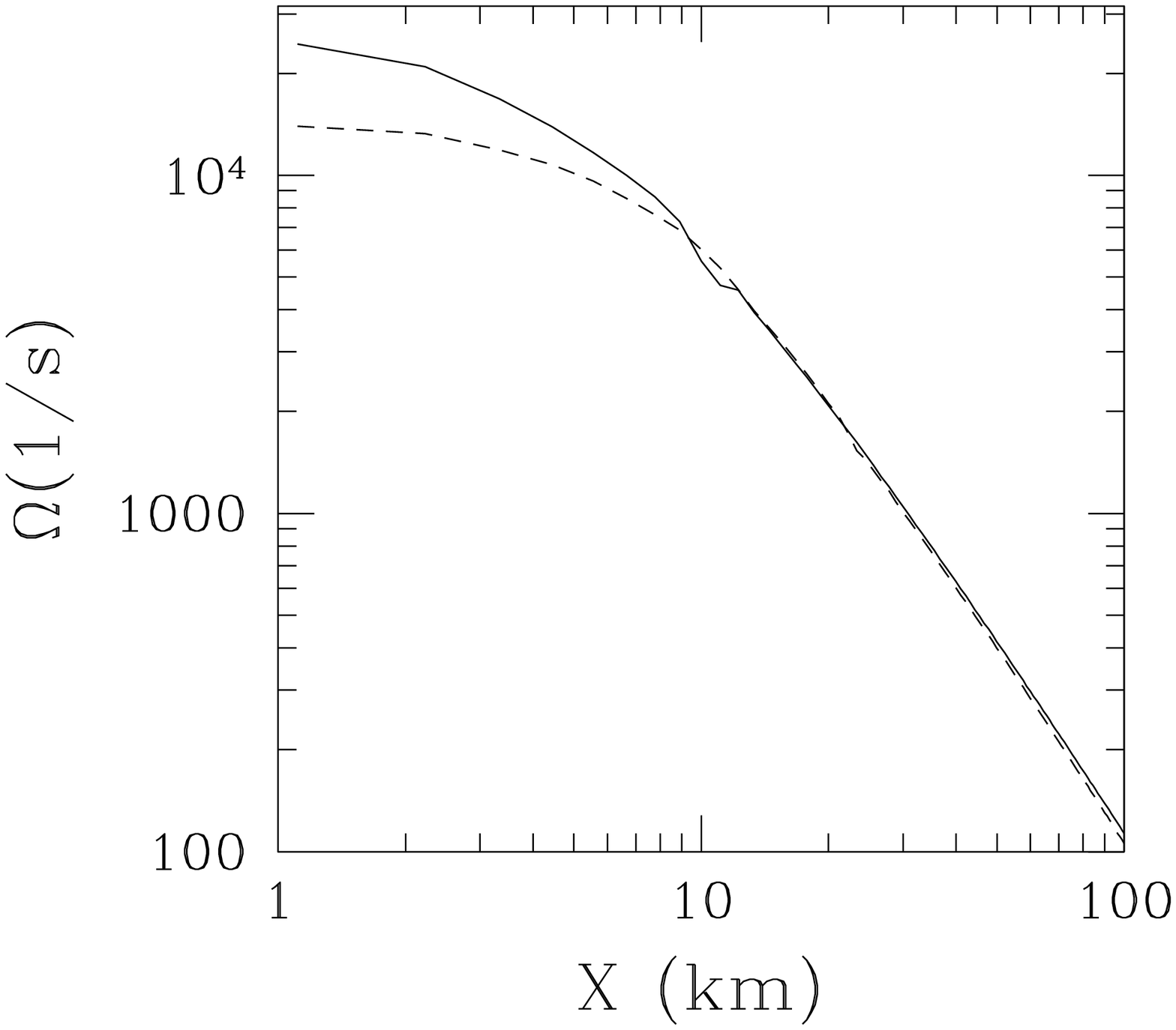}
%\vspace*{-4mm}
\caption{The same as Fig. 3(a)--(d) but for model M7c3 
at $t=114.9$ and 120.4 msec. 
The solid and dashed curves in panels (c) and (d) are drawn
for the corresponding time slices, respectively. 
\label{FIG4}}
\end{center}
\end{figure}

\begin{figure}[htb]
\vspace*{-5mm}
\begin{center}
\epsfxsize=2.8in
\leavevmode
(a)\epsffile{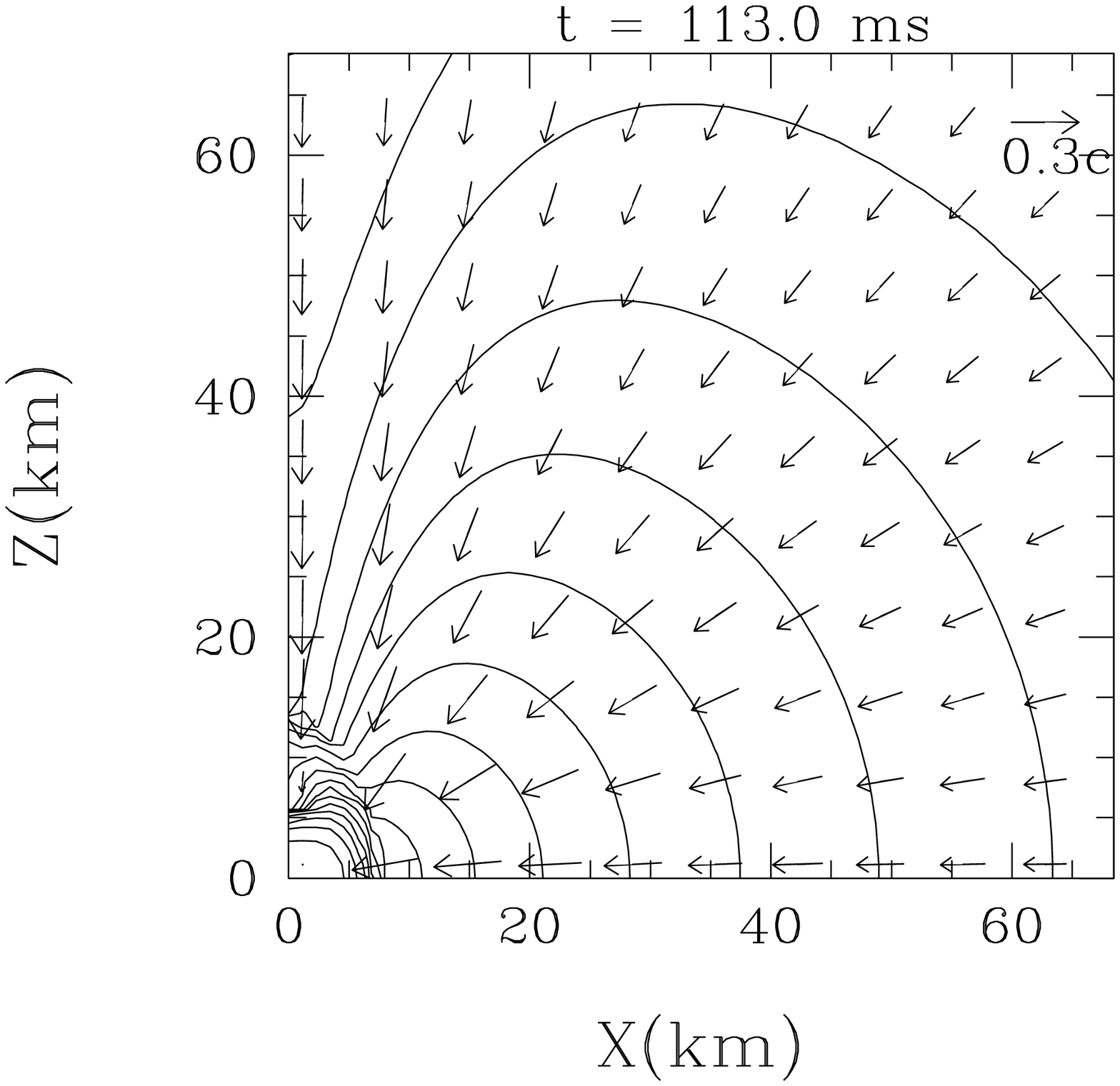}
\epsfxsize=2.8in
\leavevmode
(b)\epsffile{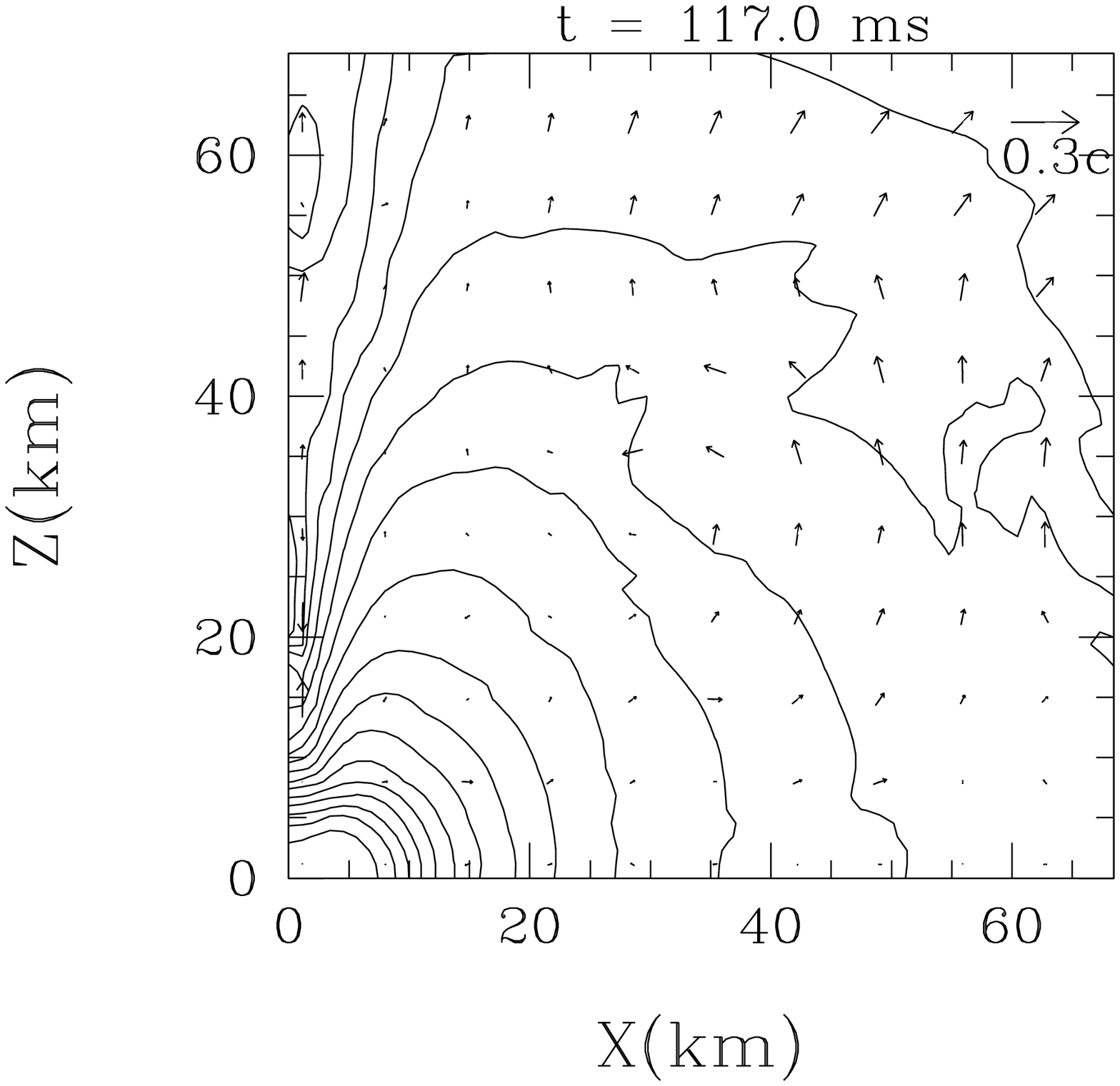}\\
\vspace*{-4mm}
\epsfxsize=2.8in
\leavevmode
(c)\epsffile{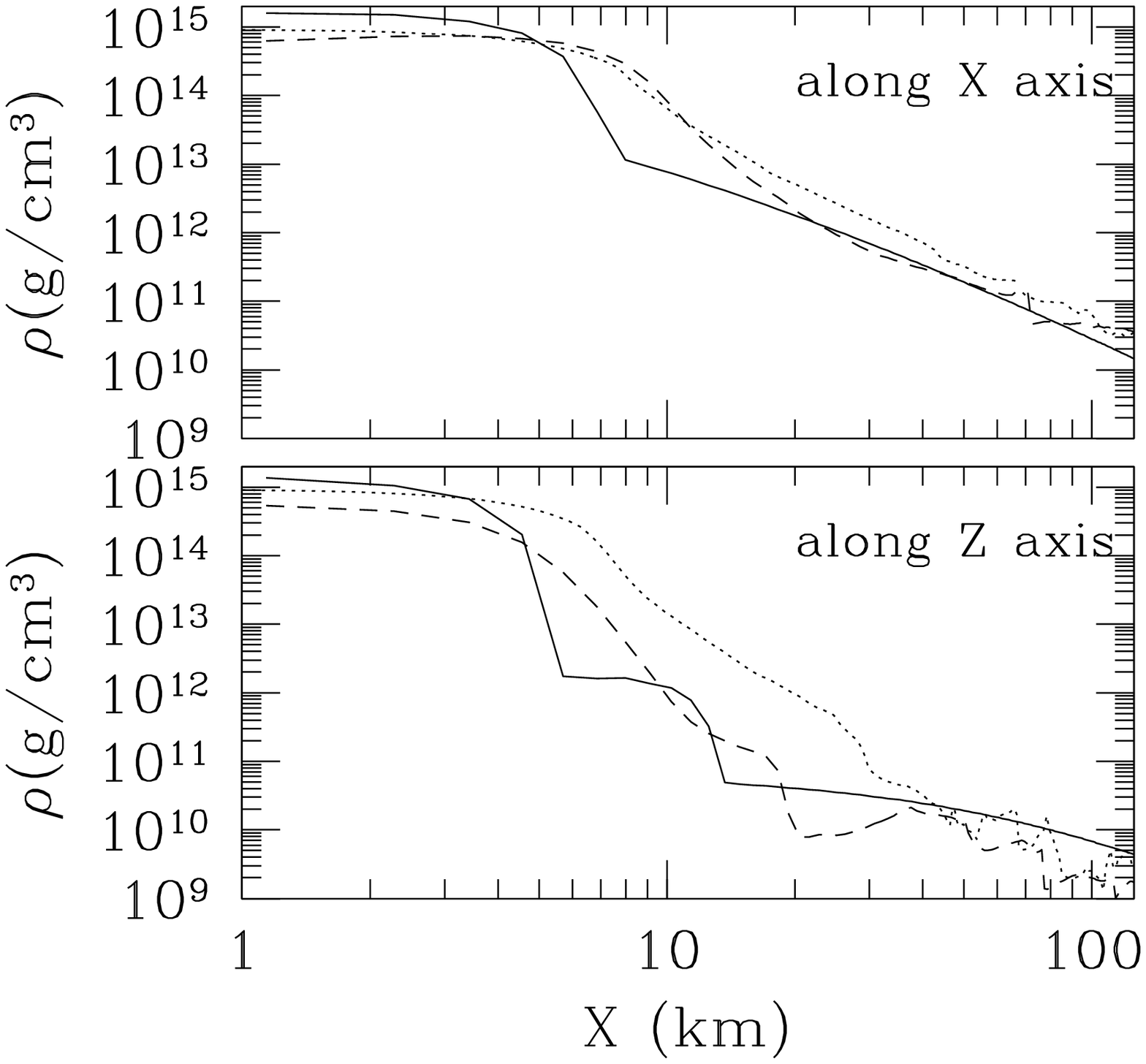}
\epsfxsize=2.8in
\leavevmode
(d)\epsffile{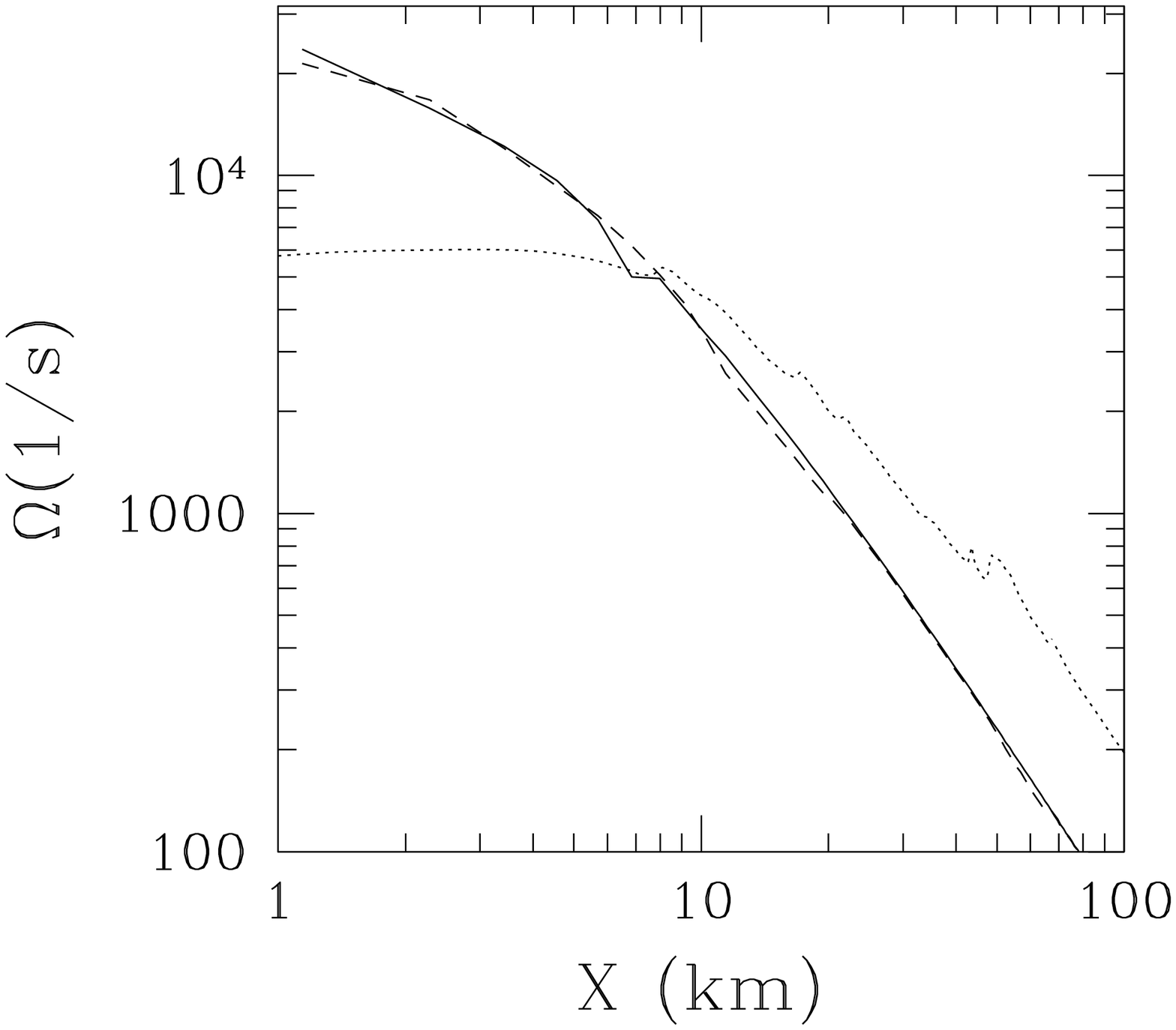}
%\vspace*{-4mm}
\caption{The same as Fig. 3(a)--(d) but for model M7c6 
at $t=113.0$ and 117.0 msec. 
The solid and dashed curves in panels (c) and (d) are drawn
for the corresponding time slices, respectively. 
For comparison, the results for model M7a2 at $t=121.3$ msec
are shown (dotted curves). 
\label{FIG5}}
\end{center}
\end{figure}

Figures 1 and 2 indicate that with the increase of $\beta_{\rm init}$, 
$\alpha_{\rm Min}$ ($\rho_{\rm Max}$) 
increases (decreases). The value of $\beta_{\rm max}$ 
is larger for the larger value of 
$\beta_{\rm init}$ as far as $\beta_{\rm init} \alt 0.02$. However, 
the amplification factor $\beta_{\rm max}/\beta_{\rm init}$ is smaller
for the larger value of $\beta_{\rm init}$. This is because in the collapse 
with the large values of $\beta_{\rm init}$, strong 
centrifugal force prevents the maximum value of compactness
(or maximum density or maximum value of the gravitational potential)
from being increased by a large factor.
Spin can be increased by a larger factor for 
a star which gains a larger compactness. 
Therefore, for stars of approximately identical mass,
the amplification factor $\beta_{\rm max}/\beta_{\rm init}$ should be
smaller for the larger value of $\beta_{\rm init}$. 

The typical value of $\beta_{\rm max}/\beta_{\rm init}$ is 10--20 for
rigidly rotating case and for differentially rotating cases
with $\beta_{\rm init} \agt 0.015$. Naive estimation predicts that
$T \propto J^2/(MR^2)$ and $W \propto M^2/R$, and hence,
$\beta \propto J^2/(M^3R)$. Thus, $\beta$ seems to 
be proportional to the inverse of 
the stellar radius in the condition that the mass and the angular
momentum of the system are conserved. Since the characteristic 
stellar radius changes by a factor of $\sim 100$  
during the collapse, we may predict that $\beta$ also 
increases by two orders of magnitude. However, this does not occur. 
The reason for the rigidly rotating case is that although the core radius 
decreases to $\sim 10$ km in the collapse, 
the outer region of the star which possesses a large fraction of
the angular momentum does not collapse to such a small radius
due to the strong centrifugal force. The reason for the 
highly differentially rotating cases with a high value of
$\beta_{\rm init} \agt 0.015$ is that the centrifugal force near the 
rotational axis is so strong that the collapse is halted before the 
stellar radius becomes $\sim 10$ km.
For low values of $\beta_{\rm init} \alt 0.01$, 
the rotational velocity in the outer region is small, 
and also, the centrifugal force in the central region is not as
strong as that for $\beta_{\rm init} \agt 0.015$. As a result, the
stellar components that enclose a large fraction of the angular momentum 
can collapse to small radii, and hence, $\beta$ can increase
by a factor of $\agt 30$. 

For given values of $K_0$ and $\beta_{\rm init}$, the value of 
$\beta_{\rm max}$ is larger for higher degrees 
of differential rotation. This suggests that stellar cores 
with a higher degree of differential rotation may be 
more subject to nonaxisymmetric dynamical instabilities. 
For the identical value of $\beta_{\rm init}$, $\beta_{\rm max}$ is larger 
for higher-mass stellar cores with a larger value of $K_0$
in the case of highly differentially rotating cores. 
(Compare, e.g., the solid triangle, the open triangle, and the cross
in Fig. 2(b); more specifically, compare the results for 
models M5c1, M7c2, and M8c3, for which the values of $\beta_{\rm init}$
are approximately the same as $0.0177$, but $\beta_{\rm max}$ is larger
for larger mass.) The reason for this behavior 
is that the stars of higher mass can reach a more 
compact state during the collapse, and as a result,
their spins can be increased by a larger factor and so can be $\beta$. 
On the other hand, for rigidly rotating cases, this feature is not
very outstanding. 

An interesting point is that the value of $\beta_{\rm max}$
has a maximum around $\beta_{\rm init} \sim 0.018$ for $A=0.1$ and 0.25: 
For $\beta_{\rm init} \alt 0.018$, $\beta_{\rm max}$ is an increase 
function of $\beta_{\rm init}$, reflecting the initial magnitude of the spin.
However, for $\beta_{\rm init} \agt 0.018$, $\beta_{\rm max}$ is
a decrease function. The reason is that the centrifugal force of 
the rotating stars with $\beta_{\rm init} \agt 0.018$ is so strong that the 
collapse is halted before the stellar core becomes compact enough. 
This feature is also reflected in Fig. 2(a) from which we find that
the value of $\alpha_{\rm Min}$ is an increase function of
$\beta_{\rm init}$. 

As reviewed in Sec. I, nonaxisymmetric dynamical instabilities 
of {\em rotating stars in equilibrium} set in when the value of
$\beta$ becomes larger than $\sim 0.27$. If we assume that 
the collapsing stars with $\beta_{\rm max} \agt 0.27$ are dynamically
unstable, Fig. 2(b) suggests that the conditions for the onset of the
instabilities will be the following: 
(i) the progenitor of the collapse should be
highly differentially rotating with $A \alt 0.25$; 
(ii) the progenitor has to be moderately 
rapidly rotating with $0.01 \alt \beta_{\rm init} \alt 0.02$; 
(iii) the progenitor star has to be massive enough. 

However, it should be kept in mind that
the condition $\beta_{\rm max} > 0.27$ is satisfied 
only for a few msec during the stellar collapse.
This indicates that if the growth time scale 
of nonaxisymmetric instabilities is not as short as a few msec, the 
system may remain nearly axisymmetric. Thus, the condition
$\beta_{\rm max} \agt 0.27$ does not have to be the criterion
for the onset of nonaxisymmetric dynamical instabilities in dynamical
systems. The examples are shown in Sec. V. 

\subsection{Profiles of density and angular velocity
for $\Gamma_1=1.3$ and $\Gamma_2=2.5$}

In \cite{TH}, Tohline and Hachisu illustrated that 
the stars with toroidal density profiles are
dynamically unstable against a bar-mode perturbation even if
$\beta$ is much smaller than 0.27. Also 
in \cite{SKE}, we indicated that not only the value of $\beta$ 
but also the degree of differential rotation is a key parameter 
for determining nonaxisymmetric stability of rotating stars.
Thus, here, we focus on the profiles of the density and the
rotational angular velocity of the outcomes in the stellar collapse. 

In Figs. 3--5, we display the snapshots of the density contour
curves, velocity vectors in the $x$-$z$ plane,
density profiles along $x$ and $z$ axes, and rotational 
angular velocity profiles as a function of the radial coordinate in
the equatorial plane at the time slices that the maximum completion is
achieved and the system relaxes to an approximately quasistationary state
for models M7c1, M7c3, and M7c6. These figures
clarify how the outcomes are changed with the varying $\beta_{\rm init}$ 
for (approximately) identical values of $A$ and $M$.  
Panels (a), (b), and (c) show that for the larger values
of $\beta_{\rm init}$, the shape of the outcome is more torus-like. 
Numerical studies for nonaxisymmetric dynamical instabilities 
in rapidly rotating stars in equilibrium have illustrated that 
torus-like stars are often unstable \cite{TH,CT,SKE}. 
This indicates that the models such as M7c1--M7c3 in which 
torus-like structures are formed are candidates for the onset of 
nonaxisymmetric dynamical instabilities even when the value of $\beta$ is
smaller than $\sim 0.27$. 

The panels (d) in Figs. 3--5 show that all the outcomes of
the collapse are differentially rotating. The degree of the 
differential rotation is very large for the cylindrical radius 
$\varpi \agt 10$ km as 
$\Omega \propto \varpi^{-\delta}$ with $\delta \sim 1.9$--2.0,
reflecting the initial profile.
In the inner region of $\varpi \alt 10$ km,
the rotational angular velocity does not change 
as steeply as that for $\varpi \gg 10$ km. This also seems to reflect 
the initial rotational velocity profile for which $\Omega$ is
nearly constant for $\varpi \alt \varpi_d$. 
However, except for the very inner region, the star is
totally differentially rotating, in particular in the outer region, 
for any models of $A=O(0.1)$. 
The results indicate that the initial rotational velocity profile
is reflected in the outcome. Thus, if the progenitor of the collapse
is highly differentially rotating, the outcomes will be always so and, 
as a result, be candidates for the bar-mode dynamical
instabilities \cite{SKE}.

In Figs. 5(c) and 5(d), we display together
the profiles of the density and
the rotational angular velocity for model M7a2 (dotted curves) at
$t=121.3$ msec at which the outcome has already relaxed to a
quasistationary state. Figure 5(c) shows that the outcome is a 
spheroid not a torus-like object (i.e., the central density is highest).
This is a characteristic property in the collapse with rigidly rotating
initial conditions \cite{SS3}. Figure 5(d) shows that 
the rotational angular velocity in the inner region is approximately
flat and thus the high-density part of the protoneutron 
star is approximately rigidly rotating. The outer region of
$\varpi \agt 10$ km, on the other hand, is differentially rotating, but the
rotational angular velocity falls off in proportional to
$\varpi^{-\delta}$ with $\delta \sim 1.4$--1.5; i.e., 
the profile is approximately that of Kepler's law, and hence, the degree
of differential rotation in this case is smaller than 
that for differentially rotating initial conditions.
More details about the outcomes in the rigidly rotating initial conditions
are found in \cite{SS3}. 

From the density contour curves, it is found that 
for all the differentially rotating models,
the column density integrated along the rotational
axis is much smaller than that along the equatorial plane
after shocks sweep the matter. Namely, a funnel is formed 
around the rotational axis even in the absence of a black hole.
This is due to the facts that 
the total mass around the rotational axis is initially small 
because of a high degree of 
differential rotation for the initial condition and
that the formed shocks are strongest around the rotational axis.
A current popular model for the central engine of gamma-ray bursts
is the so-called collapsar model \cite{COL}.
To escape the baryon-loading problem for the fire-ball model \cite{GRB}, 
it is often required to form a funnel in the collapsar models.
In their scenario, a rapidly rotating black hole
is formed, and subsequently, a jet emitted along the rotational axis
of the black hole ejects the matter. The present results suggest that
a high degree of differential rotation for the progenitor of the stellar
collapse helps making a funnel without relying on the formation of
a rapidly rotating black hole and subsequent jets.

\subsection{Dependence on equations of state}

\begin{table}[tb]
\begin{center}
\begin{tabular}{|c|c|c|c|c|c|c|} \hline
Model & $\Gamma_1$ & $\Gamma_2$ 
& $\alpha_{\rm Min}$ & $\rho_{\rm Max}$ & $\beta_{\rm max}$ & Outcome
\\ \hline
M5c1A & 1.3  & 2.5 & 0.79 & 3.2e14 &0.28& O-A \\\hline
M5c1C & 1.28 & 2.75& 0.80 & 3.8e14 &0.33& O-A \\\hline\hline 
M5c2A & 1.3  & 2.5 & 0.74 & 5.0e14 &0.28& O-A$\rightarrow$NS \\\hline
M5c2C & 1.28 & 2.75& 0.77 & 5.2e14 &0.31& O-A$\rightarrow$NS
\\\hline \hline
M7b3A & 1.3  & 2.5 & 0.65 & 4.8e14 &0.29& NS \\\hline
M7b3B & 1.32 & 2.25& 0.71 & 2.8e14 &0.23& O-A \\\hline
M7b3C & 1.28 & 2.75& 0.70 & 5.3e14 &0.29& NS \\\hline \hline
M7c2A & 1.3  & 2.5 & 0.61 & 6.0e14 &0.33 &O-A \\\hline
M7c2B & 1.32 & 2.25& 0.75 & 5.6e14 &0.21 &O-A \\\hline
M7c2C & 1.28 & 2.75& 0.67 & 5.6e14 &0.36 &O-A$\rightarrow$NS
\\\hline \hline
M7c3A & 1.3  & 2.5 & 0.51 & 9.2e14 &0.30 & NS \\\hline
M7c3B & 1.32 & 2.25& 0.38 & 1.2e15 &0.26 & O-A \\\hline
M7c3C & 1.28 & 2.75& 0.61 & 7.6e14 &0.33 & NS \\\hline \hline
M7c4A & 1.3  & 2.5 & 0.46 & 1.2e15 &0.27 & NS \\\hline
M7c4B & 1.32 & 2.25& ---  & ---    & --- & BH \\\hline
M7c4C & 1.28 & 2.75& 0.59 & 8.7e14 &0.30 & NS \\\hline \hline
M8c2A & 1.3  & 2.5 & 0.59 & 5.4e14 &0.34 & O-A \\\hline
M8c2B & 1.32 & 2.25& 0.86 & 2.2e14 &0.16 & O-A \\\hline
M8c2C & 1.28 & 2.75& 0.64 & 5.2e14 &0.37 & O-A \\\hline \hline
M8c4A & 1.3  & 2.5 & 0.29 & 1.5e15 &0.30 & NS \\\hline
M8c4C & 1.28 & 2.75& 0.51 & 9.4e14 &0.34 & NS \\\hline \hline
\end{tabular}
\caption{Numerical results with
different values of $\Gamma_1$ and $\Gamma_2$ for selected models.
}
\end{center}
\vspace{-5mm}
\end{table}

To clarify the dependence of the outcomes on the equations of state, 
we performed simulations varying $\Gamma_1$ and $\Gamma_2$
as listed in Table III for models M5c1, M5c2, M7b3, M7c2, M7c3, M7c4,
M8c2, and M8c4. Here, we focus only on highly differentially
rotating cases with $A=0.1$ and 0.25. 
The details for the cases of rigid rotation and 
moderate degrees of differential rotation 
with $A \sim 1$ are presented in \cite{SS3}. 
As listed in Table I, we choose three sets of $(\Gamma_1, \Gamma_2)$ as 
(i) (1.3, 2.5), (ii) (1.32, 2.25), and (iii) (1.28, 2.75).
In the following, we will refer to the models with (i), (ii), and (iii)
using the labels A, B, and C, e.g., as M5c1A, M5c1B, and M5c1C. 

\begin{figure}[htb]
\vspace*{-6mm}
\begin{center}
\epsfxsize=2.9in
\leavevmode
(a)\epsffile{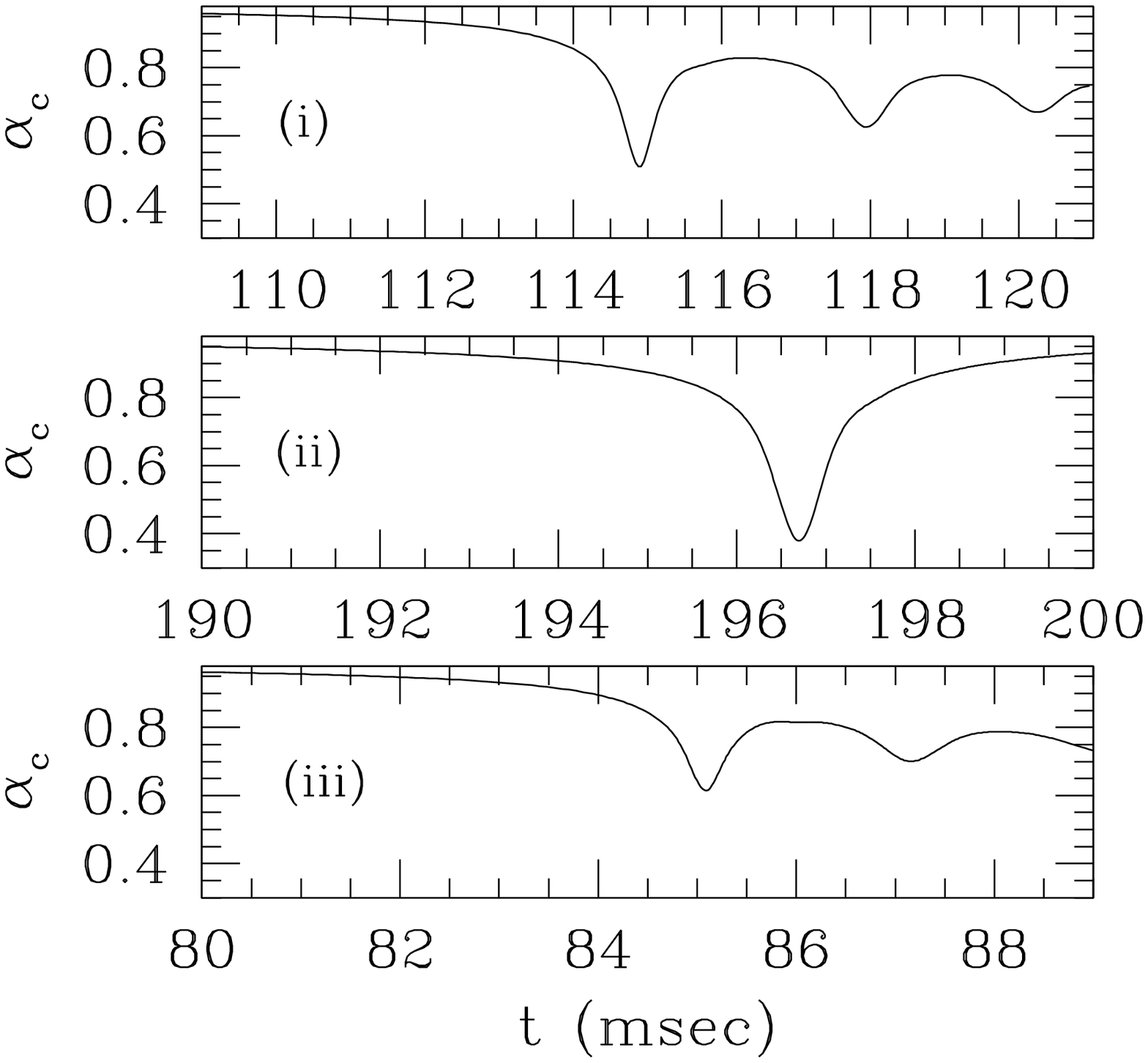}
\epsfxsize=2.9in
\leavevmode
(b)\epsffile{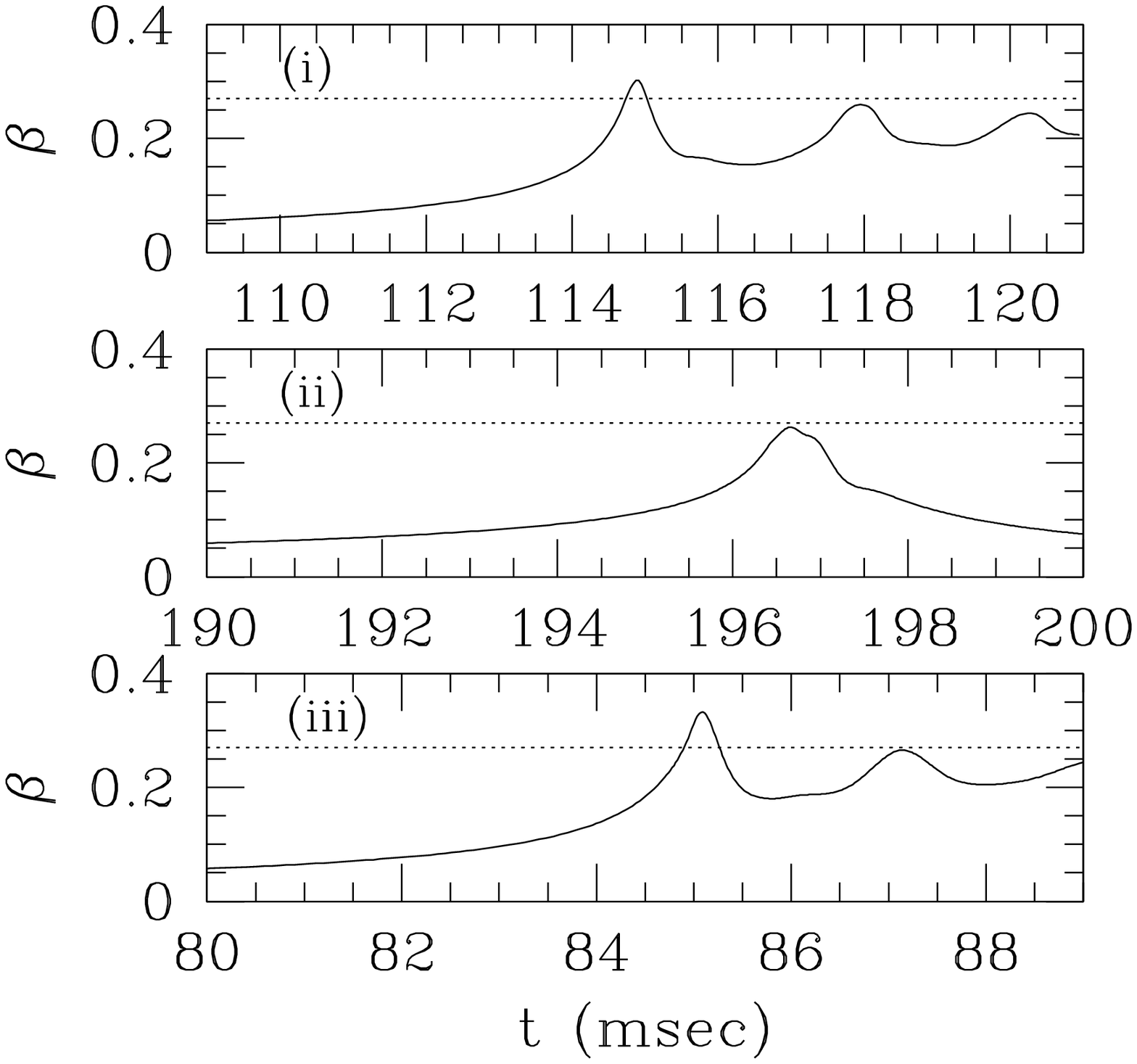}\\
\vspace*{-4mm}
\epsfxsize=2.9in
\leavevmode
(c)\epsffile{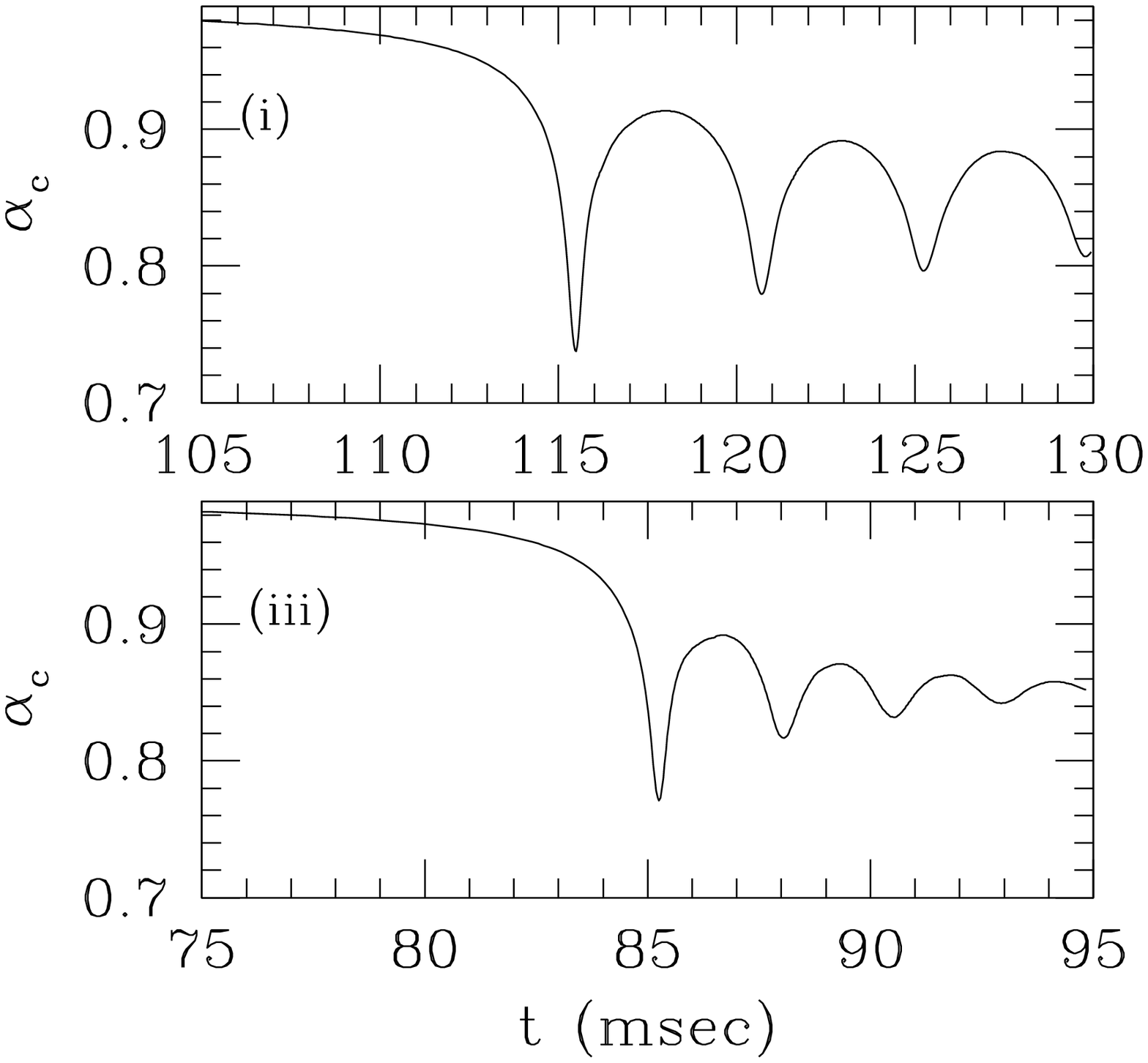}
\epsfxsize=2.9in
\leavevmode
(d)\epsffile{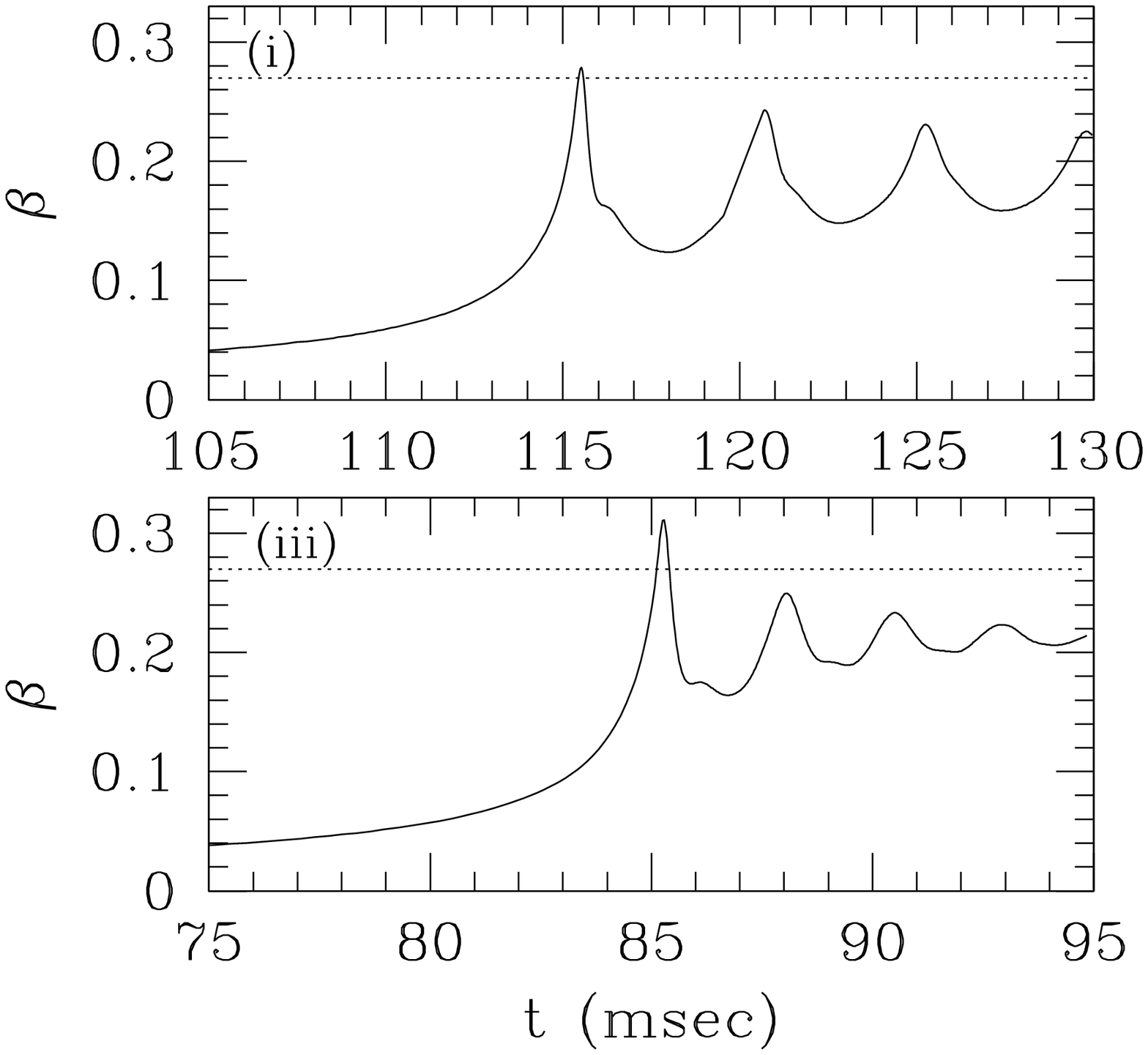}\\
%\vspace*{-4mm}
\caption{Evolution of (a) $\alpha_{c}$ and (b) $\beta$ for model M7c3 
and of (c) $\alpha_{c}$ and (d) $\beta$ for model M5c2.
Panels (i), (ii), and (iii) are results for
$(\Gamma_1, \Gamma_2)=(1.3, 2.5)$,
(1.32, 2.25), and (1.28, 2.75), respectively.
The dotted horizontal lines in panels (b) and (d) denote $\beta=0.27$. 
\label{FIG6}}
\end{center}
\end{figure}

In Fig. 6, we compare the evolutions of the central value of $\alpha$ 
and $\beta$ for models M7c3A, M7c3B, and M7c3C (Figs. 6(a) and (b)) and 
for M5c2A and M5c2C (Figs. 6(c) and (d)) as representative illustrations. 
In the previous section, we found that models M7c3 and M5c2 are possible 
candidates for the onset of nonaxisymmetric dynamical instabilities of 
$\beta_{\rm max} > 0.27$. Among these models of 
different equations of state, $\beta_{\rm max}$ for case (iii) is largest. 
On the other hand, $\beta_{\rm max}$ for case (ii) is much smaller than
those in other two cases. 
This seems to be due to the fact that for smaller values of
$\Gamma_1$, the depletion factor of the internal energy
and the pressure in an early stage of collapse in which 
$\rho \ll \rho_{\rm nuc}$ is larger. As a consequence, the
internal energy $U$ is decreased to contribute to the increase of
$\beta$, and furthermore, the matter around the rotational axis which
possesses large values of the specific angular momentum 
collapses to a more compact state, for which a spin-up 
is enhanced effectively. 
On the other hand, the values of $\alpha_c$ (compactness)
for models M7c3C and M5c2C are larger (smaller) than that for M7c3A and
M5c2A, respectively. This may be partly due to the fact that $\Gamma_2$
for case (iii) is larger than that for (i), but mainly to 
the fact that the fraction of the matter which 
simultaneously collapses is smaller for case (iii) than for (i). 
This implies that to achieve a large value of $\beta$,
it is not necessary for the whole system to become compact.
Rather, essentially needed is to accumulate the 
matter with large values of 
the specific angular momentum in the central region. 
This point is reconfirmed from the results for M7c3B. 
In this case, the value of $\beta_{\rm max}$ is much smaller than 
those for other two cases, although the value of $\alpha_c$ is smallest
among three cases. This is due to the fact that in this case,
a large fraction of the matter collapses nearly simultaneously
independent of the magnitude of the specific angular momentum. 

Table III also shows that the largest value of $\beta_{\rm max}$
is achieved for case (iii) for all the initial conditions. 
This indicates that to achieve a large value of $\beta$,
the depletion of the internal energy and the pressure
in the early stage of the collapse, which in reality will be achieved
by partial photo-dissociation of the iron to lighter elements 
and by the electron capture \cite{ST,Bethe}, should be sufficiently large
to accelerate the collapse of the central region. 

As indicated in Fig. 6, the outcomes for models M7c3 and M5c2
depend sensitively on the equations of state. 
For the small values of $\Gamma_1$ (cases (i) and (iii)), 
an oscillating protoneutron star is formed eventually.
The amplitude of the oscillation is smaller and the period is shorter 
for the smaller value of $\Gamma_1$ (case (iii)). 
As a result, the protoneutron star relaxes to a quasistationary state
more quickly. For a long period of the oscillation, the duration of
the phase, in which $\beta$ and rotational angular velocity remain
small, becomes long. This also suggests that for the onset of
nonaxisymmetric dynamical instabilities, the smaller value of $\Gamma_1$ 
may be preferable.

\begin{figure}[thb]
\vspace*{-5mm}
\begin{center}
\epsfxsize=2.8in
\leavevmode
(a)\epsffile{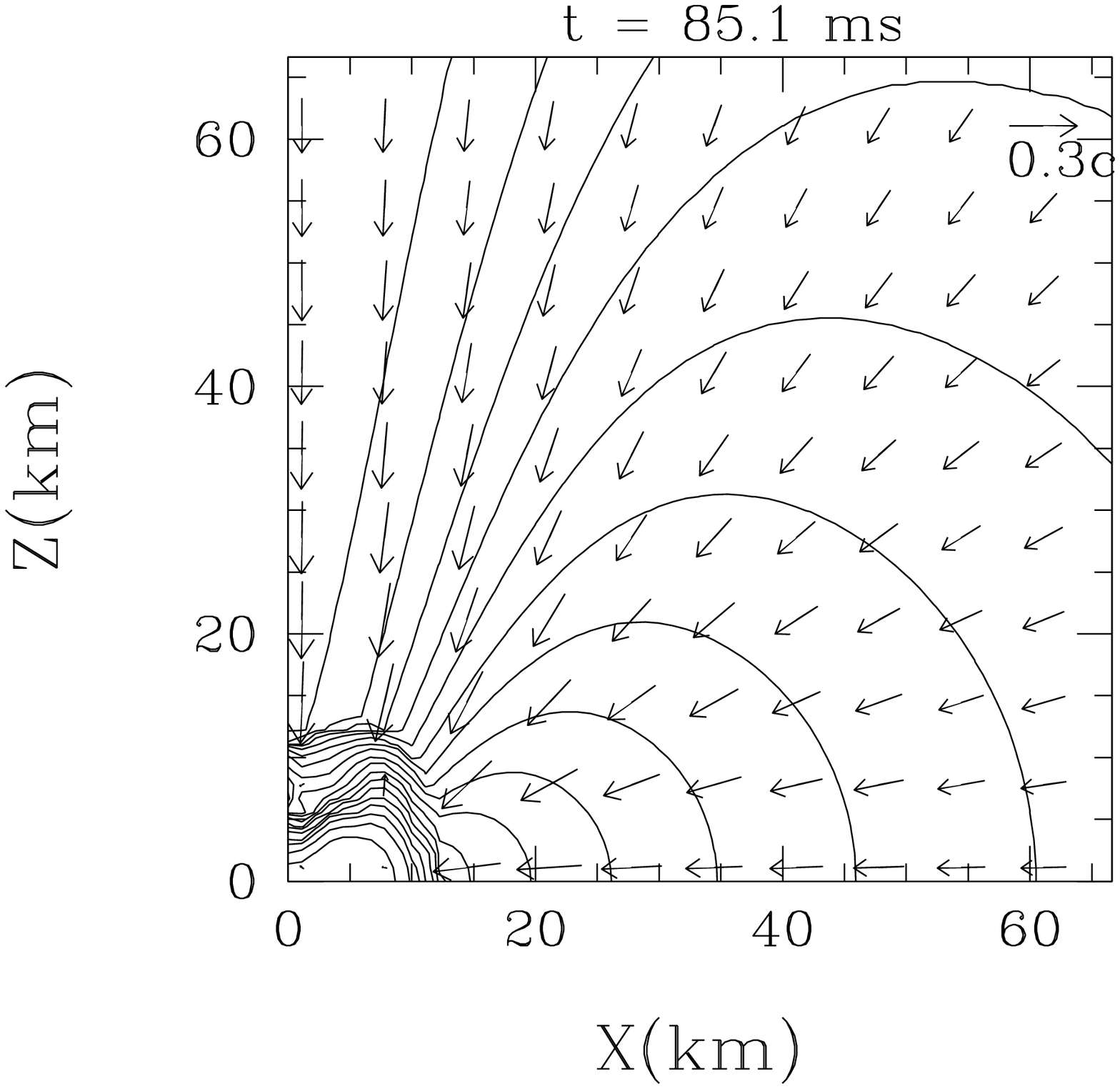}
\epsfxsize=2.8in
\leavevmode
~~~~(b)\epsffile{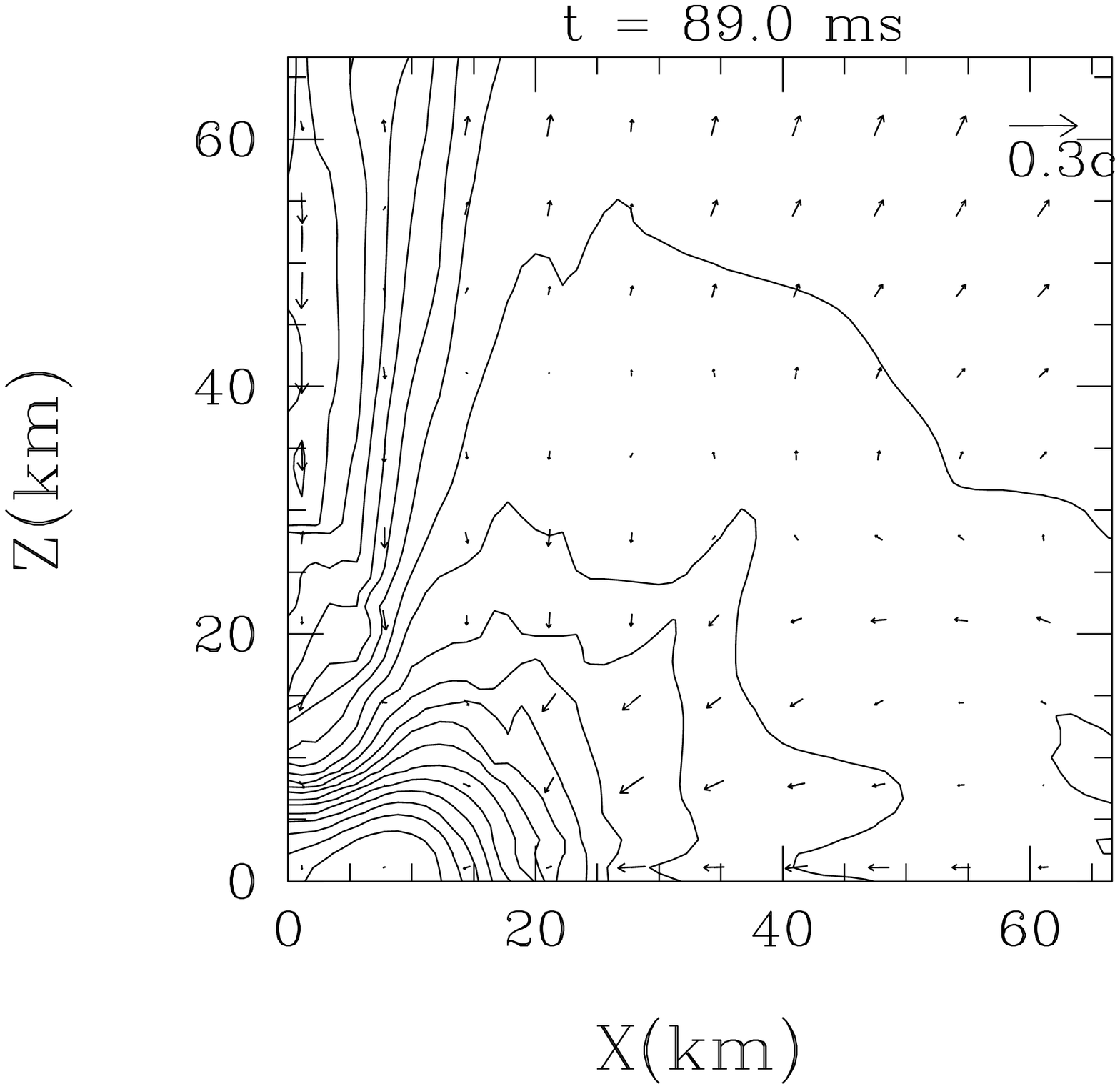}\\
\vspace*{-4mm}
\epsfxsize=2.8in
\leavevmode
(c)\epsffile{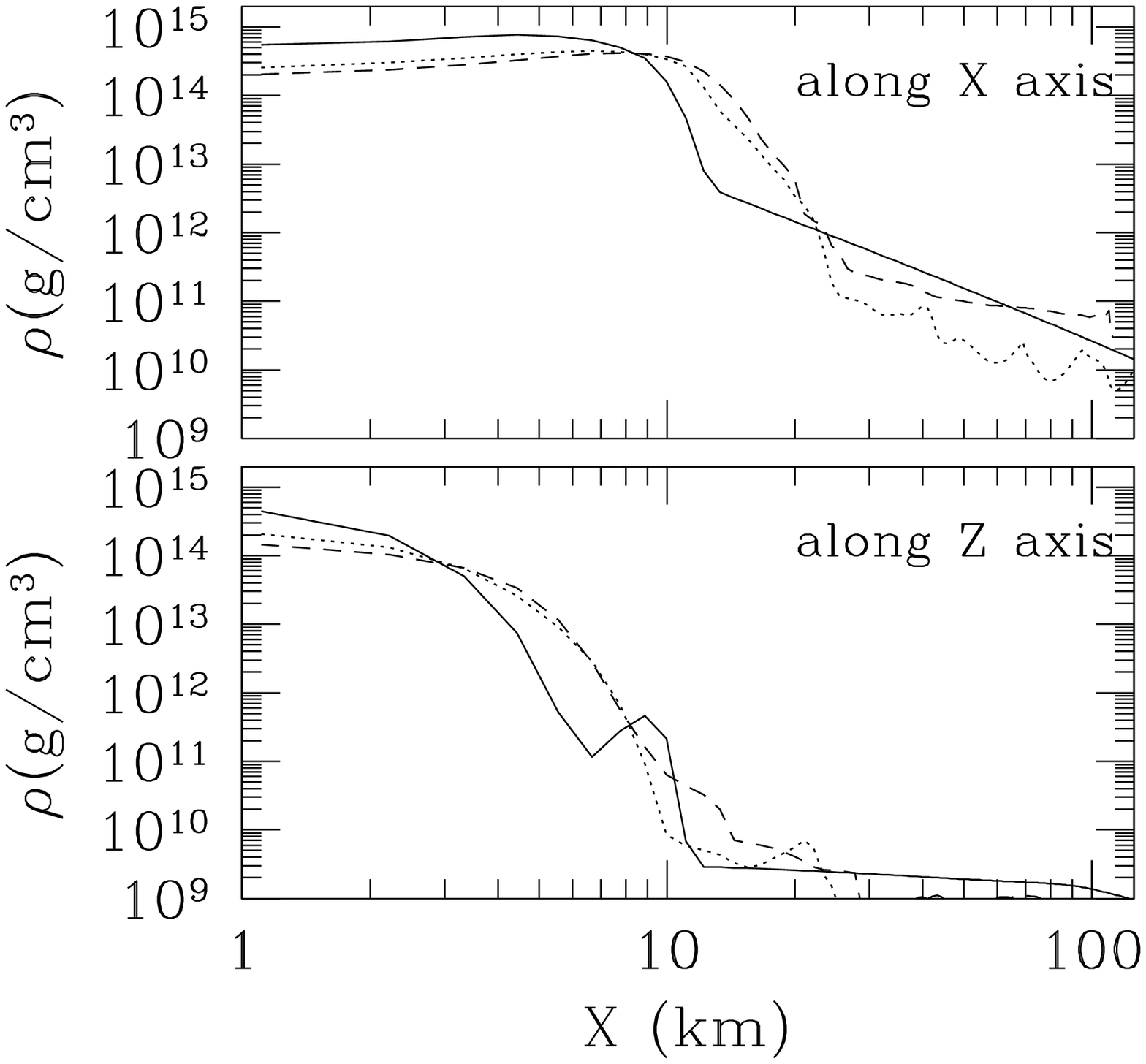}
\epsfxsize=2.8in
\leavevmode
(d)\epsffile{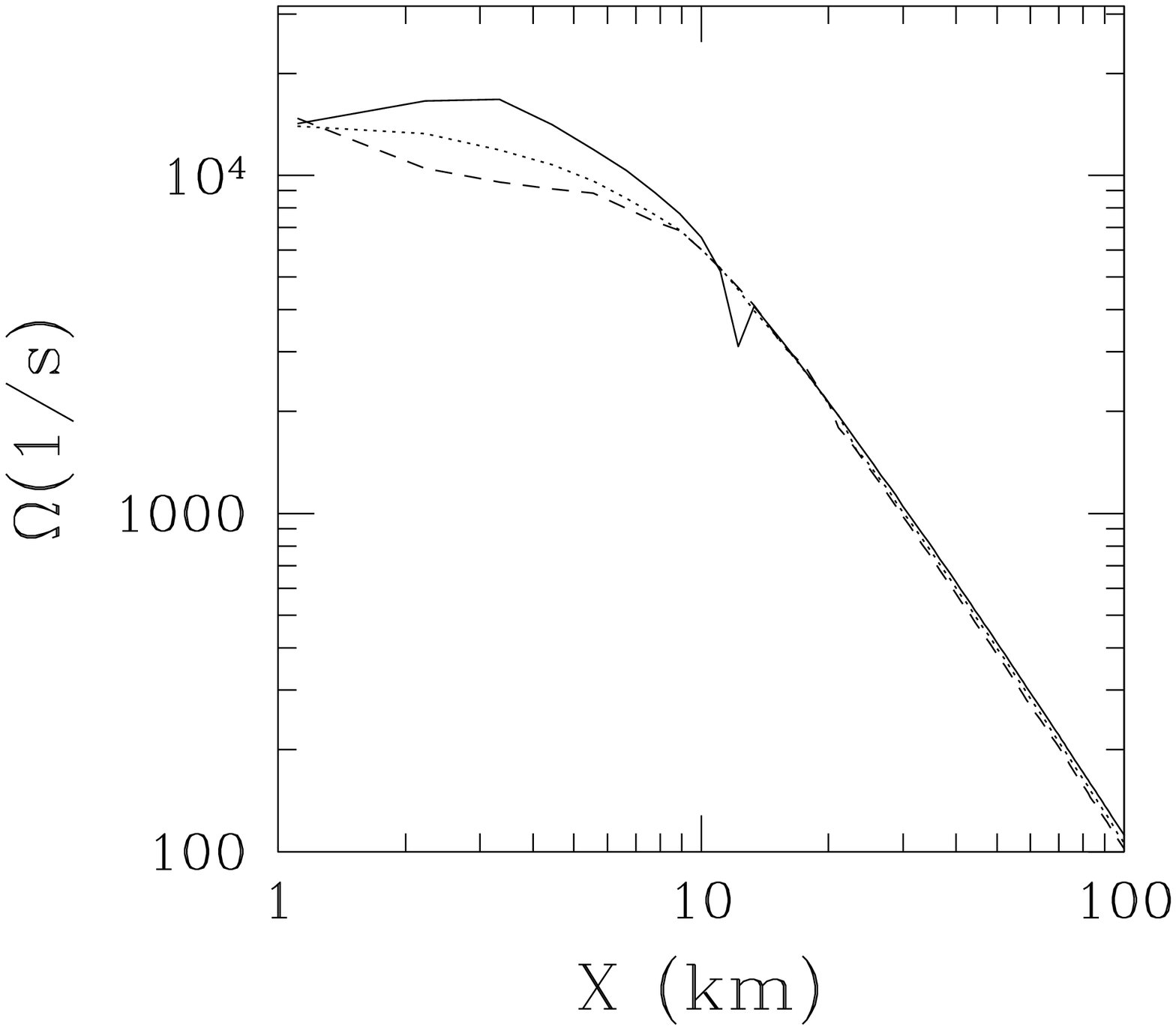}
\caption{The same as Fig. 3(a)--(d) but for model M7c3C 
at $t=85.1$ and 89.0 msec. 
The solid and dashed curves in panels (c) and (d) are drawn
for the corresponding time slices, respectively.
In panels (c) and (d), we plot the results for model M7c3A 
at $t=120.4$ msec (dotted curves) for comparison. 
\label{FIG7}}
\end{center}
\end{figure}

In Fig. 7, we display the snapshots of the density contour
curves and the velocity vectors at the time slices
that the maximum compression is
achieved and the system relaxes to an approximately quasistationary state
for model M7c3C. The density profiles and 
the angular velocity in the equatorial plane at the corresponding
time steps are shown in panels (c) and (d) together with results for 
model M7c3A (displayed in Fig. 4). It is found that the shape of the outcome
for M7c3C is more torus-like than that for M7c3A.
In addition, the degree of differential rotation for M7c3C 
is slightly higher than that for M7c3A. 
These facts indicate that the outcome for M7c3C is likely to be
more subject to nonaxisymmetric dynamical instabilities.
These properties depend weakly on the value of $K_0$ (i.e., mass of
the progenitor). Indeed, 
the outcome of M5c2C is more torus-like and the degree of
differential rotation in the central region for M5c2C is
higher than those for M5c2A. As illustrated in Sec. V, thus, 
the value of $\Gamma_1$ is one of key parameters for determining
the onset of nonaxisymmetric dynamical instabilities. 

\begin{figure}[htb]
\vspace*{-4mm}
\begin{center}
\epsfxsize=2.8in
\leavevmode
(a)\epsffile{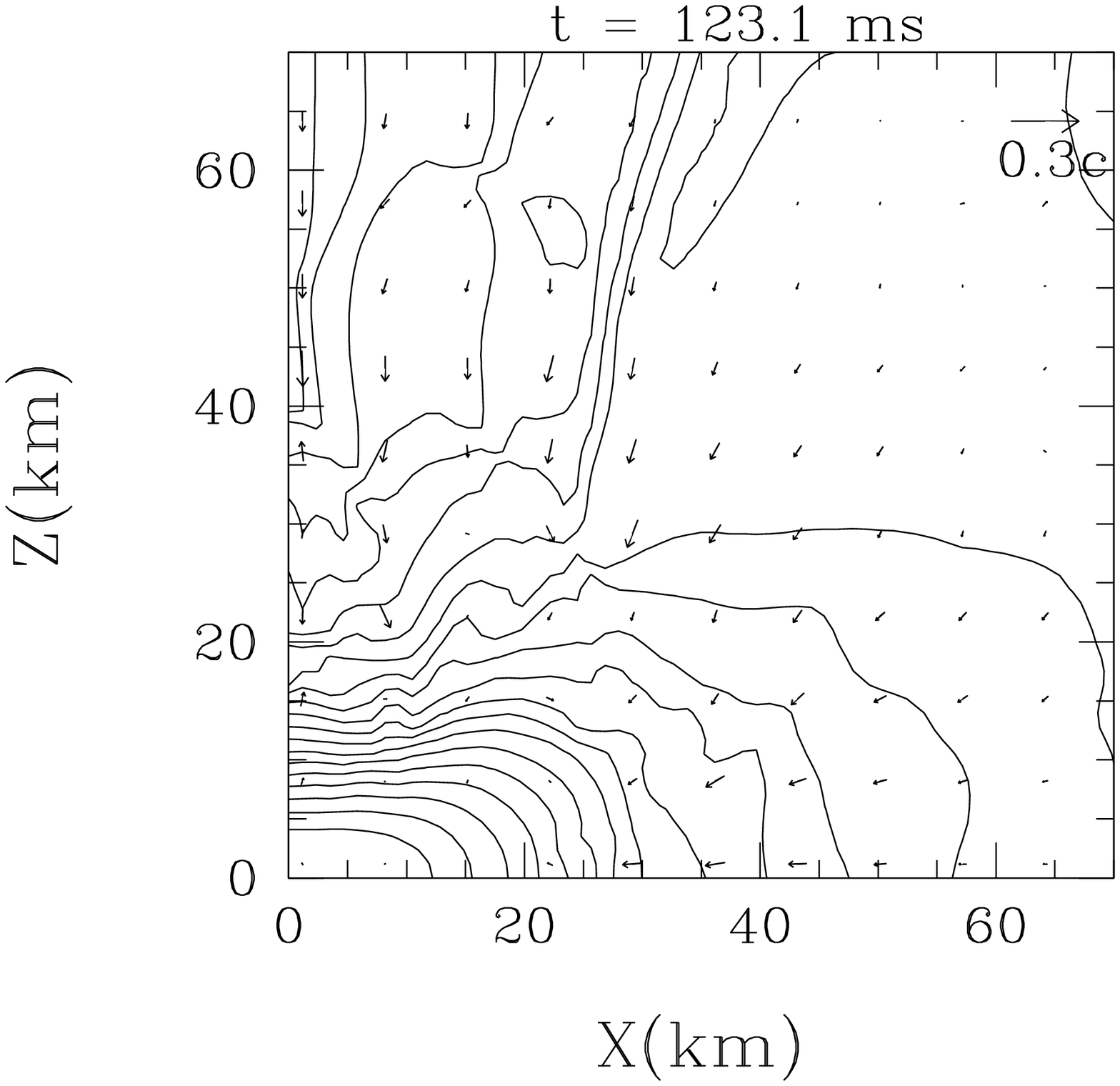}
\epsfxsize=2.8in
\leavevmode
(b)\epsffile{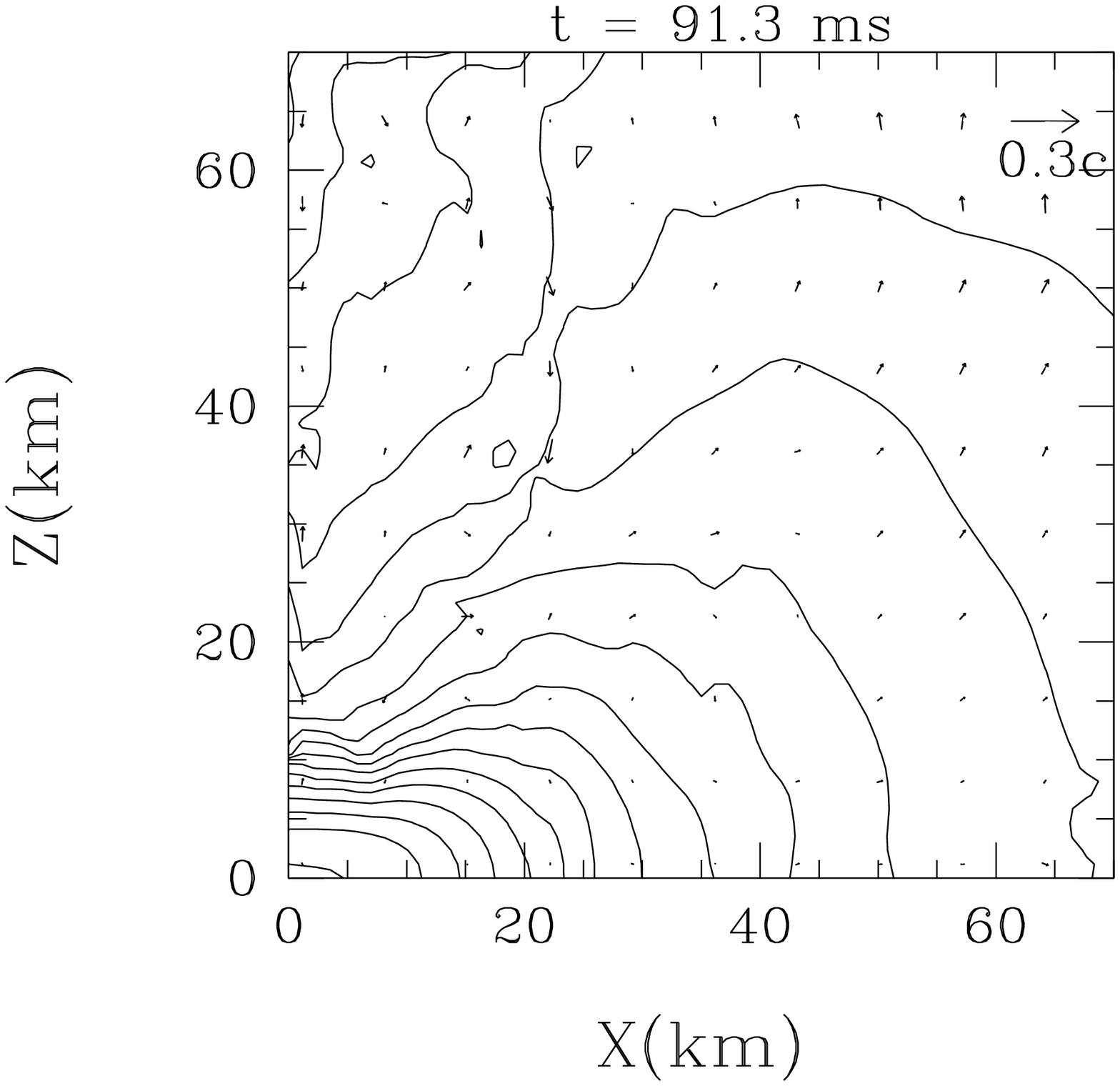} \\
\vspace*{-4mm}
\epsfxsize=2.8in
\leavevmode
(c)\epsffile{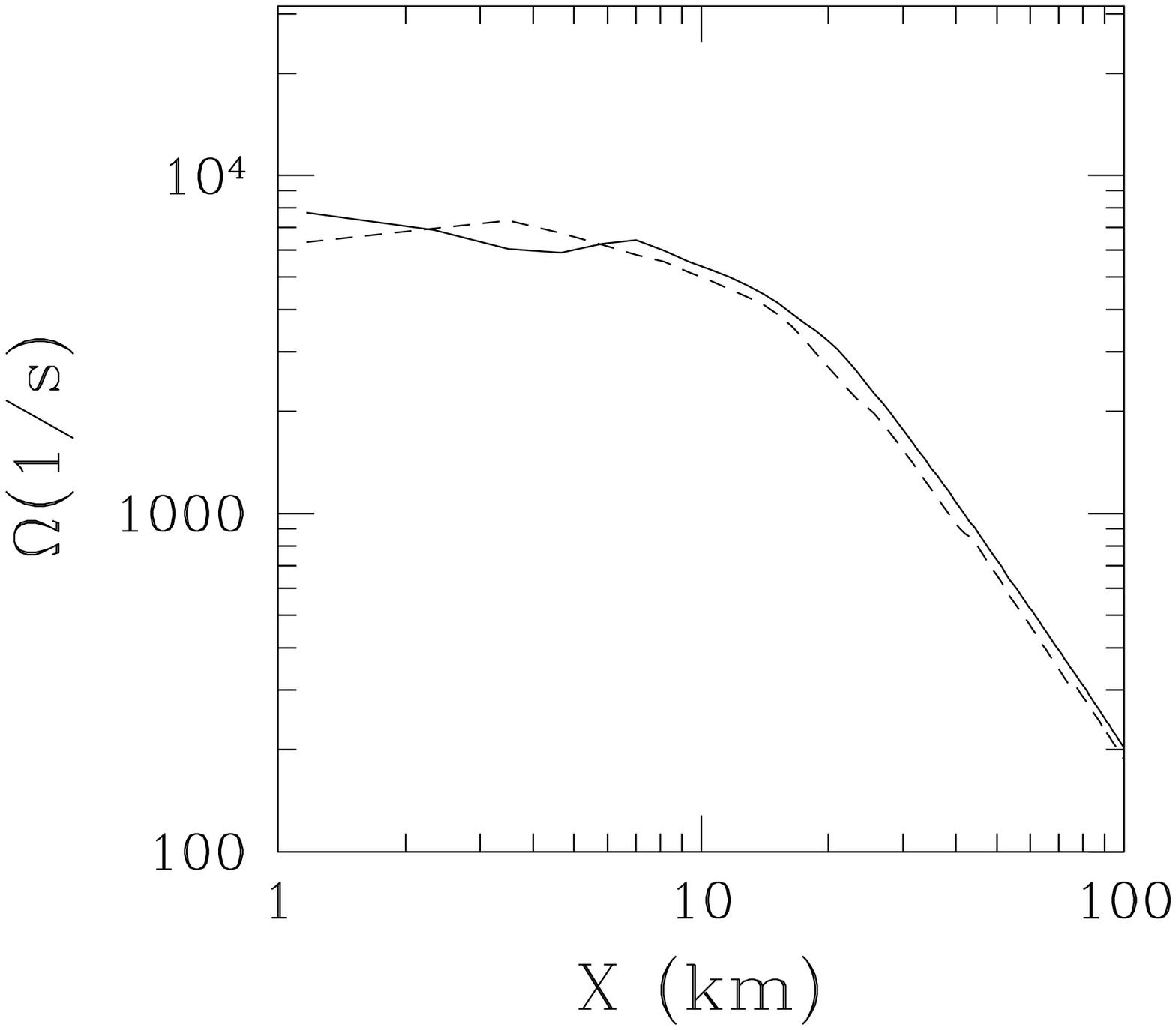}
\epsfxsize=2.8in
\leavevmode
(d)\epsffile{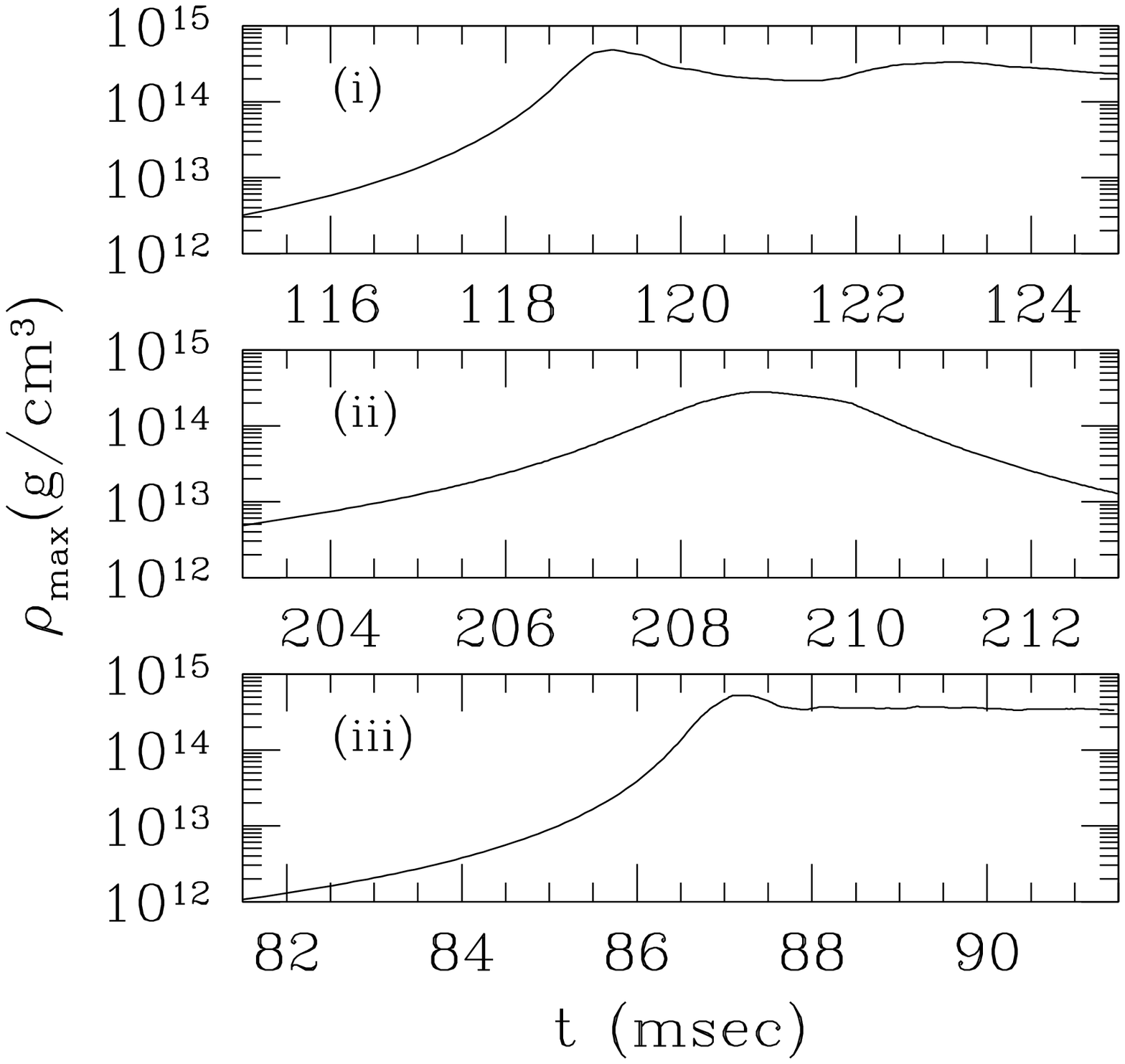}
%\vspace*{-4mm}
\caption{
(a) The density contour curves for $\rho$ for model M7b3A 
at $t=123.1$ msec. The contour curves and vectors are drawn 
in the same method as in Fig. 3.
(b) The same as panel (a) but for model M7b3C at $t=91.3$ msec.
(c) Angular velocity along the radial coordinate in the 
equatorial plane for models M7b3A at $t=123.1$ msec (solid curve) and 
M7b3C at $t=91.3$ msec (dashed curve).
(d) $\rho_{\rm max}$ as a function of time for models M7b3A, M7b3B,
and M7b3C. 
\label{FIG8}}
\end{center}
\end{figure}

We note that the properties pointed above is found only for $A=0.1$.
For $A=0.25$, the value of $\beta_{\rm max}$
for two equations of state (i) and (iii) are not very different.
Also, the density profile and 
the angular velocity profile of the formed neutron stars 
are similar (see Fig. 8). This indicates that for 
large values of $A$, the larger depletion factor of the
internal energy (smaller value of $\Gamma_1$) in the early
stage of the collapse does not play an important role
for accumulating the matter of large specific angular momentum
in the central region. This result suggests that
the stability property against nonaxisymmetric deformation will not
depend on the choice of $\Gamma_1$ and $\Gamma_2$ for
$A=0.25$ as strongly as for $A=0.1$ as long as $\Gamma_1 \leq 1.3$.
On the other hand, for $\Gamma_1 = 1.32$, the outcomes are
completely different from those for other two cases as in the case
of $A=0.1$.

\subsection{Candidates of nonaxisymmetric dynamical instabilities}

As reviewed in Sec. I, nonaxisymmetric dynamical instabilities 
of rotating stars in isolated equilibrium may 
set in when the value of $\beta$ becomes larger than $\sim 0.27$ or
when the degree of differential rotation is sufficiently high. 
It is found that to achieve $\beta_{\rm max} \agt 0.27$, 
the following conditions are necessary; 
(i) the progenitor of the collapse should be
highly differentially rotating with $A \alt 0.25$; 
(ii) the progenitor has to be moderately 
rapidly rotating with $0.01 \alt \beta_{\rm init} \alt 0.02$; 
(iii) the progenitor should be massive enough
to make a compact core for which an efficient spin-up is possible.

As indicated in \cite{TH,SKE}, 
even in the case of $\beta < 0.27$, nonaxisymmetric dynamical 
instabilities may set in if the degree of differential rotation
is sufficiently large. To achieve such a state, the conditions (i) and
(ii) are also necessary. In addition, the following condition
is required: (iv) the depletion factor of the
internal energy and the pressure in an early stage of
collapse during which $\rho \ll \rho_{\rm nuc}$
is large enough to induce a significant collapse
in the central region for making a torus-like structure and 
a steep profile of rotational angular velocity. 
In the next section, we present numerical results of the 
three-dimensional simulations and illustrate that 
the condition (iv) plays an important role for the onset of
nonaxisymmetric dynamical instabilities.

\section{Results in three-dimensional simulations}

\subsection{Features of nonaxisymmetric dynamical instabilities}

As described in Sec. IV, there are several candidate models
for which nonaxisymmetric dynamical instabilities may set in. 
We performed the three-dimensional simulations focusing on the candidates 
in which $\beta_{\rm max} \agt 0.27$ and $\beta \sim 0.2$ 
after bounce. In this paper, however, we do not pay attention to the models 
in which oscillating stars with the period $\agt 10$ msec are
formed since the simulations for such models take too much computational
time. 

In the simulations, 
we initially superimposed a nonaxisymmetric bar-mode density perturbation
as defined in Eq. (\ref{barmode}). 
Specifically, we picked up models M5c1, M5c2, M7b3, M7c2, M7c3, and M7c4 
with $(\Gamma_1,\Gamma_2)=(1.3,2.5)$ and (1.28, 2.75)
(referred to them, e.g., as M5c1A and M5c1C) as listed in Table IV. 
Since the matter in the outer region is discarded in 
preparing the initial conditions for the three-dimensional 
simulations according to Eq. (\ref{eq54}),
the mass and the angular momentum are smaller than those 
in the corresponding axisymmetric 
simulations by 10--20\%. As a consequence, the numerical results 
deviate slightly from those obtained by the axisymmetric 
simulations even in the case that nonaxisymmetric deformation is small. 
However, qualitative differences between two results are not found
and also the quantitative disagreement is small (see below). 

\begin{table}[tb]
\begin{center}
\begin{tabular}{|c|c|c|c|c|c|c|c|c|c|} \hline
Model & $A$ & $K_0$ & $\Gamma_1$ & $\Gamma_2$ & $M_* $ & $M$ & $J/M^2$ 
& $L$ & Stability\\ \hline
M5c1C &0.1 &5e14 &1.28 &2.75& 1.287 &1.288  & 1.380  
& 144 & Unstable \\\hline
M5c2A &0.1 &5e14 &1.3  &2.5 & 1.384 &1.384  & 1.103  
& 147 & Stable \\\hline
M5c2C &0.1 &5e14 &1.28 &2.75& 1.224 &1.235  & 1.204  
& 147 & Unstable \\\hline
M7b3A &0.25&7e14 & 1.3 & 2.5 & 2.234 & 2.235 & 1.154 
& 182 & Stable \\\hline
M7b3C &0.25&7e14 &1.28 &2.75& 1.944 & 1.946 & 1.195 
& 182 & Stable \\\hline
M7c2A &0.1 &7e14& 1.3 & 2.5 & 2.347 & 2.348 & 1.081 
& 170 & Stable \\\hline
M7c2C &0.1 &7e14& 1.28 &2.75 &2.106 & 2.108 & 1.172 
& 170 & Unstable \\\hline
M7c3A &0.1 &7e14& 1.3 & 2.5 & 2.263 & 2.265 & 0.939 
&171  & Stable \\\hline
M7c3C &0.1 &7e14& 1.28& 2.75& 2.014 & 2.016 & 1.024 
& 171 & Unstable \\\hline
M7c4C &0.1 &7e14& 1.28& 2.75& 1.956 & 1.958 & 0.922 
&172  & Stable \\\hline
\end{tabular}
\caption{Parameters and numerical results
for three-dimensional simulations. 
$K_0$, $M_* (M)$, $\rho_{\rm max}$, and $L$ 
are listed in units of cgs, $M_{\odot}$, ${\rm g/cm^3}$, and km,
respectively. The ADM mass $M$ is still nearly equal to
the baryon rest-mass $M_*$. 
In the last column, the stability against the bar-mode is shown. 
}
\end{center}
\vspace{-5mm}
\end{table}

\begin{figure}[htb]
\vspace*{-6mm}
\begin{center}
\epsfxsize=2.8in
\leavevmode
(a)\epsffile{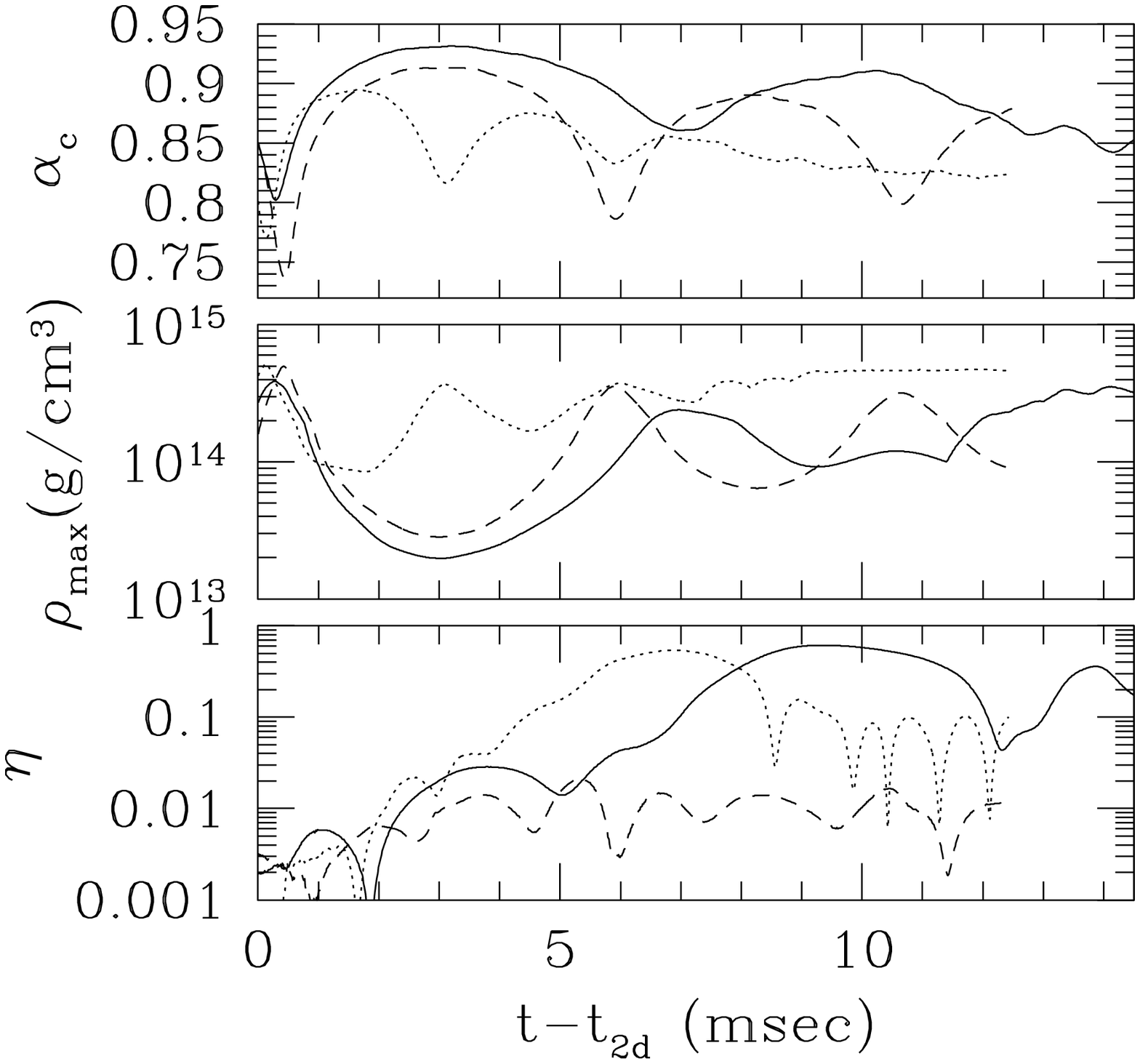}
\epsfxsize=2.8in
\leavevmode
(b)\epsffile{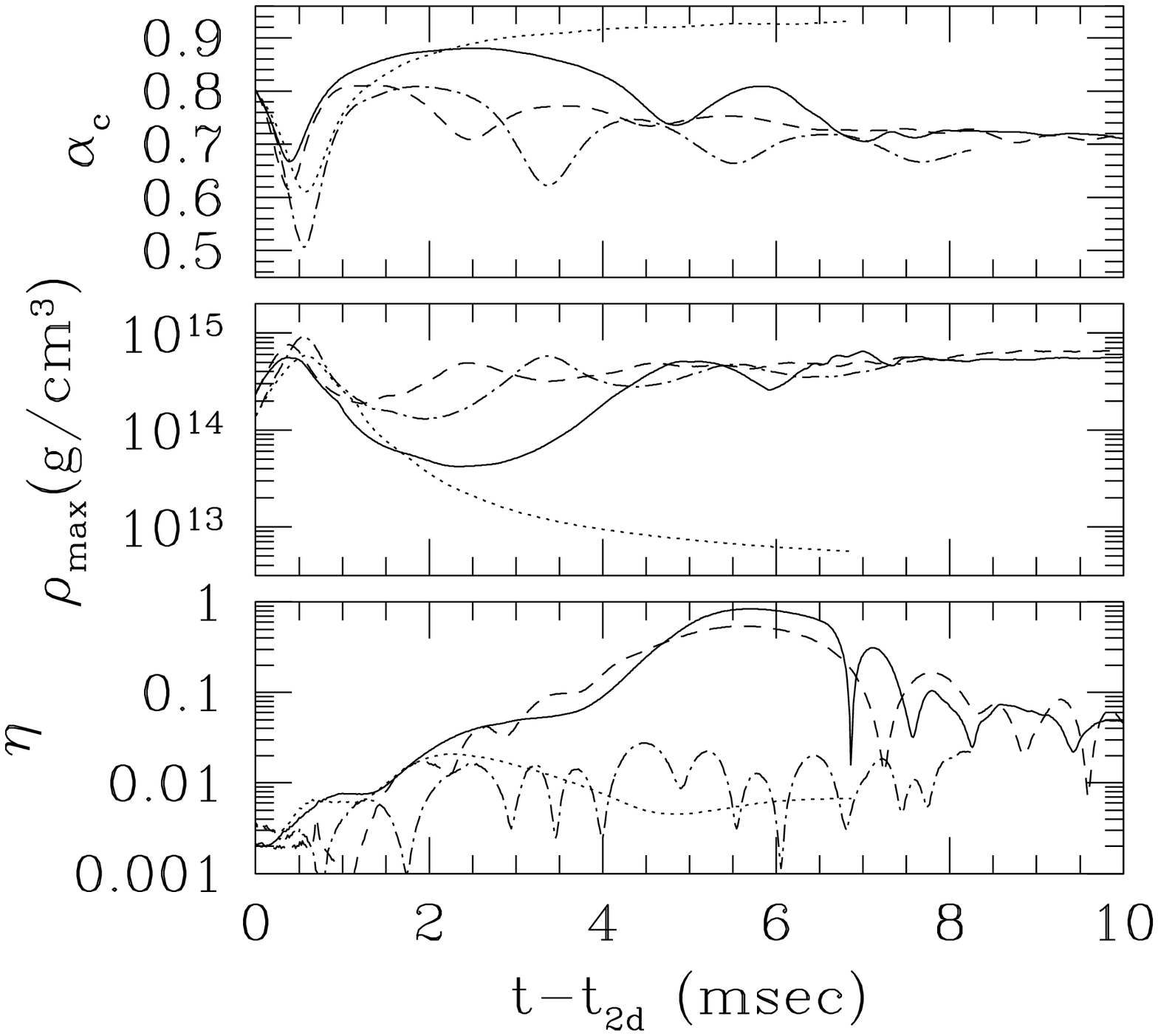}\\
\vspace*{-4mm}
\epsfxsize=2.8in
\leavevmode
(c)\epsffile{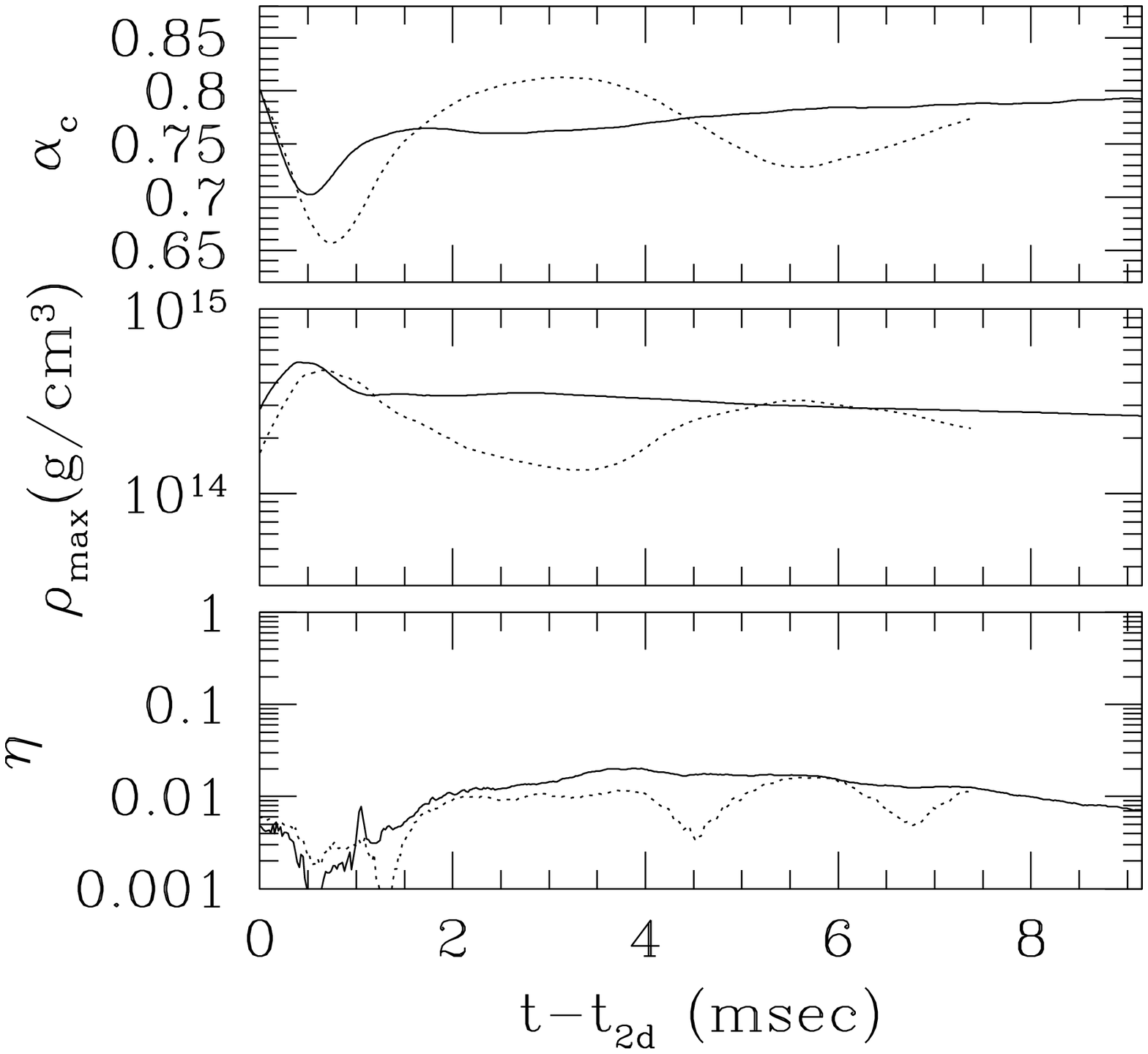}
%\vspace*{-4mm}
\caption{Evolution of $\rho_{\rm max}$, $\alpha_c$, and $\eta$ 
in the three-dimensional simulations 
(a) for models M5c1C (solid curves), M5c2A (dashed curves),
and M5c2C (dotted curves), 
(b) for models M7c2A (dotted curves),
M7c2C (solid curves), M7c3A (dotted-dashed curves),
and M7c3C (dashed curves), and
(c) for models M7b3A (dotted curves) and M7b3C (solid curves). 
Here, $t_{\rm 2d}$ denotes the time at which we change to the
three-dimensional code; $t_{\rm 2d}=$85.4, 115.1, 
85.1, 114.9, 85.1, 121.2, 84.7, 118.6, and 86.7 msec for M5c1C, M5c2A, 
M5c2C, M7c2A, M7c2C, M7c3A, M7c3C, M7b3A, and M7b3C, respectively. 
\label{FIG9}}
\end{center}
\end{figure}

In this paper, the dynamical stability against 
bar-mode deformation is analyzed 
using a distortion parameter defined by 
\beqn
\eta \equiv ( \eta_{+}^2 + \eta_{\times}^2)^{1/2}, 
\eeqn
where 
\beqn
\eta_+ \equiv {Q_{xx} - Q_{yy} \over Q_{xx} + Q_{yy}},~~~
\eta_{\times} \equiv {2Q_{xy} \over Q_{xx} + Q_{yy}},
\eeqn
and 
\beqn
Q_{ij}=\int_{\rho > \rho_{\rm cut}}  \rho_* x^i x^j d^3 x. 
\eeqn
Here, the integration is carried out only for $\rho \geq \rho_{\rm cut}$
where $\rho_{\rm cut}$ is a selected cutoff density. In this paper, 
we chose as $\rho_{\rm cut}=\rho_{\rm max}/100$ 
to focus on the high-density region. 
For comparison, we also chose the cutoff density as zero 
(i.e., the distortion parameter is defined in terms of $I_{ij}$).
In this case, the distortion parameter is denoted as $\eta_0$.
In the following, we primarily adopt $\eta$, and 
if the value of $\eta$ grows exponentially, we judge that
the model is dynamically unstable. 

In Fig. 9, we show the evolution of $\rho_{\rm max}$, $\alpha_c$, and $\eta$
for models M5c1C, M5c2A, M5c2C, M7b3A, M7b3C,
M7c2A, M7c2C, M7c3A, and M7c3C. For models M5c2A, M7b3A, M7b3C, 
M7c2A, M7c3A, and M7c4C, $\eta$ (and also $\eta_0$) does not increase
exponentially (cf. Table IV). This implies that these models 
are dynamically stable against bar-mode deformation. 
For models M5c1C, M5c2C, M7c2C, and M7c3C, on the other hand, 
the values of $\eta$ approximately increase in proportional to 
$e^{t/\tau}$ where $\tau$ denotes a characteristic growth time. 
Thus, these models are unstable. The growth time $\tau$ of
these models is approximately 0.8--1 msec. This is the same order
of magnitude as the dynamical time scale $\rho_c^{-1/2}$. 
Thus, the instabilities found here are indeed the dynamical instabilities. 

\begin{figure}[t]
\begin{center}
\epsfxsize=2.4in
\leavevmode
\epsffile{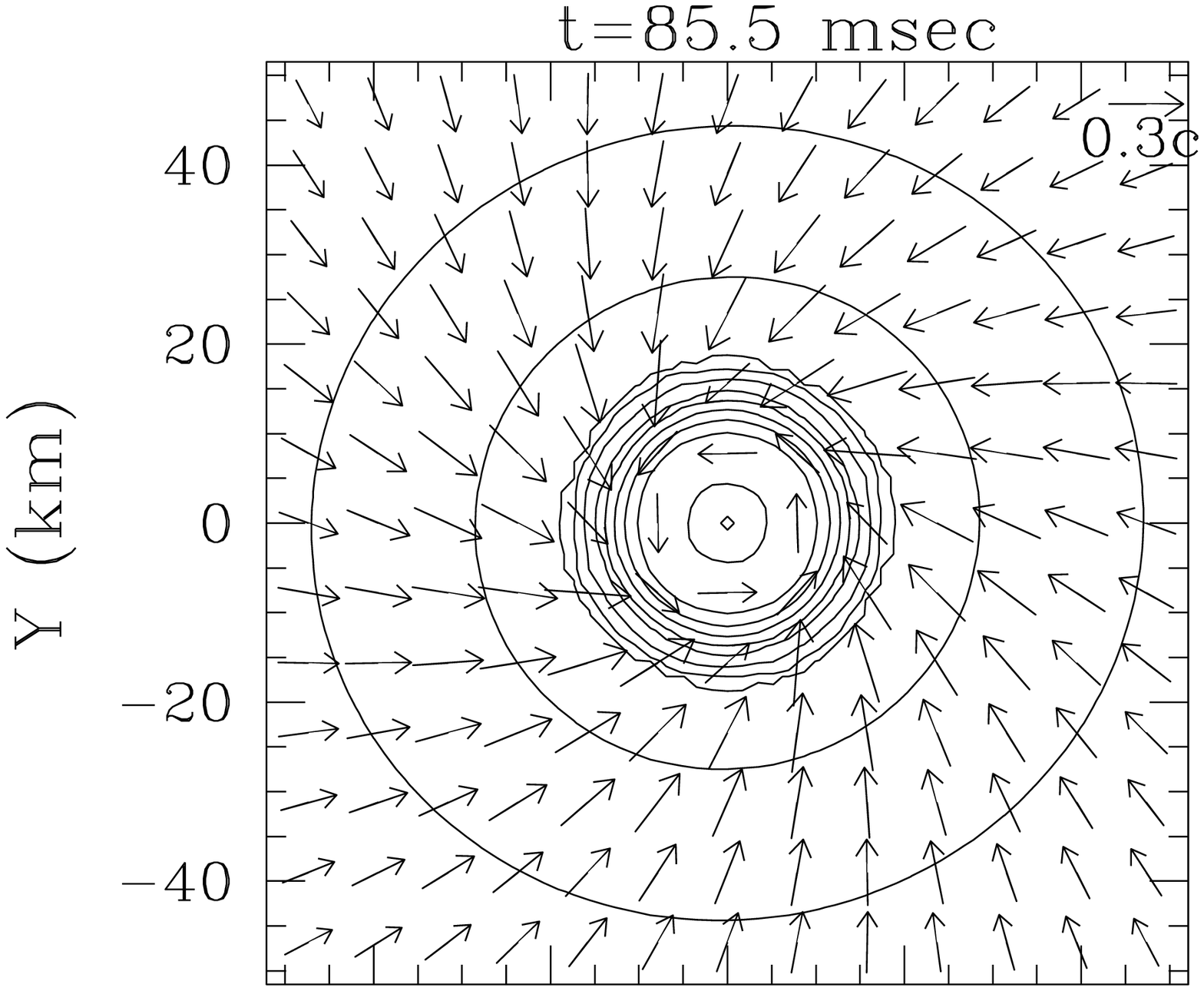}
\epsfxsize=2.4in
\leavevmode
\hspace{-1.2cm}\epsffile{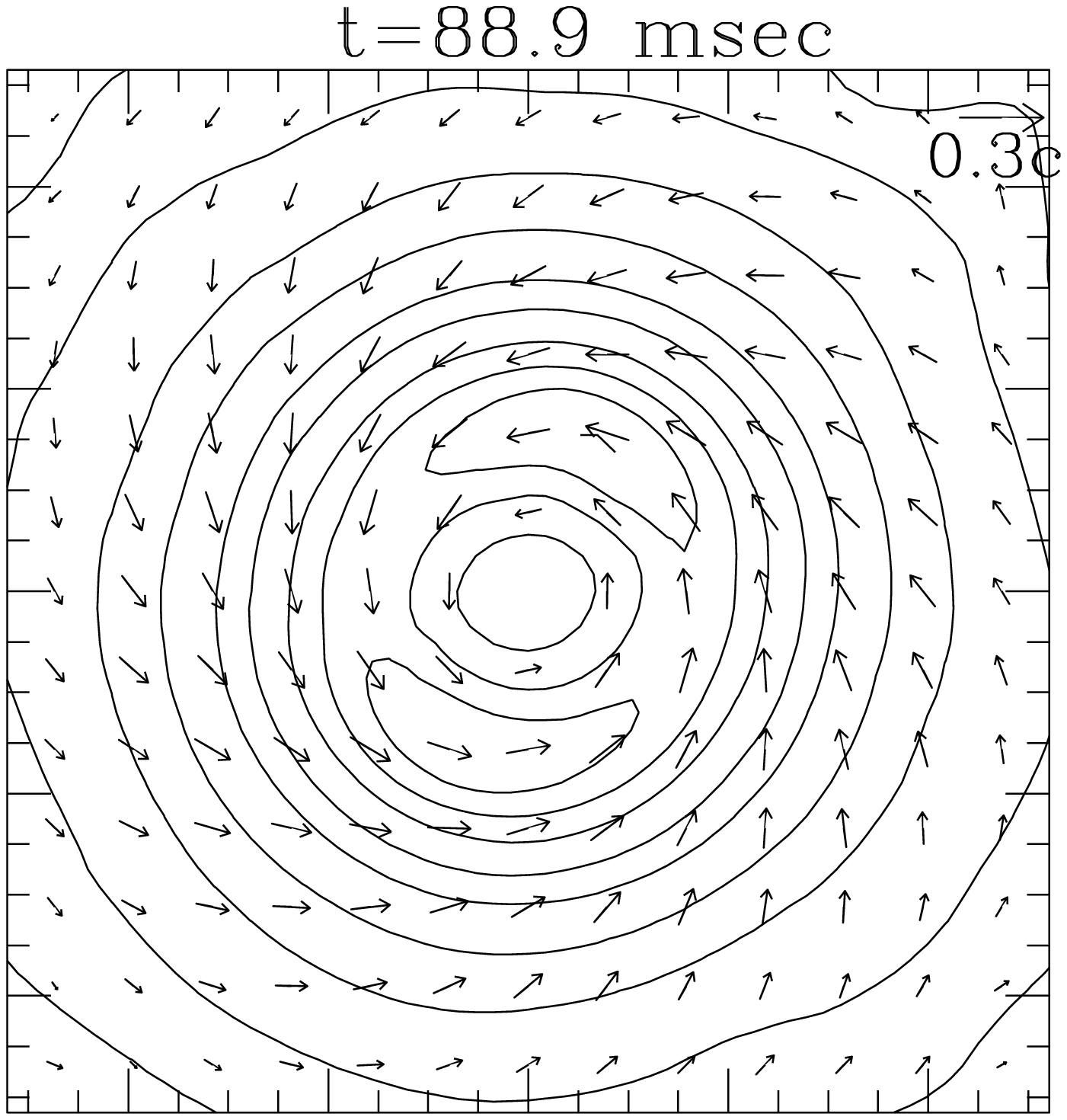} 
\epsfxsize=2.4in
\leavevmode
\hspace{-1.2cm}\epsffile{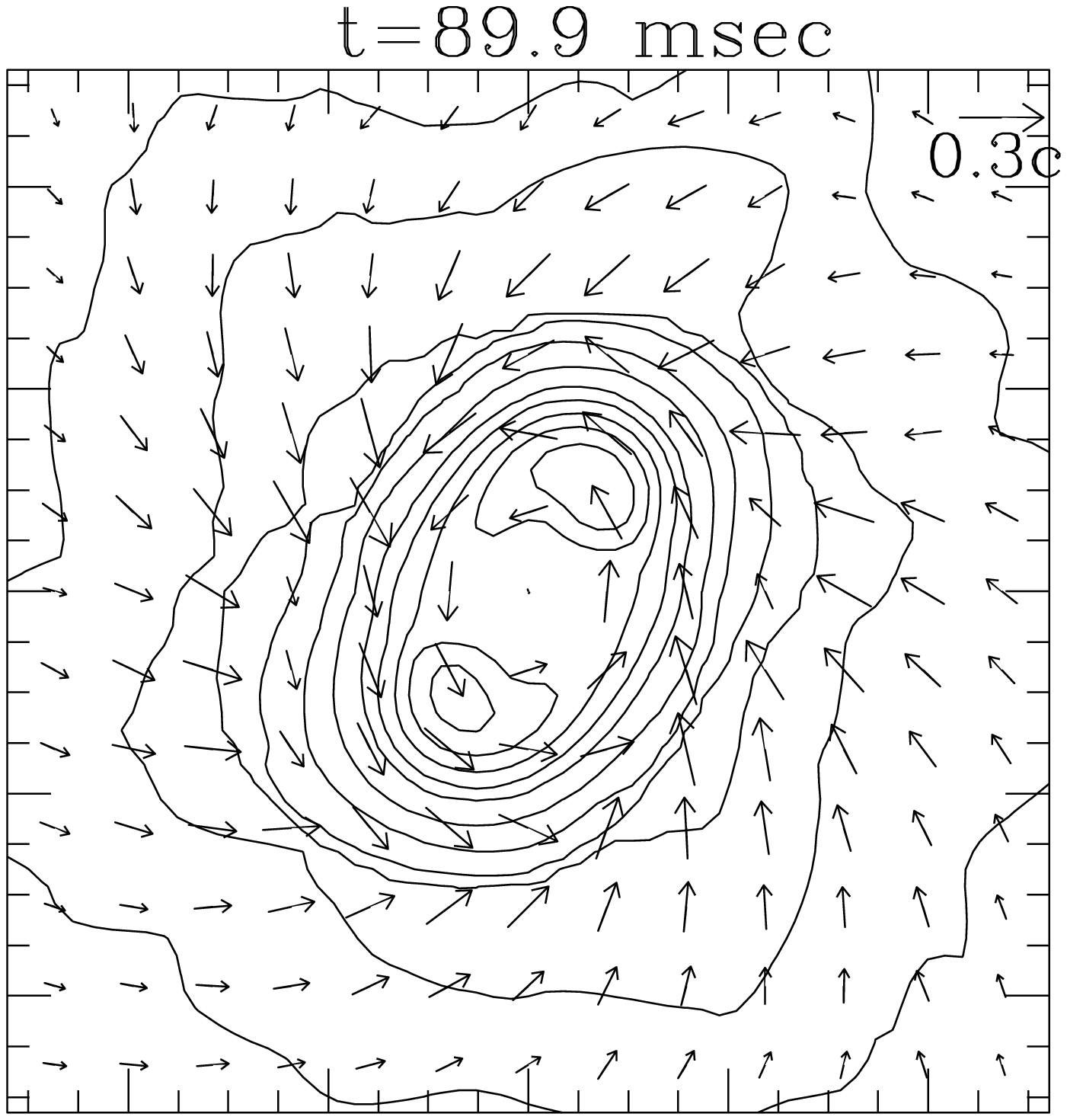} \\
\vspace{-1.2cm}
\epsfxsize=2.4in
\leavevmode
\epsffile{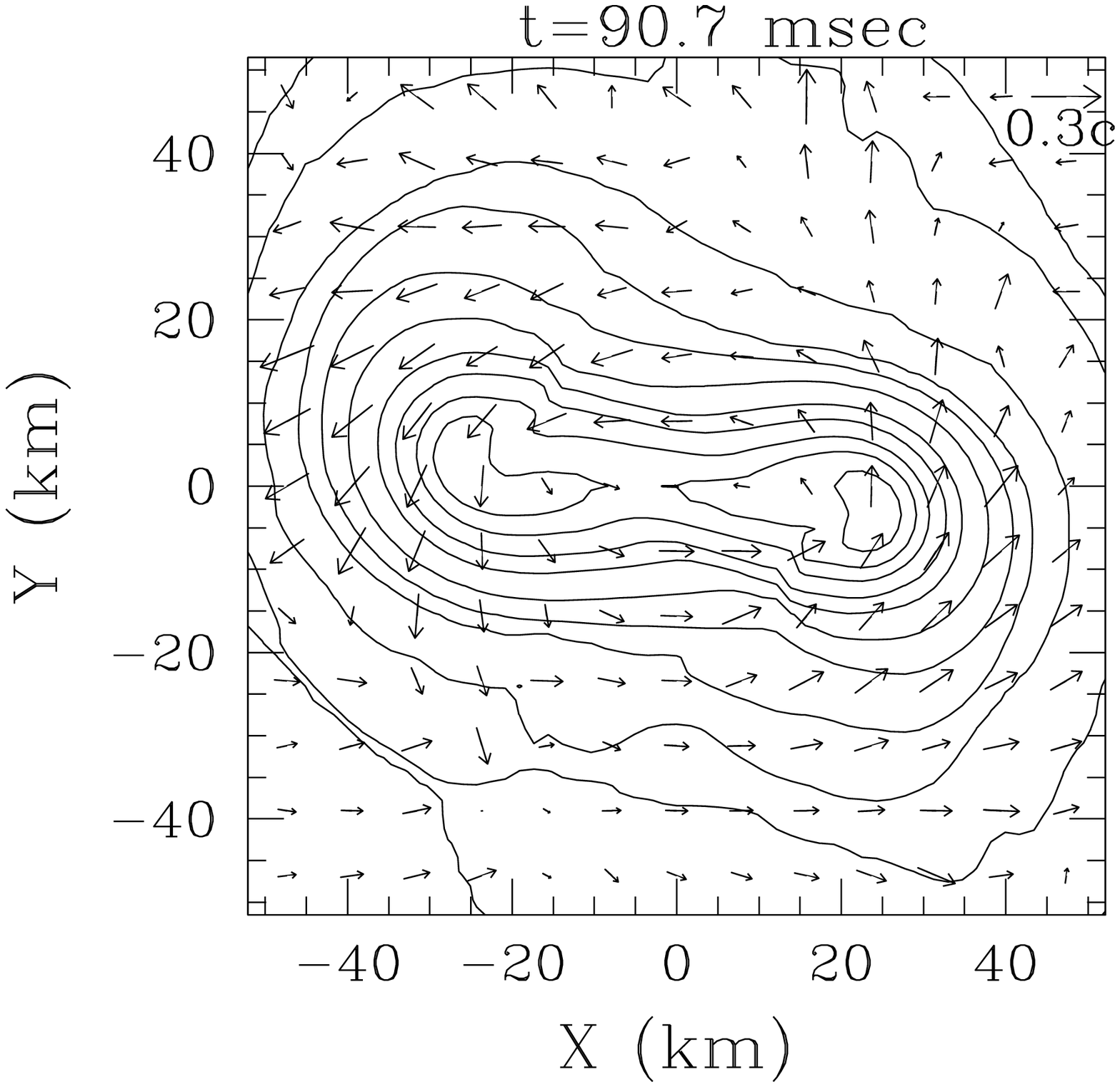} 
\epsfxsize=2.4in
\leavevmode
\hspace{-1.2cm}\epsffile{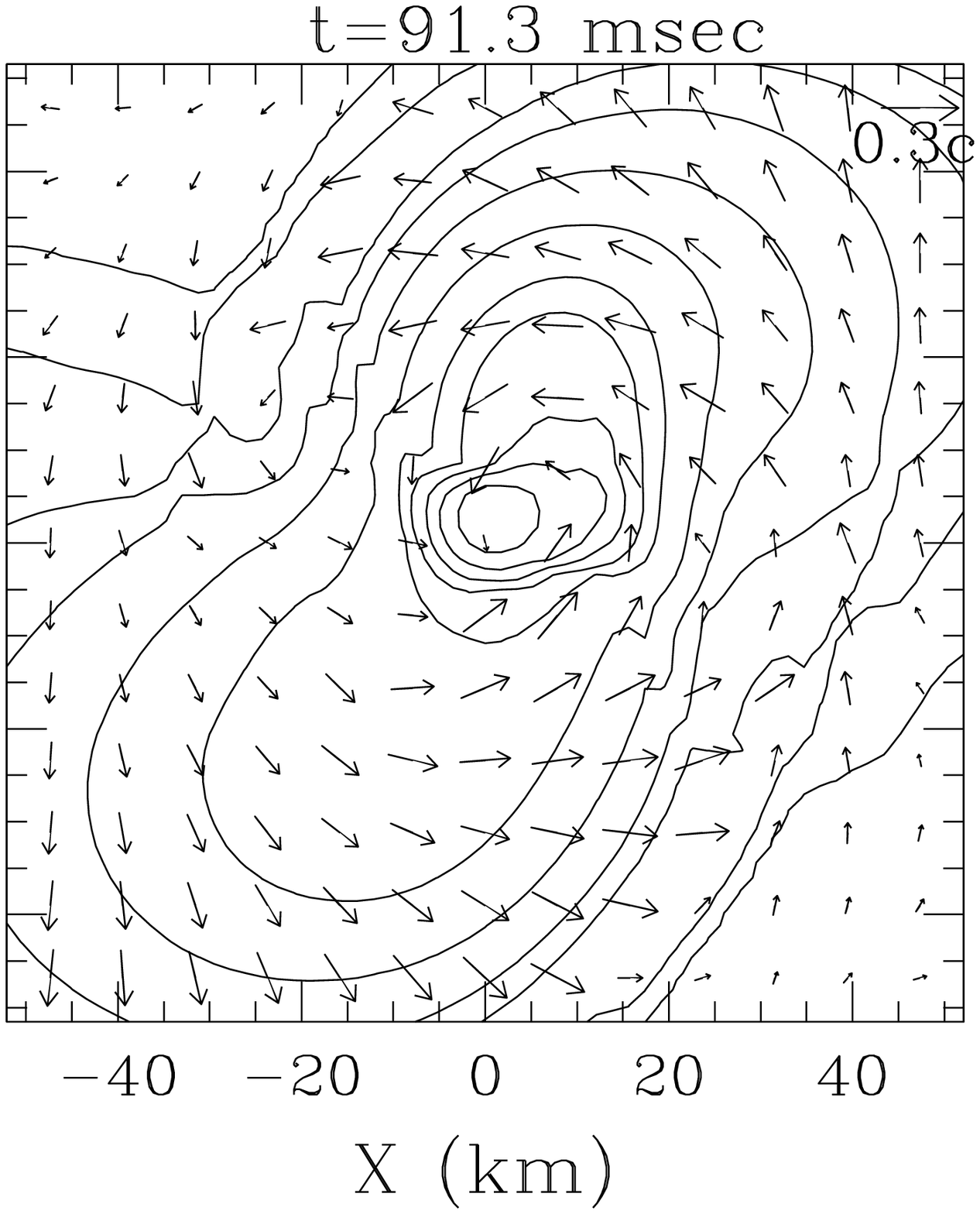}
\epsfxsize=2.4in
\leavevmode
\hspace{-1.2cm}\epsffile{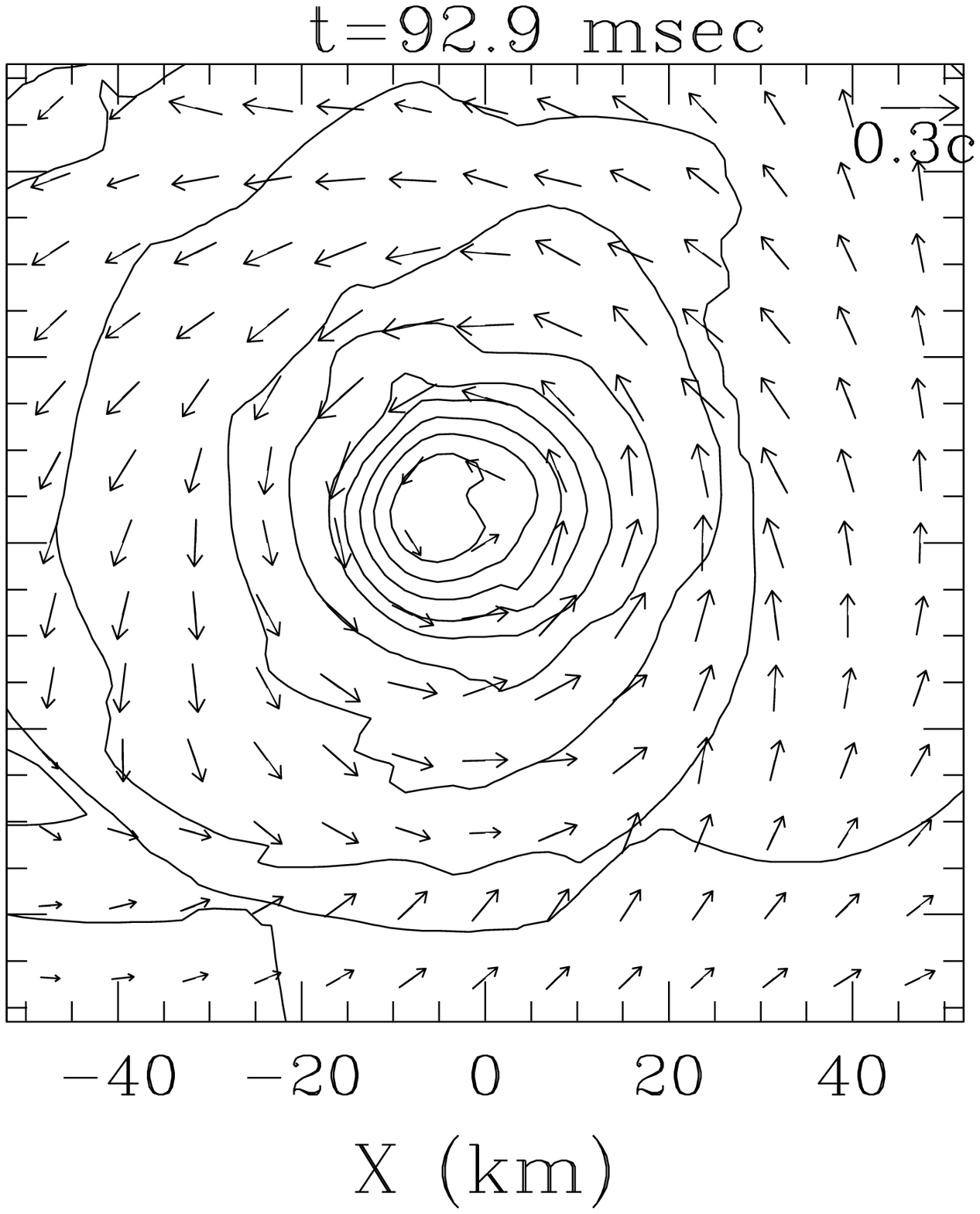} 
%\vspace{-1.2cm}
%\epsfxsize=2.4in
%\leavevmode
%\epsffile{fig10g.ps} 
%\epsfxsize=2.4in
%\leavevmode
%\hspace{-1.2cm}\epsffile{fig10h.ps}
%\epsfxsize=2.4in
%\leavevmode
%\hspace{-1.2cm}\epsffile{fig10i.ps}
\end{center}
\vspace{-8mm}
\caption{
Snapshots of the density contour curves for $\rho$ 
in the equatorial plane for model M7c2C. 
The solid contour curves are drawn for
$\rho/\rho_{\rm max}=1$, 0.8, 0.6, 0.4, 0.2, and $10^{-j/2}$ 
for $j=2,\cdots,8$. 
Vectors indicate the local velocity field $(v^x,v^y)$, and the scale 
is shown in the upper right-hand corner. 
\label{FIG10} }
\end{figure}

\begin{figure}[t]
\begin{center}
\epsfxsize=2.4in
\leavevmode
\epsffile{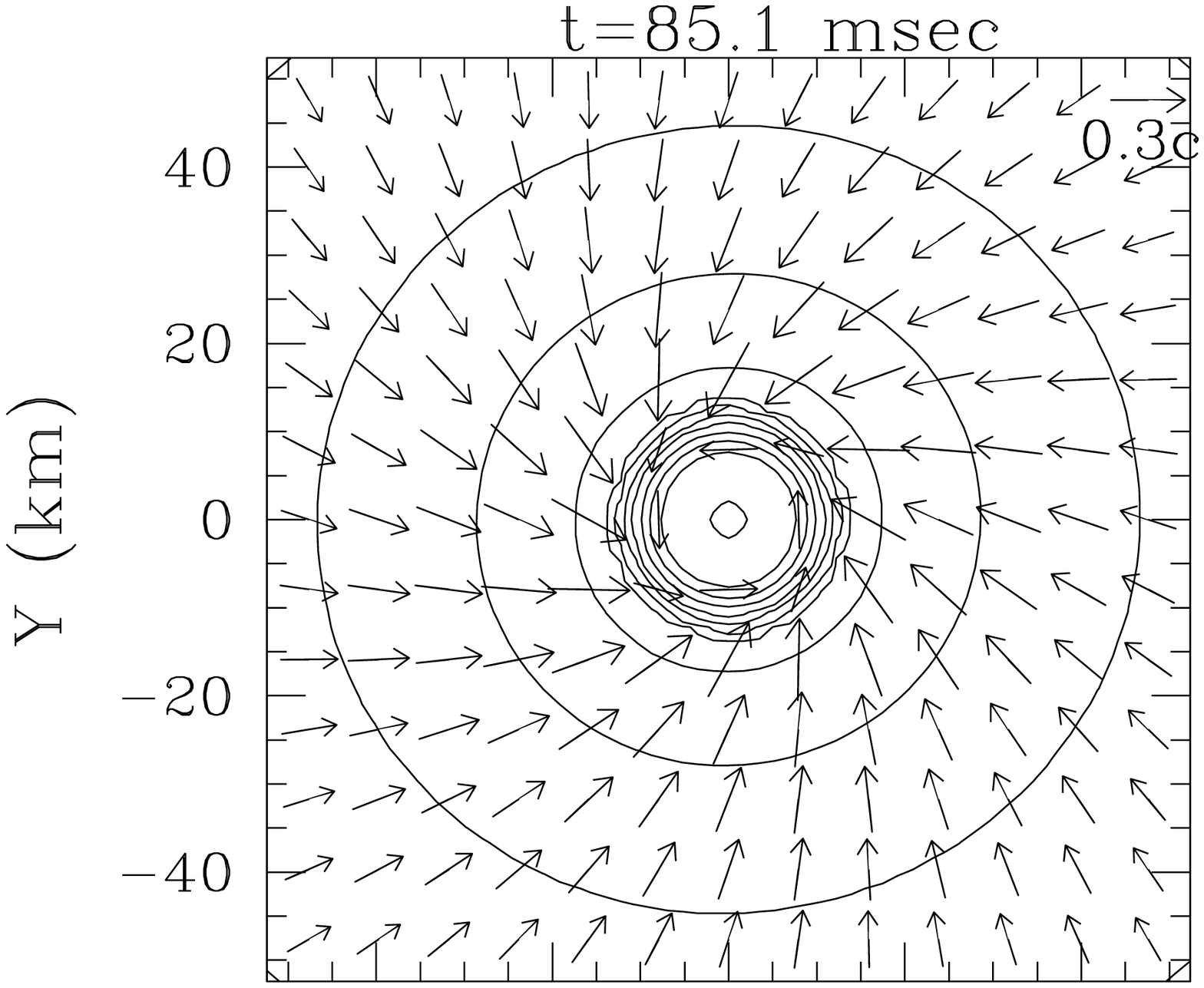}
\epsfxsize=2.4in
\leavevmode
\hspace{-1.2cm}\epsffile{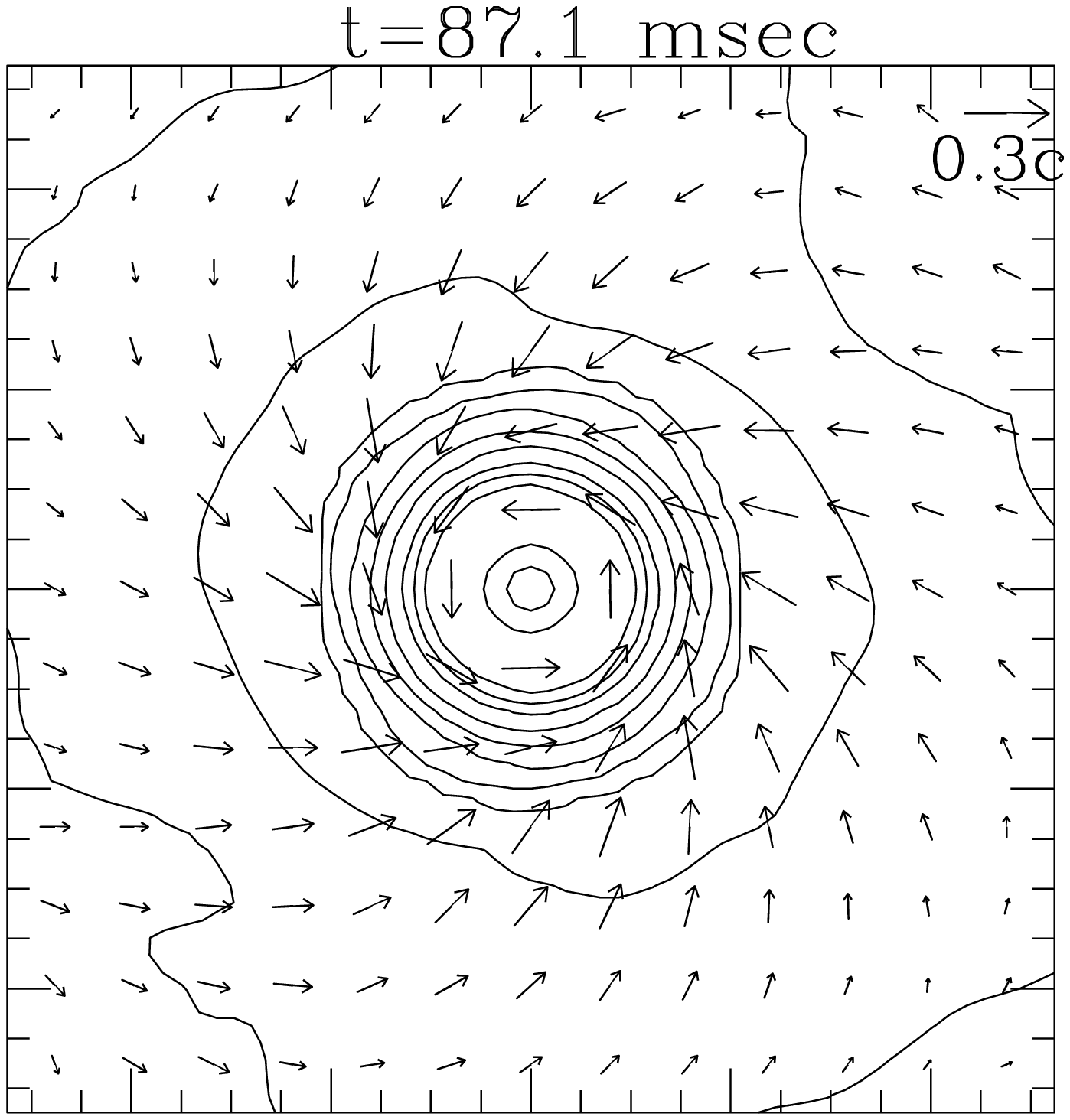} 
\epsfxsize=2.4in
\leavevmode
\hspace{-1.2cm}\epsffile{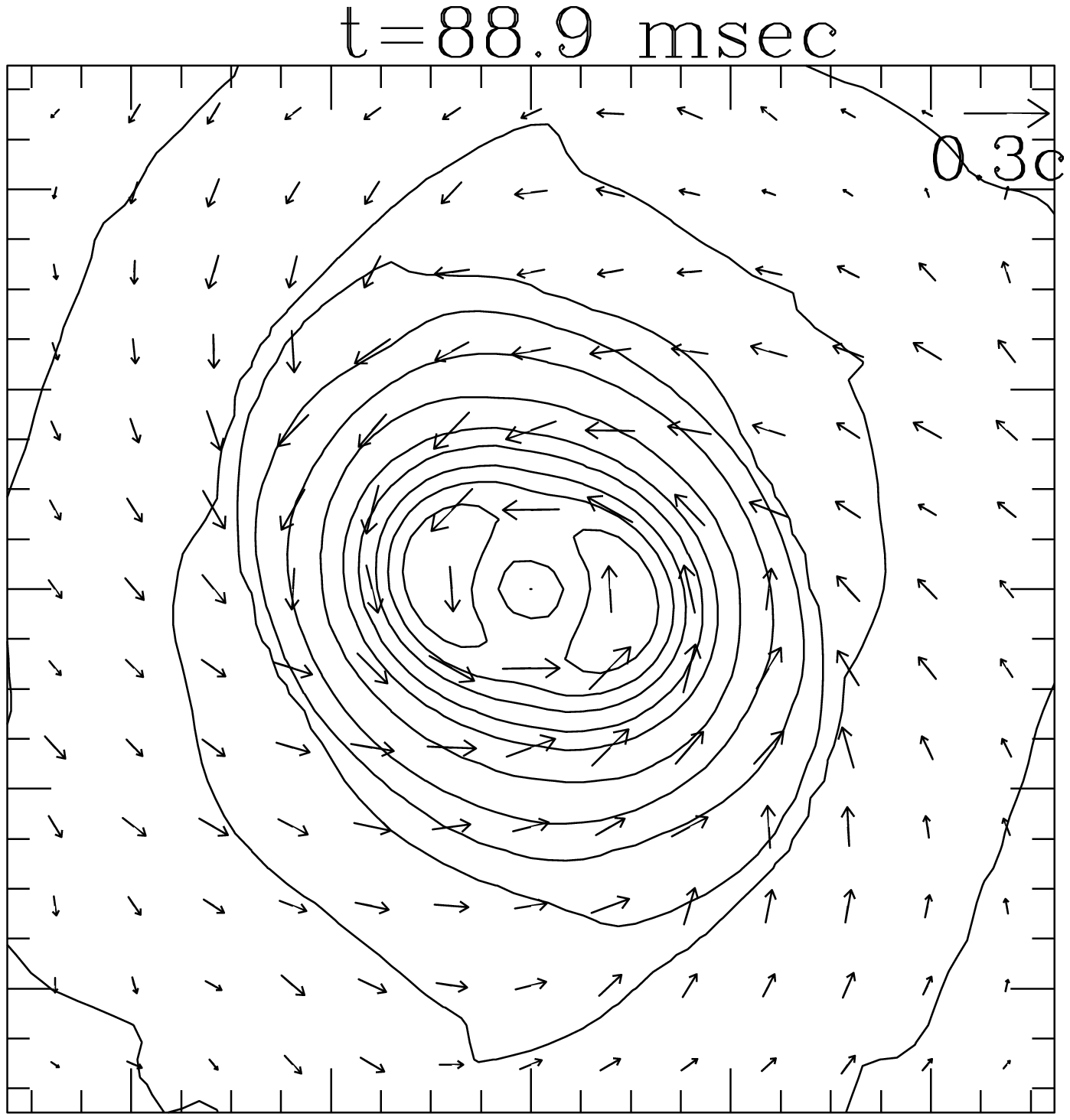} \\
\vspace{-1.2cm}
\epsfxsize=2.4in
\leavevmode
\epsffile{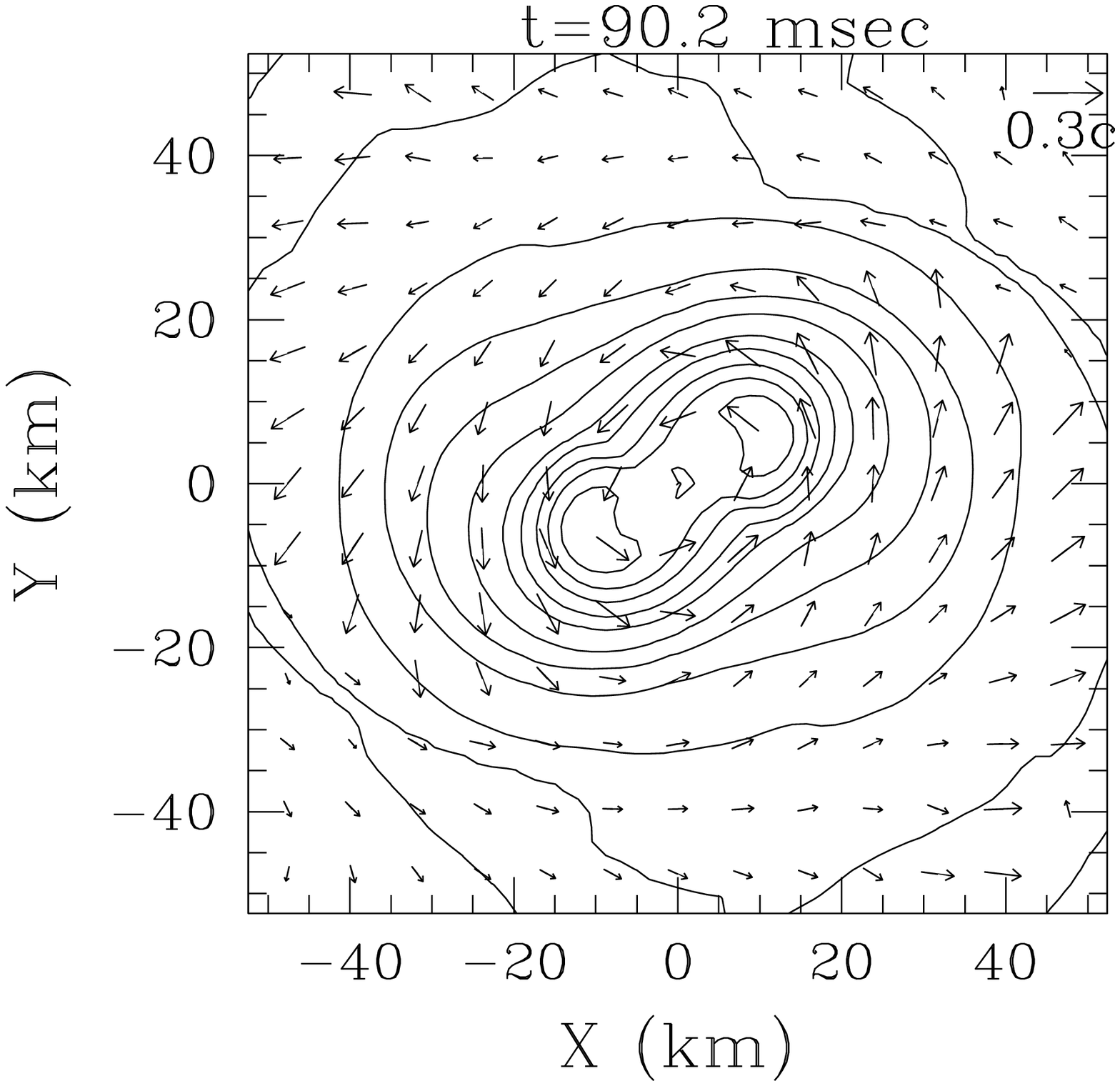} 
\epsfxsize=2.4in
\leavevmode
\hspace{-1.2cm}\epsffile{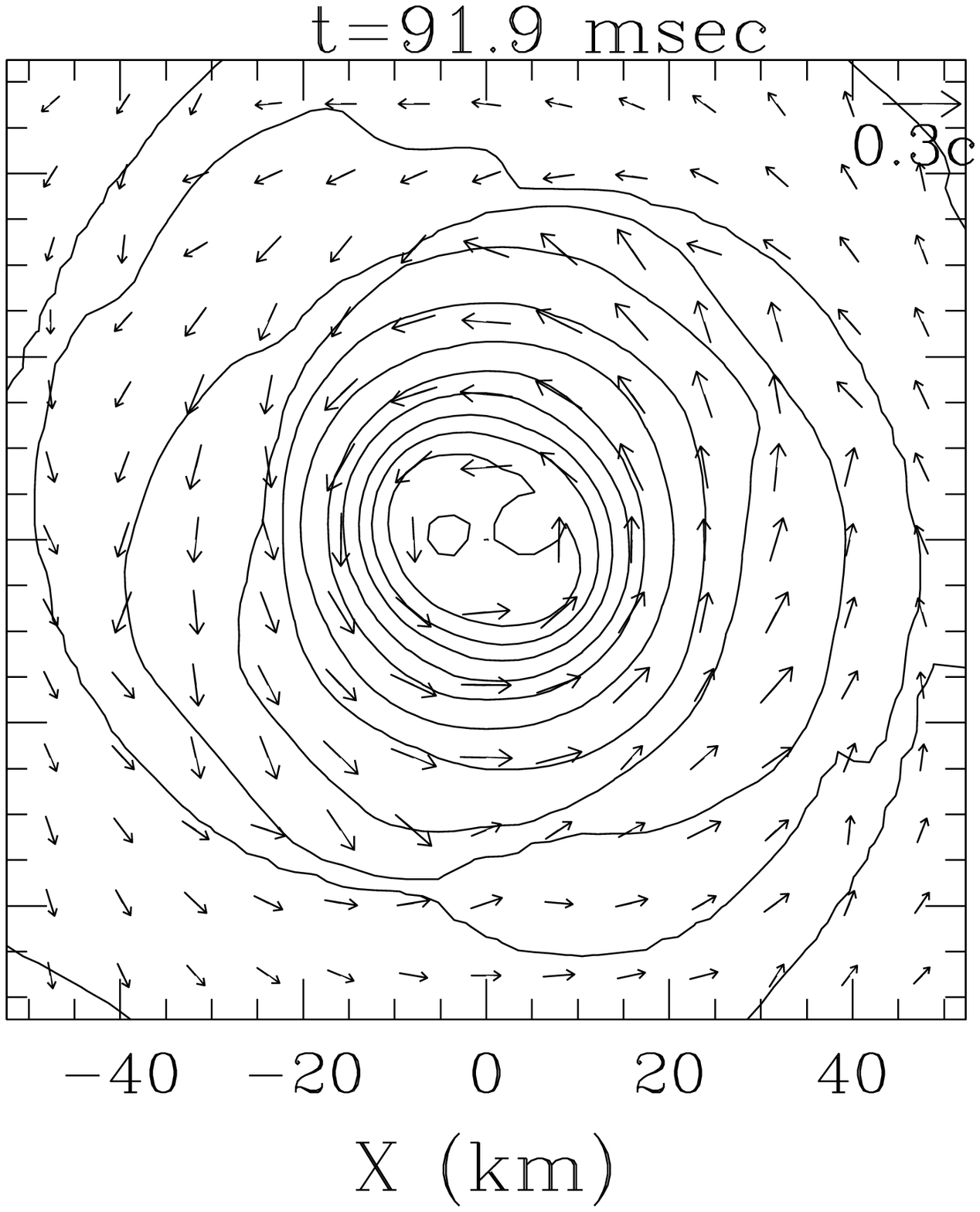}
\epsfxsize=2.4in
\leavevmode
\hspace{-1.2cm}\epsffile{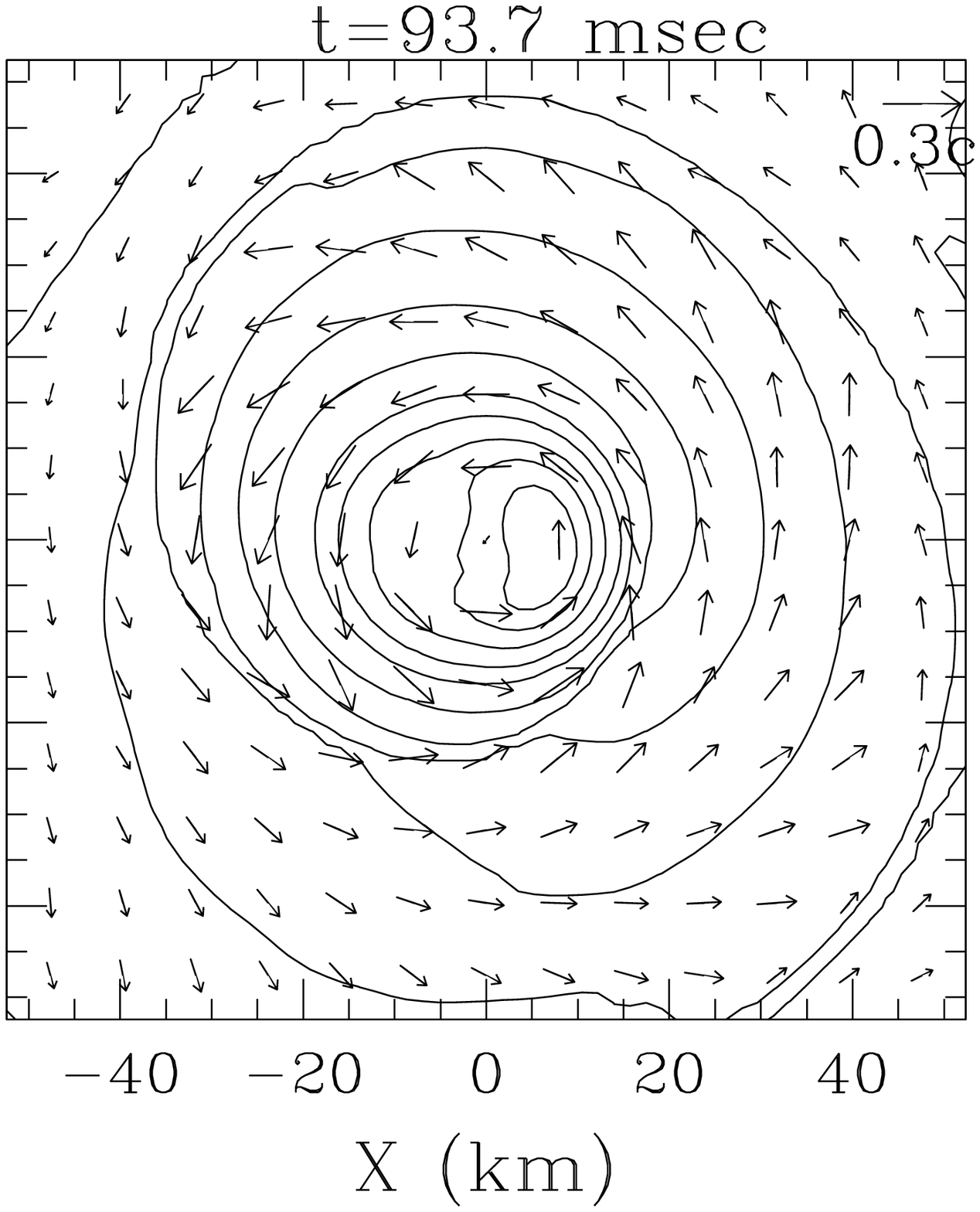} 
\end{center}
\vspace{-8mm}
\caption{The same as Fig. 10 but for model M7c3C. 
\label{FIG11} }
\end{figure}

\begin{figure}[t]
\begin{center}
\epsfxsize=2.4in
\leavevmode
\epsffile{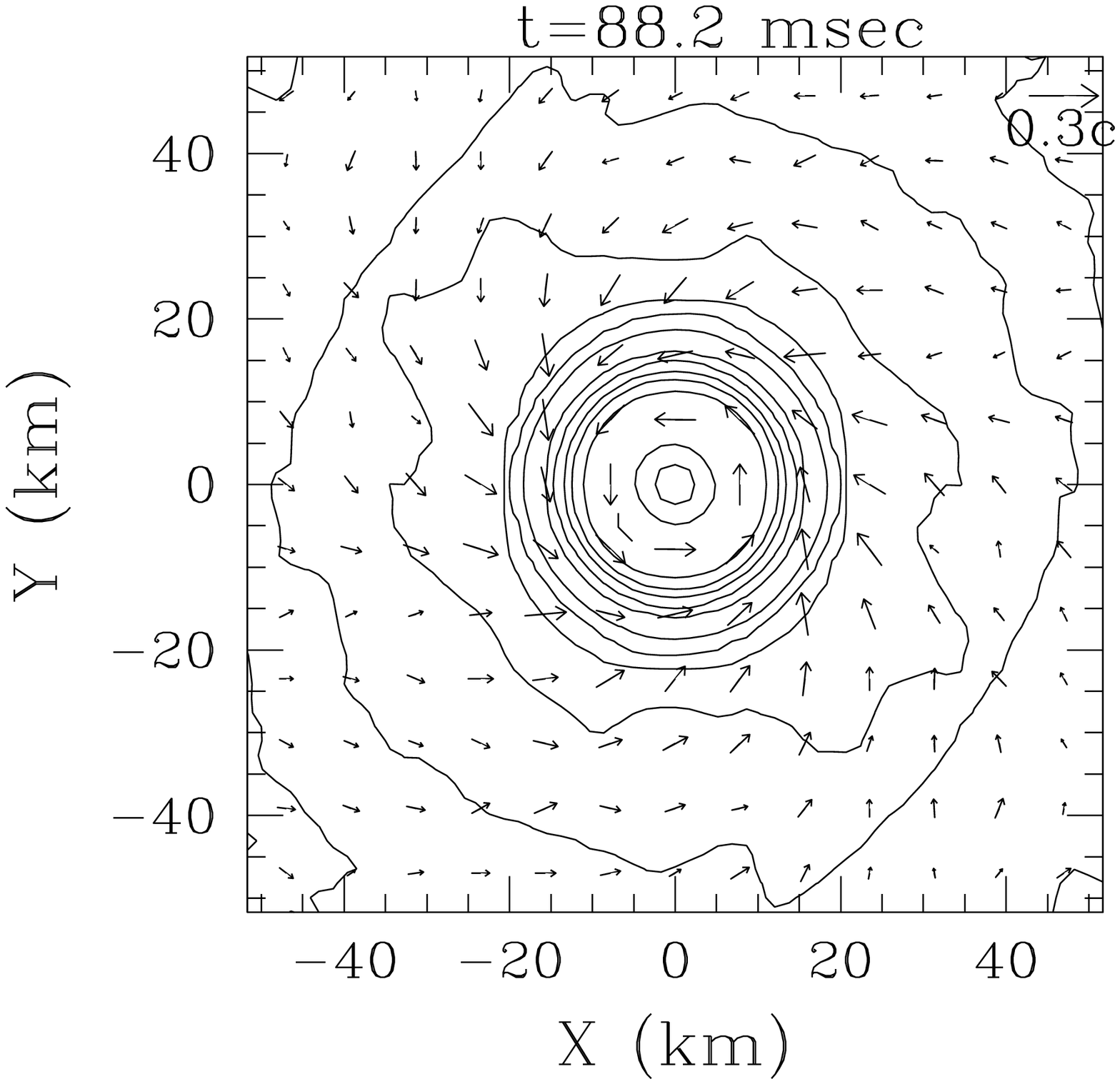}
\epsfxsize=2.4in
\leavevmode
\hspace{-1.2cm}\epsffile{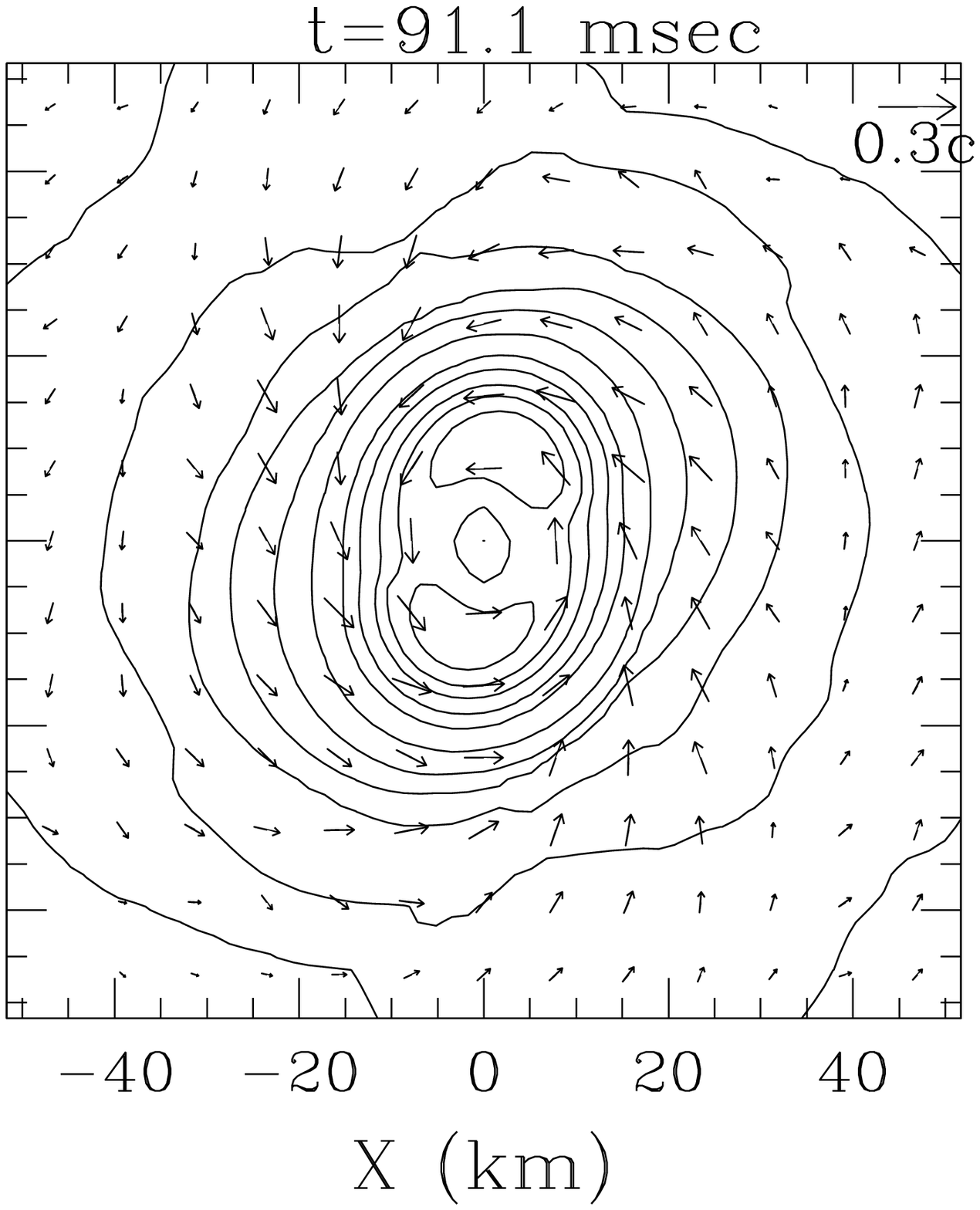} 
\epsfxsize=2.4in
\leavevmode
\hspace{-1.2cm}\epsffile{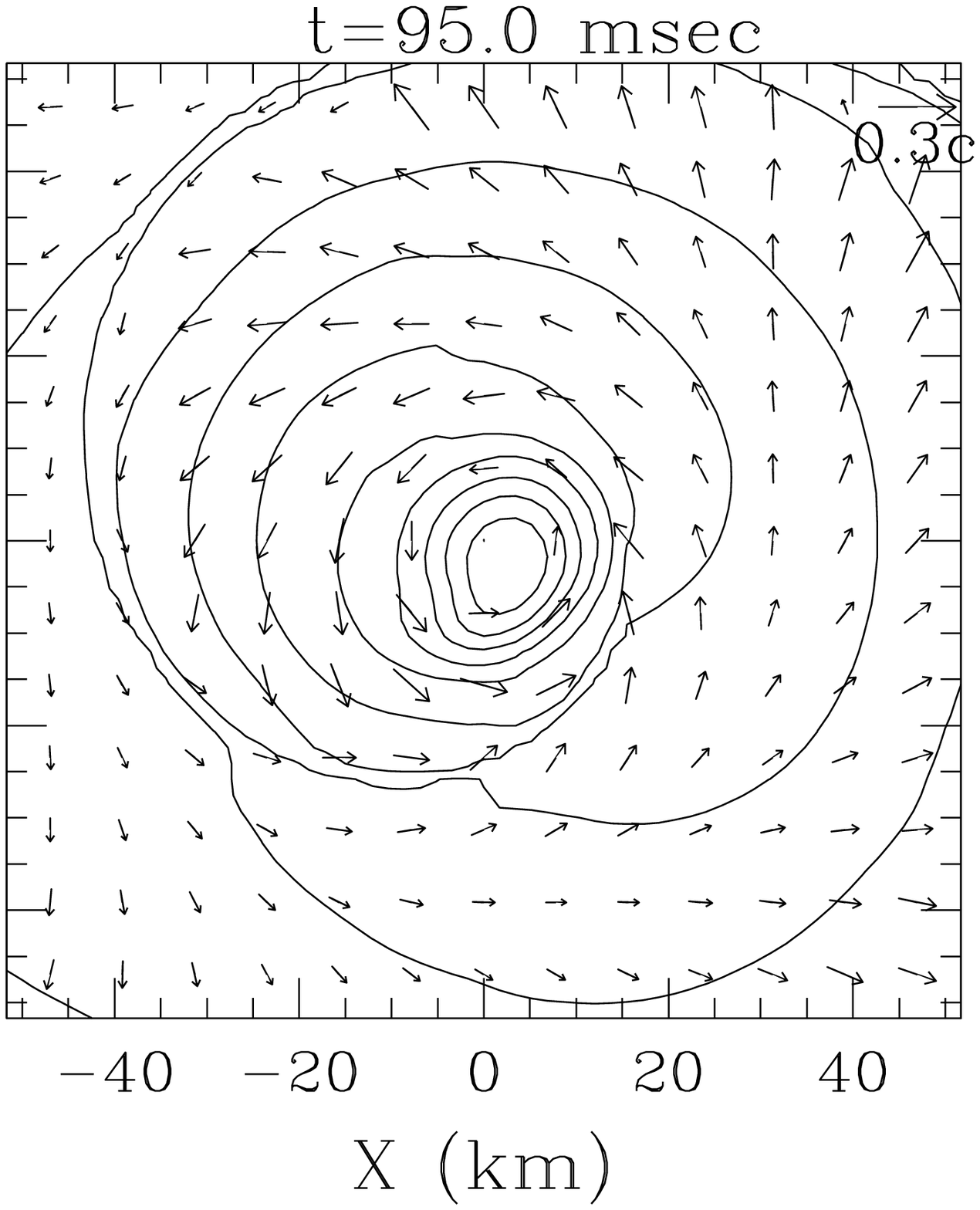}
\end{center}
\vspace{-8mm}
\caption{The same as Fig. 10 but for model M5c2C. 
\label{FIG12} }
\end{figure}

In Figs. 10--12, we display the snapshots of the density contour
curves and velocity vectors for the selected time slices 
for models M7c2C, M7c3C, and M5c2C.
In all the models, the collapse proceeds in approximately
axisymmetric manner throughout the initial collapse to the first bounce, 
forming a torus-like structure. 
For M7c2C, after the first bounce, the formed core expands by a
large factor, and then, collapses again. In this second collapse,
nonaxisymmetric instabilities grow significantly: In the torus-like
high-density region, two density peaks are formed 
(third panel of Fig. 10). Then, the separation of the
density peaks increases, and a bar-like structure is formed (fourth panel), 
developing spiral arms in the outer region.
Subsequently, the separation decreases, and
they eventually merge and form a single peak (fifth and sixth panels).
In the outer region, spiral arms are developed, 
which play a role for transferring the angular momentum
of the formed core to the outer region. Because of this angular momentum
transfer as well as the dynamical friction force
to the bar from the surrounding matter, the nonaxisymmetric structure of 
the central core is quickly erased and the protoneutron star 
eventually relaxes to a slightly nonaxisymmetric quasistationary state. 

For model M7c3C, the nonaxisymmetric instabilities grow
in a similar manner to that of M7c2C. In this case, however,
the maximum value of $\eta$, which denotes
the achieved maximum degree of nonaxisymmetric deformation, 
is slightly smaller. This seems to reflect the fact that the angular momentum
is not as large as that of M7c2C.
For model M5c2C, the evolution is very similar to that of M7c3C.
However, the growth rate of $\eta$ for M5c2C is slightly smaller than 
for M7c2C. The reason is that the mass and the compactness of the 
outcome formed after the collapse are smaller, and hence,
the growth time of the nonaxisymmetric dynamical instabilities,
which is approximately proportional to the dynamical time scale,
becomes longer. 

A noteworthy feature for the unstable models is that 
in the late phase in which the bar-mode perturbation damps,
the $m=1$ mode grows gradually and becomes a dominate mode
eventually.
With the growth of this mode, small one-armed spiral arm is formed
(see, e.g.,  the last panels of Figs. 10--12).
The excitation of this mode is probably due to the fact that 
the formed star is highly differentially rotating \cite{M1,M11}.
However, the effect of its growth is not very 
outstanding since the amplitude of the perturbation is not
very large and fairly quickly damps due to the angular momentum
transfer to the outer region. Therefore, we conclude that the onset of 
$m=1$ mode instabilities is not as important as that of the bar-mode
for the evolution of the system, 
although the density configuration of the 
formed protoneutron star becomes asymmetric due to it. 

The formation of the bar and subsequent 
outward transfer of the angular momentum change the
density profile of the protoneutron stars. In Fig. 13, 
we show the evolution of $\alpha_c$, $\rho_{\rm max}$, and $\eta$ 
for models M7c2C and M7c3C in the three-dimensional simulations
as well as in the axisymmetric ones.
To illustrate that the convergence is approximately achieved,
the three-dimensional results with $N=156$, 188, and 220 are shown together. 
In the early stage of the evolution ($t \alt 88$ msec) in which
the amplitude of the bar-mode perturbation is small,
the results of the three-dimensional and axisymmetric simulations
are in good agreement: i.e., $\alpha_c$ and $\rho_{\rm max}$ are 
simply in a damped oscillation.
Slight disagreement between the results of the three-dimensional and 
axisymmetric simulations is likely due to the fact that we 
discard the matter located in the outer region in the three-dimensional
simulations.
In a stage in which the system is approximately axisymmetric,
shock dissipation which damps the oscillation is only
the mechanism for modifying the density profile. On the other hand, 
in the late stage with $\eta \agt 0.1$ (see Figs. 13 (c) and (d)),
$\rho_{\rm max}$ ($\alpha_c$) gradually increases (decreases) with
time in the three-dimensional simulations.
This reflects the effect of the angular momentum transfer 
by which the centrifugal force in the inner region is weaken 
and the formed object becomes more compact than the outcome 
in the axisymmetric simulations. 
In particular, the effect is remarkable for model M7c2C. In this case, 
an oscillating (type O-A) star (not a protoneutron star
in the definition of this paper) is formed in the early phase of
the axisymmetric simulation, while in the three-dimensional simulation,
the protoneutron star is promptly formed because of the quick contraction
due to the outward transfer of the angular momentum. 
For more massive cases with $M \agt 3M_{\odot}$,
protoneutron stars which are supported by strong
differential rotation may be formed first \cite{BSS}, but 
the angular momentum transfer may trigger black hole formation. 
This effect may also play an important role in stellar collapse of 
very massive ($M \agt 250 M_{\odot}$) stars (population III stars) 
which is triggered by the electron-positron pair creation instability
\cite{Fryer}. Very massive stars are likely to be rapidly rotating
\cite{Bond}, and the collapse may not result directly
in a black hole but in very massive self-gravitating disks \cite{S03b}.
The disks will be dynamically unstable against nonaxisymmetric 
deformation and the resulting angular momentum transfer by a 
nonaxisymmetric structure may induce black hole formation. 

Figure 13 also shows that with increasing the value of $N$,
the numerical results achieve a convergence. 
The results of $N=188$ and 220 agree well 
(except for those in the very late time for which the numerical error
is accumulated too much), implying that 
a convergent result is obtained with $N \sim 200$.
By the way, the period of the quasiradial 
oscillation becomes spuriously longer due to
the larger numerical dissipation with the smaller value of $N$. Such 
spurious effect may lead to underestimation of 
the growth rate and the achieved maximum value of $\eta$. 
The convergent test carried out
here gives us a caution that we have to guarantee 
a sufficient grid resolution in this problem. 

In the lower panels of Fig. 13(c) and (d), 
we display the evolution of $\eta_0$ to compare with that of $\eta$.
For model M7c2C, $\eta_0$ increases to be much larger than the initial value.
However, the value of $\eta_0$ is smaller than that of $\eta$ even when
the bar-mode grows to a nonlinear regime. This implies that the bar
structure is formed mainly in the central region.
This feature is more outstanding for model M7c3C in which
the increase of $\eta_0$ from the initial value
is not seen. Thus, we conclude that the bar-mode perturbation
is amplified only in the central region. This is reasonable since
in the models with $A=0.1$, the outcomes are rapidly 
rotating only in the central region. 

\begin{figure}[htb]
\vspace*{-6mm}
\begin{center}
\epsfxsize=2.8in
\leavevmode
(a)\epsffile{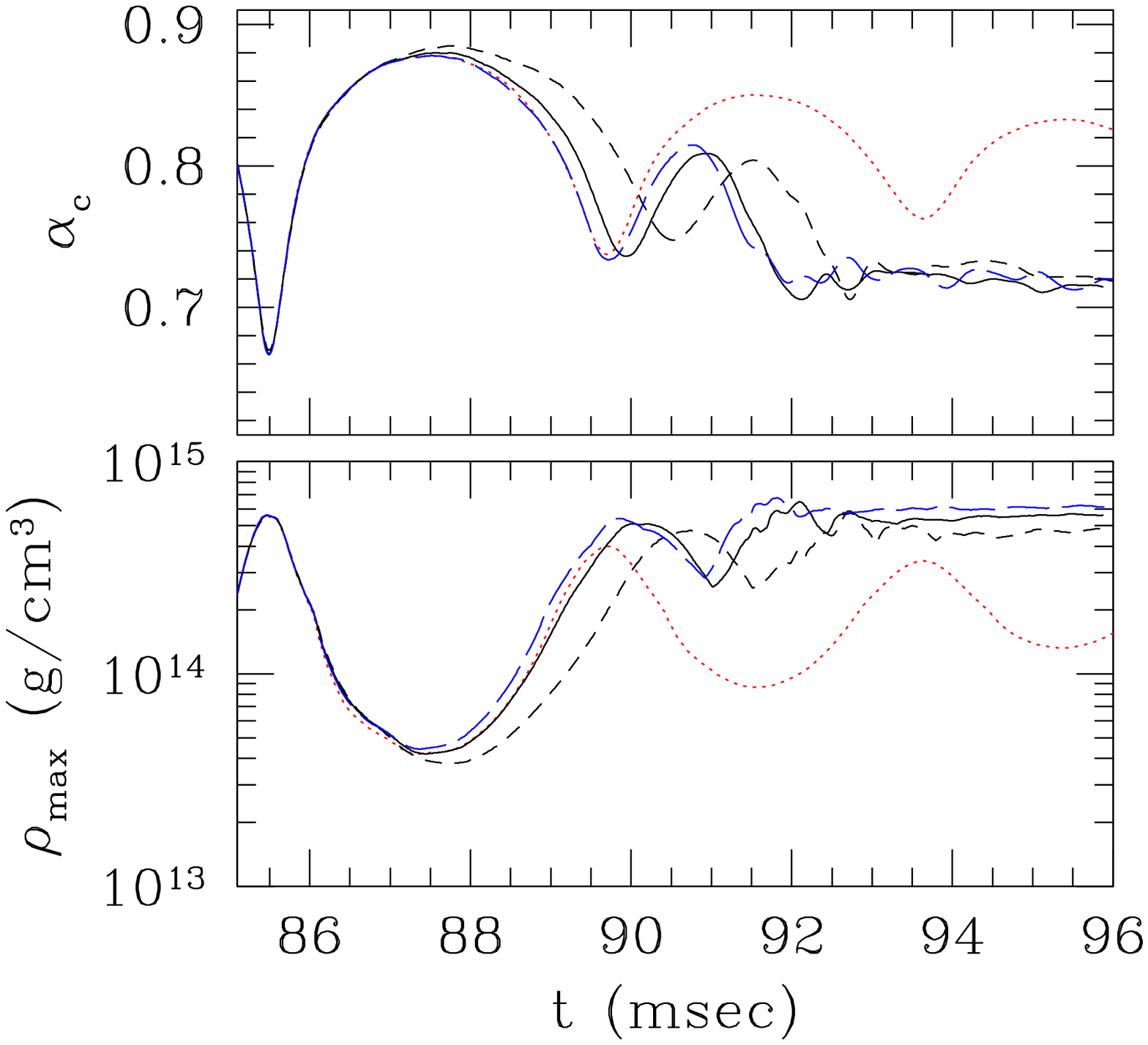}
\epsfxsize=2.8in
\leavevmode
~~~(b)\epsffile{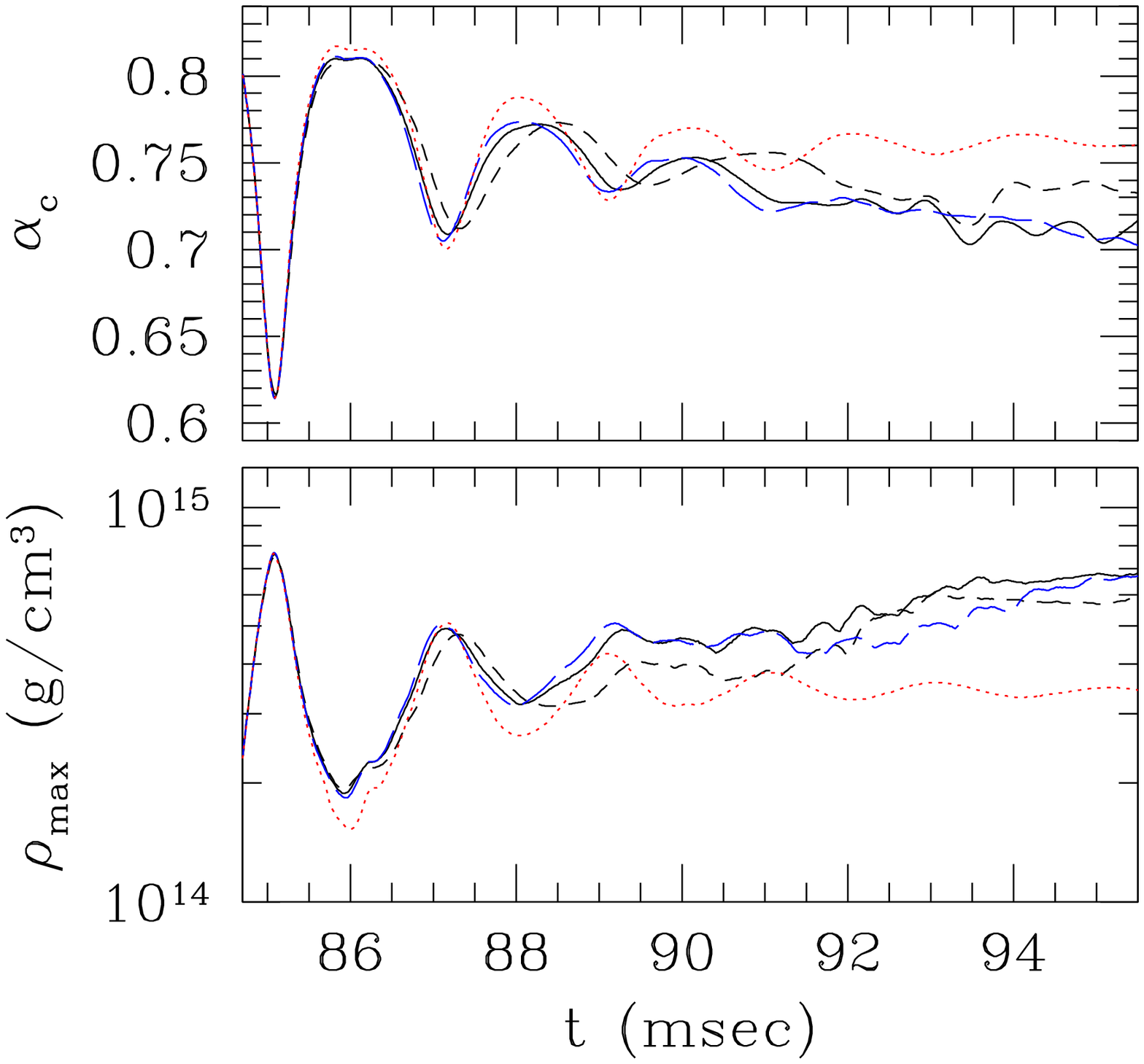}\\
\vspace*{-4mm}
\epsfxsize=2.8in
\leavevmode
(c)\epsffile{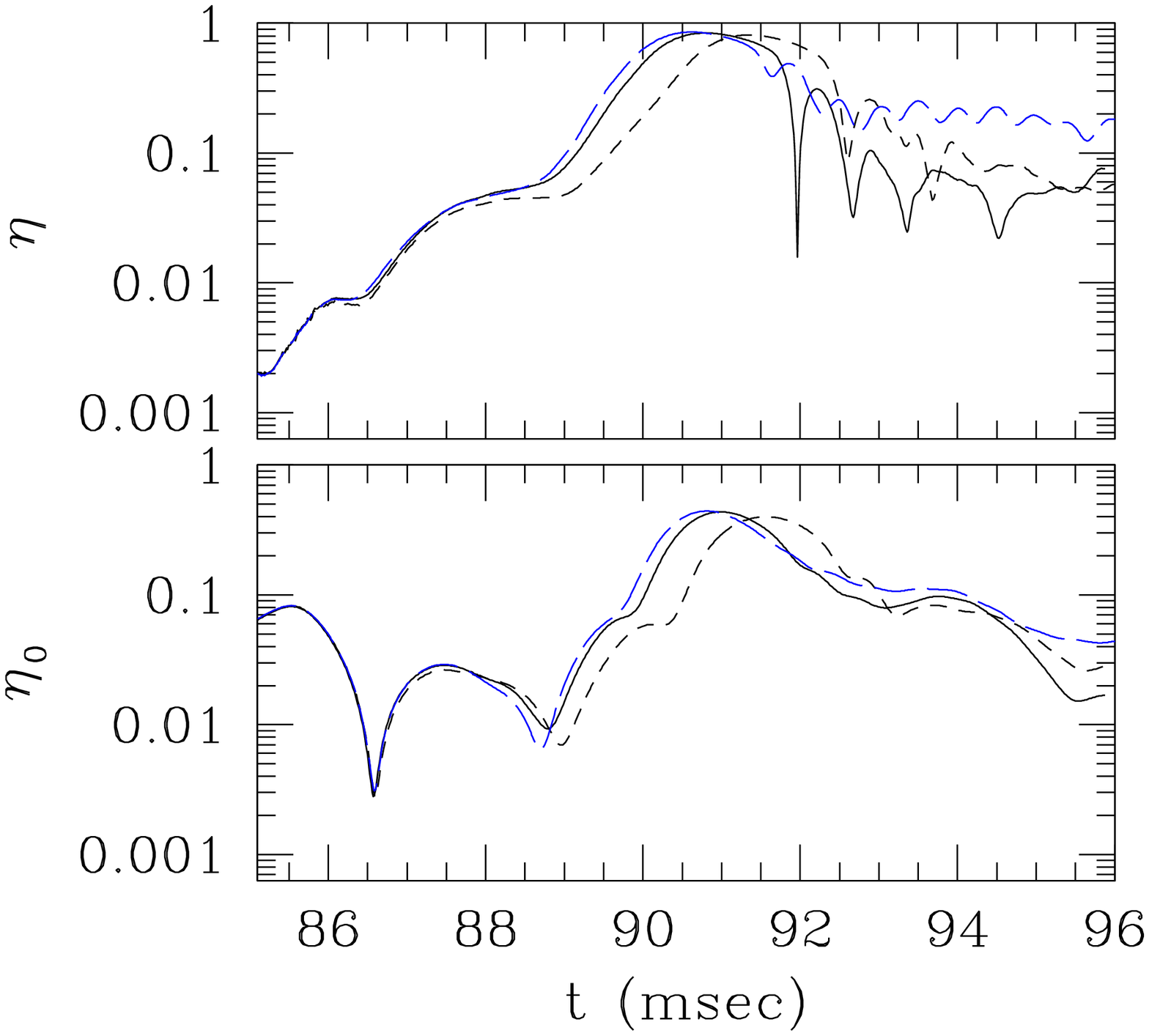}
\epsfxsize=2.8in
\leavevmode
~~~(d)\epsffile{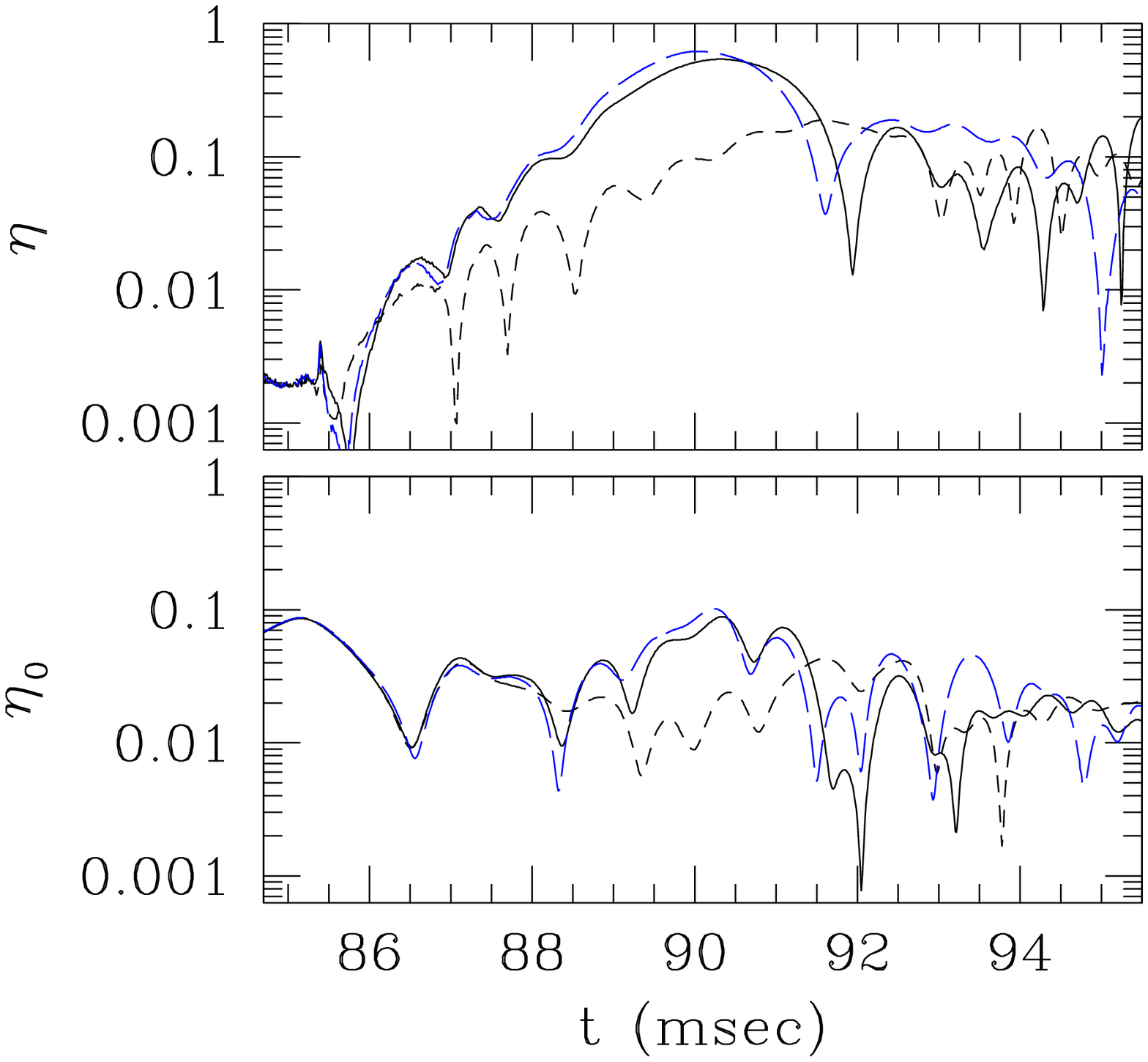}
%\vspace*{-4mm}
\caption{Evolution of $\alpha_c$ and $\rho_{\rm max}$
(a) for model M7c2C and (b) for model M7c3C in three-dimensional simulations
with $N=156$ (dashed curves), 188 (solid curves), and 220 (long-dashed
curves) as well as in axisymmetric simulation (dotted curves).
Evolution of $\eta$ and $\eta_0$
(c) for model M7c2C and (d) for model M7c3C
in three-dimensional simulations with varying grid resolution. 
\label{FIG13}}
\end{center}
\end{figure}

\subsection{Criterion for the onset of nonaxisymmetric dynamical instabilities}

\begin{figure}[htb]
\vspace*{-4mm}
\begin{center}
\epsfxsize=2.8in
\leavevmode
\epsffile{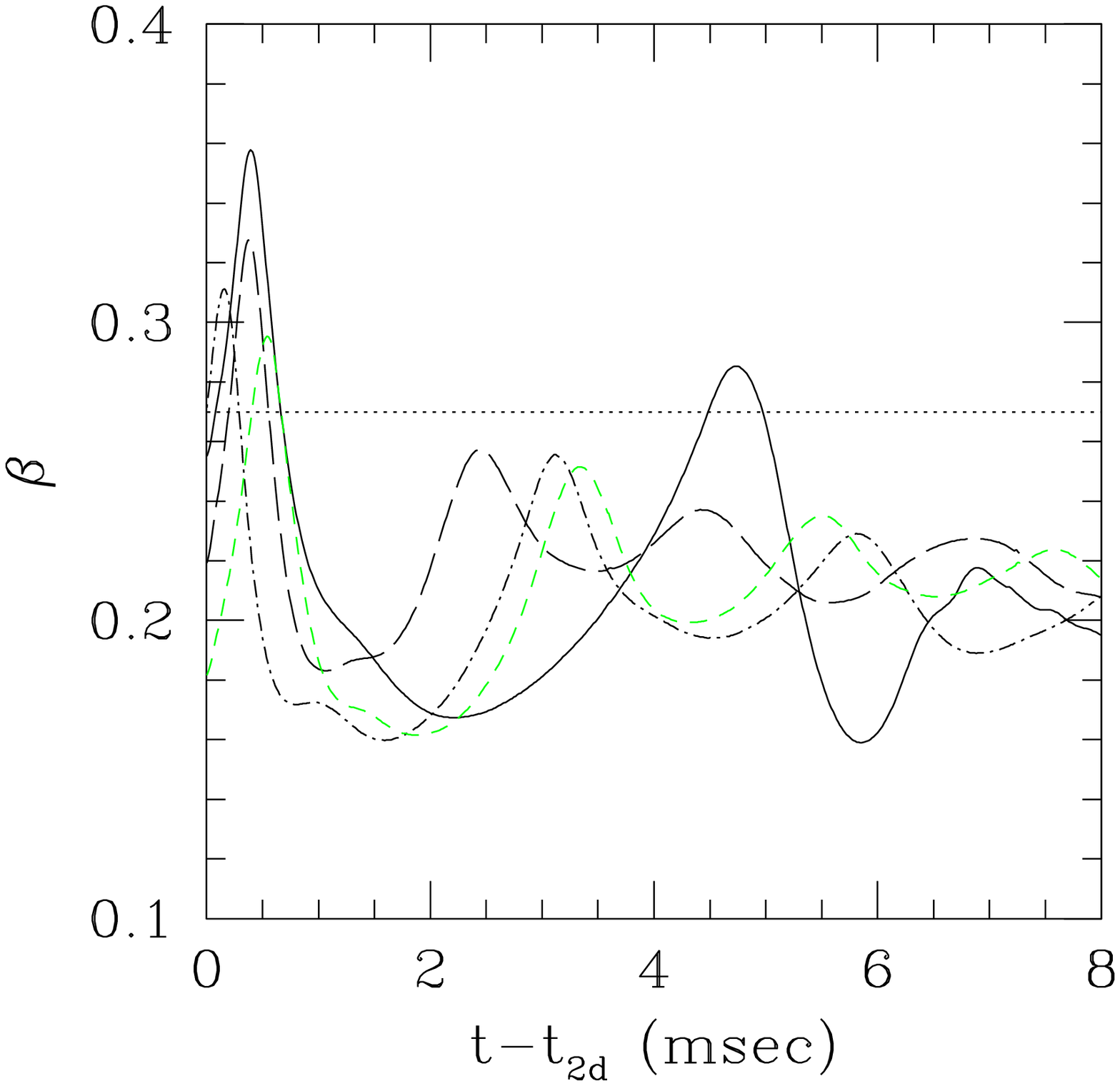}
\vspace*{-6mm}
\caption{
Evolution of $\beta$ 
for models M7c2C (solid curve), M7c3C (long-dashed curve), M7c3A
(dashed curve), and M5c2C (dotted-dashed curves) in three-dimensional
simulations. The values of $t_{\rm 2d}$ are listed in the caption of Fig. 9. 
The dotted line denotes $\beta=0.27$. 
\label{FIG14}}
\end{center}
\end{figure}

Models M7c2C and M7c3C are dynamically unstable
against bar-mode and $m=1$ mode deformation, while model M7c4C is stable
for both modes. This implies that for the onset of dynamical
nonaxisymmetric instabilities, 
high values of $\beta$ are necessary for given values of $\Gamma_1$ and $M$.

All the models with $\Gamma_1=1.3$ that we picked up here are stable.
On the other hand, the models with $\Gamma_1=1.28$ are much more
prone to be unstable. This suggests that only for a sufficiently
small value of $\Gamma_1 \alt 1.28$, 
the collapsing star can be unstable. As mentioned in Sec. IV, 
for the smaller value of $\Gamma_1$
(for the larger depletion factor of the internal energy and
the pressure at the onset of collapse), the outcomes are more 
torus-like than those for other values of $\Gamma_1$, 
and also, the degree of differential rotation is larger. 
These facts are likely reasons that models with $\Gamma_1 =1.28$
are more subject to the dynamical instabilities. 

Even with $M \approx 1.5M_{\odot}$ (models M5c1C and M5c2C),
nonaxisymmetric bar-mode instabilities set in, although
the maximum values of $\beta$ and compactness for the outcomes 
are smaller than for models M7c2C and M7c3C. 
This indicates that the mass and compactness achieved in the collapse
are not very important parameters for triggering the bar-mode
instabilities as far as $M$ is larger than $\sim 1.5M_{\odot}$.
However, it should be noted that 
general relativistic effects certainly help making a compact outcome.
Thus, if the mass is much smaller than $\sim 1.5M_{\odot}$,
nonaxisymmetric instabilities may not set in. 

Although a high value of $\beta$ is necessary, 
the onset of the nonaxisymmetric dynamical instabilities is not simply
determined by the value of $\beta$; i.e., although
a large value of $\beta \agt 0.27$
is preferable for the onset, it is neither the necessary 
nor the sufficient conditions. The first evidence for this statement is
that the values of $\eta$ for the unstable 
models do not increase at the first bounce at which the value of
$\beta$ becomes maximum with $\beta_{\rm max} \agt 0.27$.
The growth of the perturbation
is significantly induced in the subsequent bounce stages.
Also, model M7c2A is dynamically stable
although $\beta_{\rm max}\approx 0.33$. 
These show that even if $\beta$ exceeds $\sim 0.27$, 
the nonaxisymmetric perturbations do not grow. Probably, 
the duration of the phase for which $\beta > 0.27$ would have
to be much longer than 
the dynamical time scale for the onset of the dynamical instabilities. 

Second, the value of $\beta$ during the growth of the bar-mode
perturbation is smaller than 0.27 for any unstable model. 
In Fig. 14, we show the time evolution of $\beta$ for models
M7c2C, M7c3A, M7c3C, and M5c2C. It shows that during the growth of 
the perturbation, $\beta$ for models M7c3C and M5c2C 
is at most $\sim 0.25$ and in average $\sim 0.2$,
which is much smaller than the widely-believed critical value $\sim 0.27$.
There are at least three possible reasons that $\beta$ may be 
smaller than 0.27 for the onset of the
nonaxisymmetric dynamical instabilities. 
The first one is that the effective value of $\beta$ in the high-density
region may be larger than the global value, and may be large enough 
for the onset of the nonaxisymmetric dynamical instabilities. 
This is likely to be the case in particular 
for the highly differentially rotating collapse 
since the effective value of $\beta$ in the central high-density
region where the nonaxisymmetric perturbation grows dominantly 
is larger than the whole value for such cases. 
The second possibility is that the onset of the nonaxisymmetric 
dynamical instabilities is due to the high degree of 
differential rotation as indicated in a Newtonian simulation \cite{SKE}. 
In this case, the high value of $\beta > 0.27$ is not necessary. 
The third possibility is that general relativistic effects
reduce the critical value of $\beta$ below 0.27. 
Indeed, in \cite{SBS}, we showed that 
the critical value of $\beta$ can be decreased by $\sim 10\%$
due to the general relativistic effects for compact stars
with the compactness $\sim 0.1$--0.2. 
All these possibilities show that the critical value of $\beta$
for the onset of the nonaxisymmetric dynamical instabilities may be
smaller than 0.27 depending sensitively 
on several parameters, and thus, it cannot be uniquely determined. 

Figure 14 also shows that the values of $\beta$ for models
M7c3A and M7c3C are not very different during the oscillation phase 
although they are stable and unstable against bar-mode deformation,
respectively. This also illustrates that the value of $\beta$ does not
uniquely determine the dynamical stability. 
As shown in Fig. 7, on the other hand, the profiles of the density and
the rotational angular velocity in the central region
are different between two models. Thus, in this case, the degree of
differential rotation and the steepness of the density profile
play an important role for determining the stability. 

The case with $\Gamma_1=1.28$, in which 
the depletion of the pressure and the internal energy 
in an early stage of collapse with 
$\rho \ll \rho_{\rm nuc}$ is largest 
among the three cases, is more subject to the nonaxisymmetric dynamical 
instabilities. This indicates that a large depletion of the internal
energy and the pressure in the early stage is an essential element for
the onset of the nonaxisymmetric dynamical 
instabilities. The reason is that for the larger depletion
factor, the collapse in the central region proceeds significantly
to make a compact core, and hence, to increase the spin of
the central region as illustrated in the axisymmetric
simulations (cf. Fig. 7). 
In a realistic phenomena, the depletion of the pressure and the 
internal energy in the early stage is determined by the partial
photo-dissociation of the iron to 
lighter elements, by the electron capture, and by the
neutronization \cite{ST,Bethe}. Since the 
depletion factor is a crucial parameter, an appropriate modeling 
for such microphysical processes will be necessary for 
a more detailed study on the nonaxisymmetric dynamical instabilities
in the future. 

No evidence for fragmentation of protoneutron stars is found in the
first $\sim 10$--20 msec after the bounce in the present numerical
simulations. Previous studies in the field of protostar formation from
collapsing gas clouds (e.g., \cite{MHN}) show that the fragmentation
occurs when the thermal energy at an initial stage of the collapse is
much smaller than the gravitational potential energy: In the case of
the small thermal energy, a torus-like or a disk-like structure is
formed as a result of the collapse and subsequently the fragmentation
takes place. This indicates that if the value of $\Gamma_1$ is much
smaller than 1.28 (i.e., if the fraction of the depletion of the
internal energy and the pressure in an early stage of collapse is much
larger than $16\%$), the fragmentation may occur during the stellar
core collapse. However, such an extremely small value of $\Gamma_1$
(an extremely large value of the depletion factor) is unlikely to be
achieved in the stellar core collapse \cite{ST,Bethe}, and therefore,
we infer that the fragmentation of protoneutron stars would not occur
in nature, at least in a few 10 msec after the stellar collapse.

To summarize, the nonaxisymmetric dynamical instabilities set in only
for the case that the following conditions are satisfied: (i) the
progenitor of the stellar core collapse is rapidly rotating with the
initial value of $\beta \agt 0.01$, (ii) the degree of differential
rotation for the velocity profile of the initial condition is very
high with $A \alt 0.1$, (iii) the depletion factor of the pressure and
the internal energy in an early stage of collapse in which $\rho \ll
\rho_{\rm nuc}$ should be large enough to induce a rapid collapse in
the central region of the stellar core and for an efficient
spin-up. With the increase of stellar core mass, the maximum value of
$\beta$ achieved during the collapse is increased, but this does not
significantly change the stability property as far as $M$ is larger
than $\sim 1.5M_{\odot}$.  It is also found that the value of $\beta$
does not uniquely determine the property of the dynamical stabilities.

\subsection{Gravitational waveforms from nonaxisymmetrically
deformed stars}

\begin{figure}[htb]
\vspace*{-4mm}
\begin{center}
\epsfxsize=2.8in
\leavevmode
(a)\epsffile{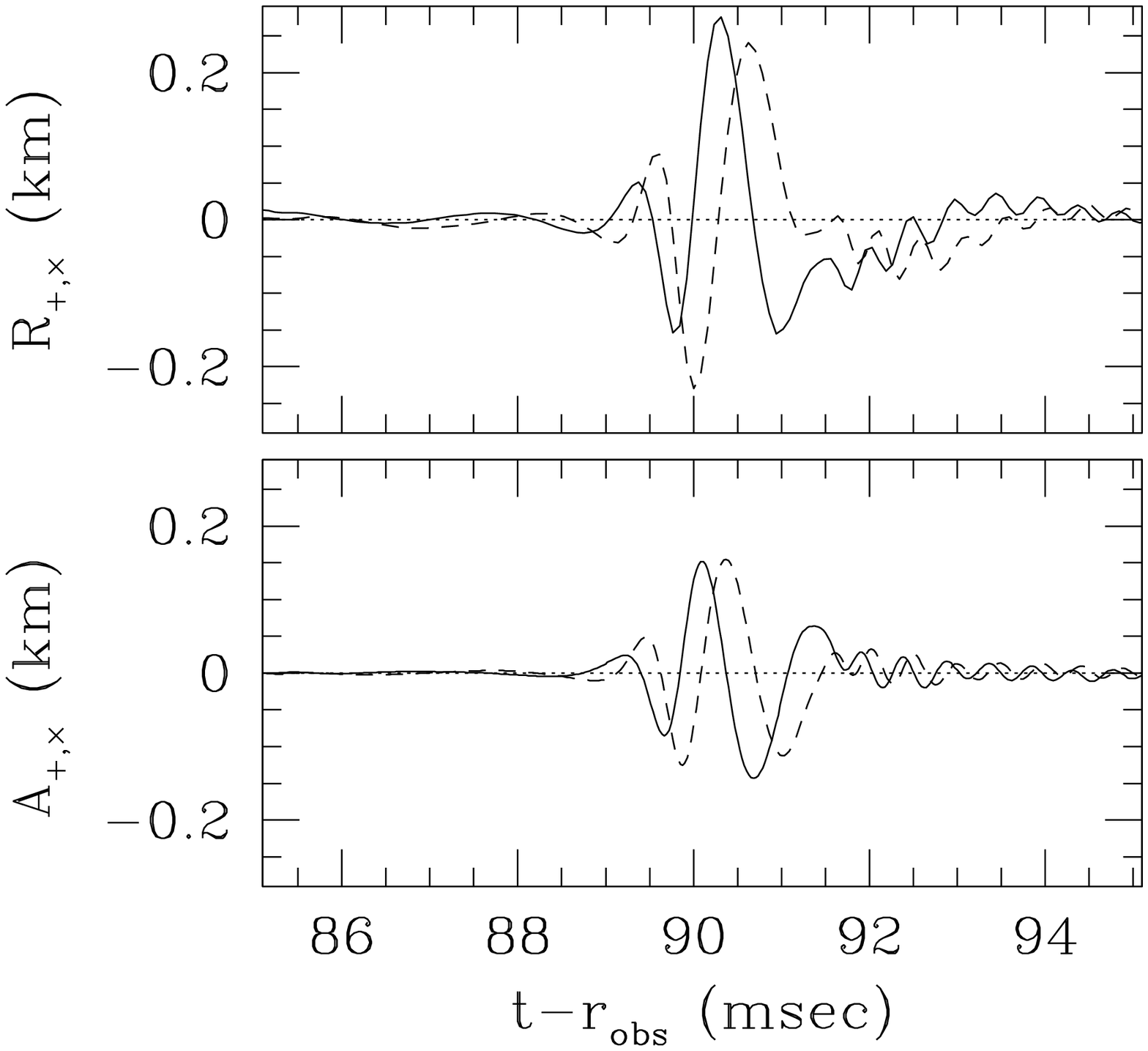}
\epsfxsize=2.8in
\leavevmode
~~~(b)\epsffile{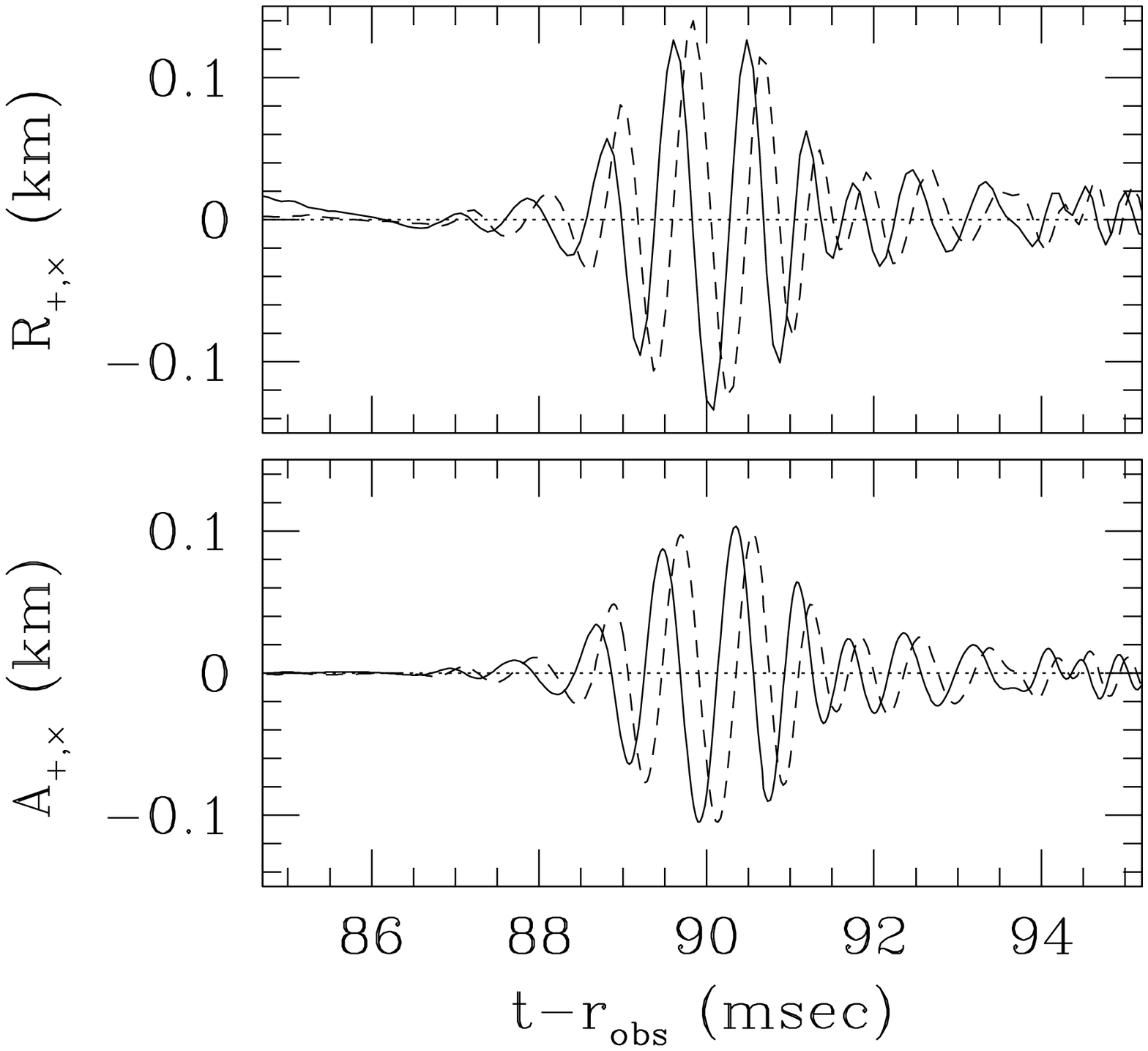} \\
\vspace*{-4mm}
\epsfxsize=2.8in
\leavevmode
(c)\epsffile{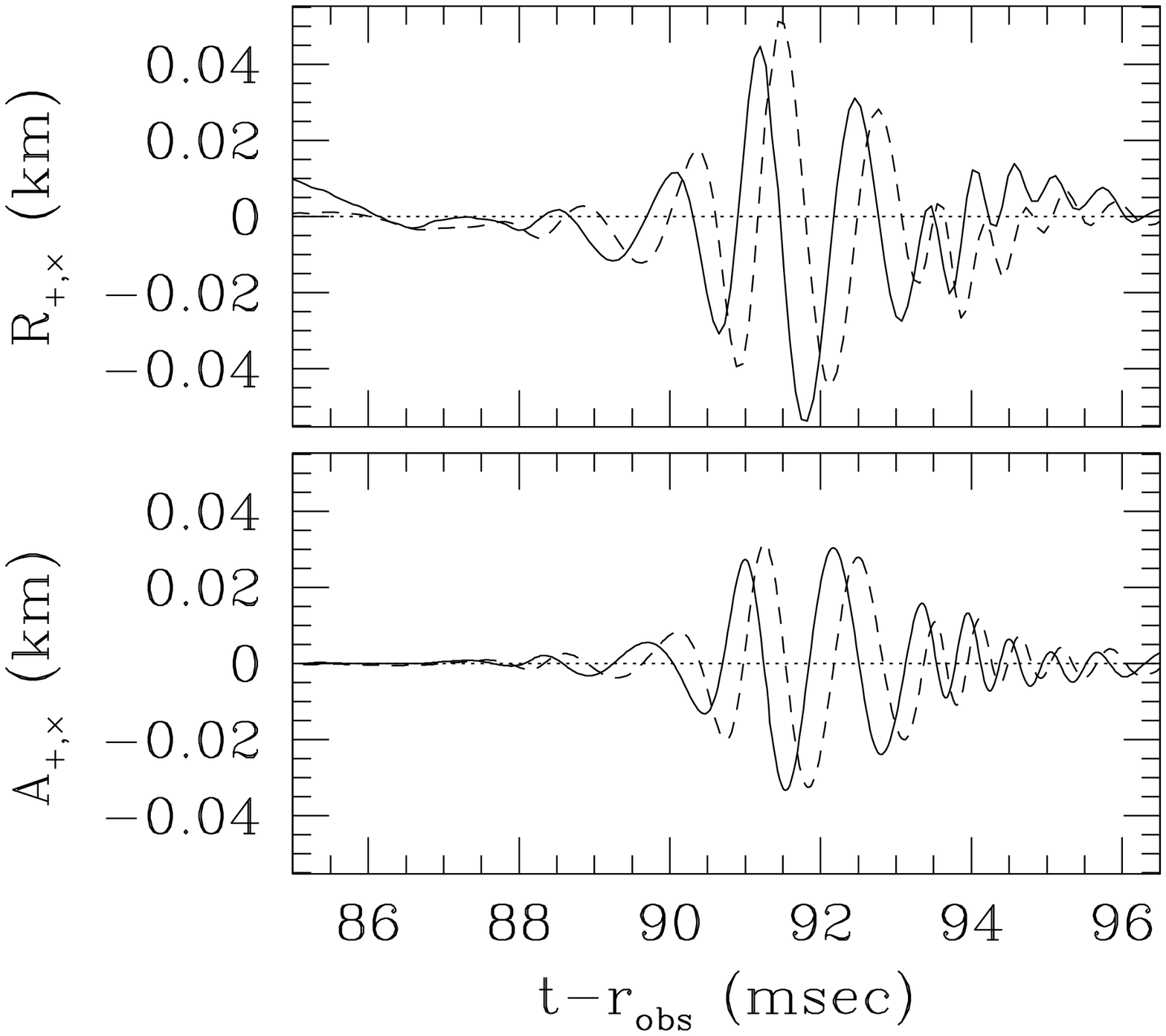}
\epsfxsize=2.8in
\leavevmode
~~~(d)\epsffile{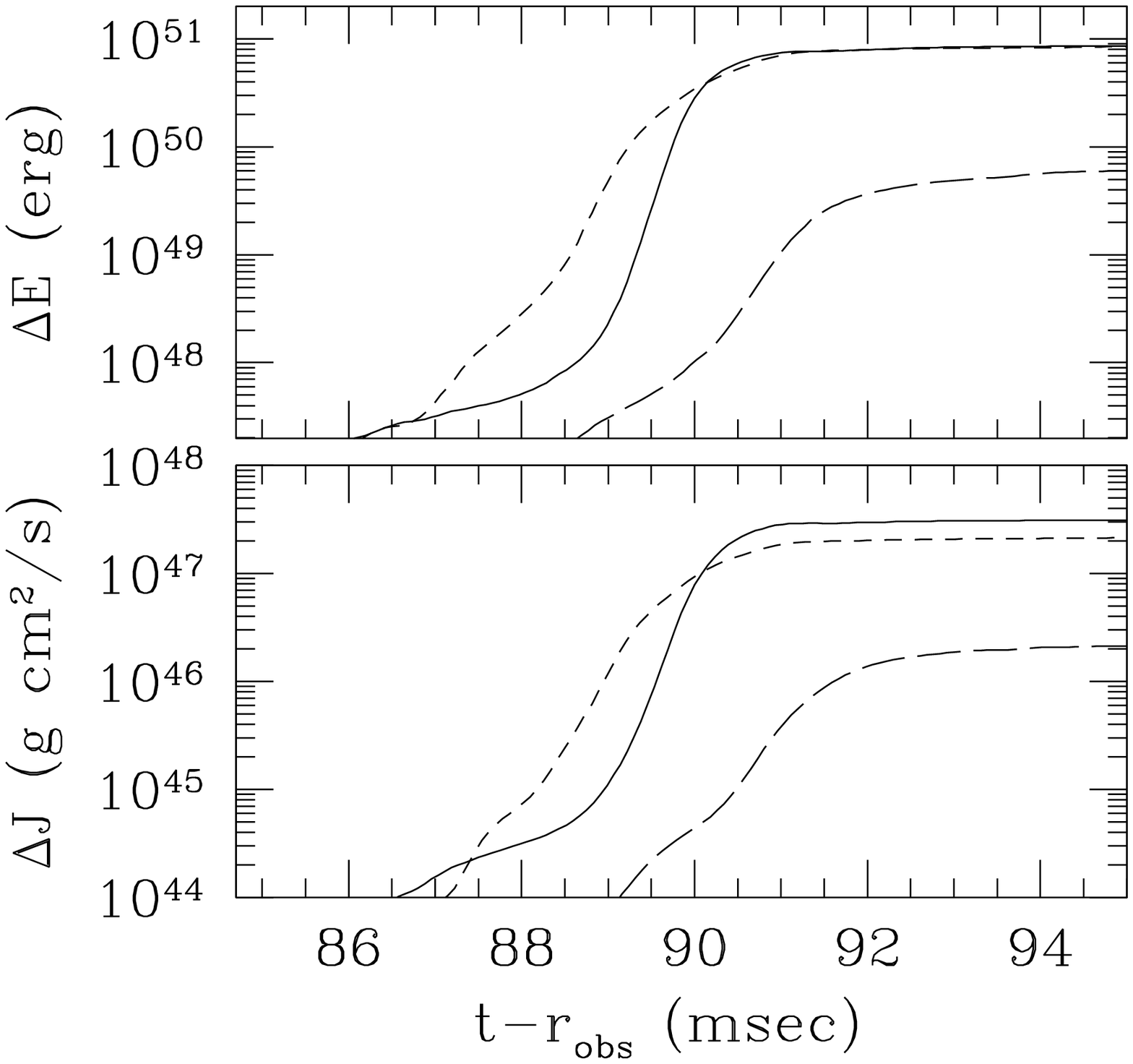}
%\vspace*{-4mm}
\caption{
Gravitational waveforms in the gauge-invariant wave
extraction method ($R_{+,\times}$)
and in the quadrupole formula ($A_{+,\times}$) (a) for M7c2C, 
(b) for M7c3C, and (c) for M5c2C. 
(d) Total emitted energy and angular momentum as a function of retarded time 
for models M7c2C (solid curves), M7c3C (dashed curves), and
M5c2C (long-dashed curves). 
\label{FIG15}}
\end{center}
\end{figure}

In Fig. 15, we show gravitational waveforms and
total emitted energy and angular momentum as a function
of retarded time for models M7c2C, M7c3C, and M5c2C. For these models, 
nonaxisymmetric dynamical instabilities set in after the bounce 
resulting in formation of a bar and spiral arms and in
excitation of gravitational waves with $m=2$ modes. 
Gravitational waveforms are computed both by the gauge-invariant
wave extraction and by the quadrupole formula. 

Figures show that with the amplification of $\eta$, the amplitude of
gravitational waves are increased.  However, once it reaches the
maximum, the amplitude damps quickly as in the evolution of
$\eta$. This is due to the effect that the bar-mode perturbation plays
a role for transferring the angular momentum from the inner region to
the outer one. Eventually, the bar-mode perturbation damps, resulting
in a quick damping of gravitational wave amplitude.  The damping is in
particular outstanding for model M7c2C.  This is due to the fact in
this model, the amplitude of the bar-mode is largest, and hence, the angular
momentum transfer is most effective. Due to this fact, the total
emitted energy for models M7c2C and M7c3C becomes approximately
identical although the maximum amplitude of gravitational waves for
M7c2C is about twice larger.

In isolated rotating stars, once the bar-mode instabilities set in and
saturate, the amplitude of their perturbation remains approximately
constant, resulting in emission of quasiperiodic gravitational waves
in a dissipation time scale of gravitational radiation which is much
longer than the dynamical time scale (e.g., \cite{SKE}). However, in
the rotating core collapse, the amplitude of gravitational waves is
damped by the angular momentum transfer from the bar to the
surrounding matter, for which the time scale is nearly equal to the
dynamical time scale and much shorter than the emission time scale of
gravitational radiation.

The maximum amplitude of gravitational waves for model M7c2C is by a
factor of $\sim 2$ larger than that for M7c3C, reflecting that the
degree of nonaxisymmetric deformation is larger. The amplitude for
model M5c2C is by a factor of $\sim 2.5$ smaller than that of
M7c3C, although the initial value of $\beta$ is approximately
identical and the waveforms are very similar for these two models.
According to the quadrupole formula, the amplitude of gravitational
waves is approximately proportional to $M^2$ if the radius of the
formed protoneutron star is identical. Thus, the dependence 
on mass is reflected in the amplitude. 

In the evolution of models M7c2C, M7c3C, and M5c2C, the $m=1$ mode
perturbation grows in the late phase of the evolution. However, this
does not affect the amplitude of gravitational waves significantly,
since the amplitude of the perturbation is not very large and the
$m=1$ mode does not contribute to the lowest-order (mass quadrupole)
waveforms in the three-space of the reflection symmetry with respect
to the equatorial plane.

For models M7c2C and M7c3C, 
the maximum values of $R_{+,\times}$ are $\sim 0.25$ km and 0.15 km,
respectively. For M5c2C, it is even smaller $\sim 0.05$ km. 
The amplitude of gravitational waves, $h$, observed 
at a distance of $r$ along the optimistic direction ($\theta=0$)
is written as 
\beqn
h \approx 10^{-21} \biggl( {R_{+,\times} \over 0.31 {\rm km}}\biggr)
\biggl({10 {\rm Mpc} \over r}\biggr). \label{hamp}
\eeqn
This implies that the observed amplitude at 
a distance of 10 Mpc is at most $h \alt 8 \times 10^{-22}$
for initial core mass $M \sim 2.5M_{\odot}$ and
$h \sim 1.5 \times 10^{-22}$ for $M \sim 1.5M_{\odot}$. 

\begin{figure}[thb]
\vspace*{-6mm}
\begin{center}
\epsfxsize=2.8in
\leavevmode
(a)\epsffile{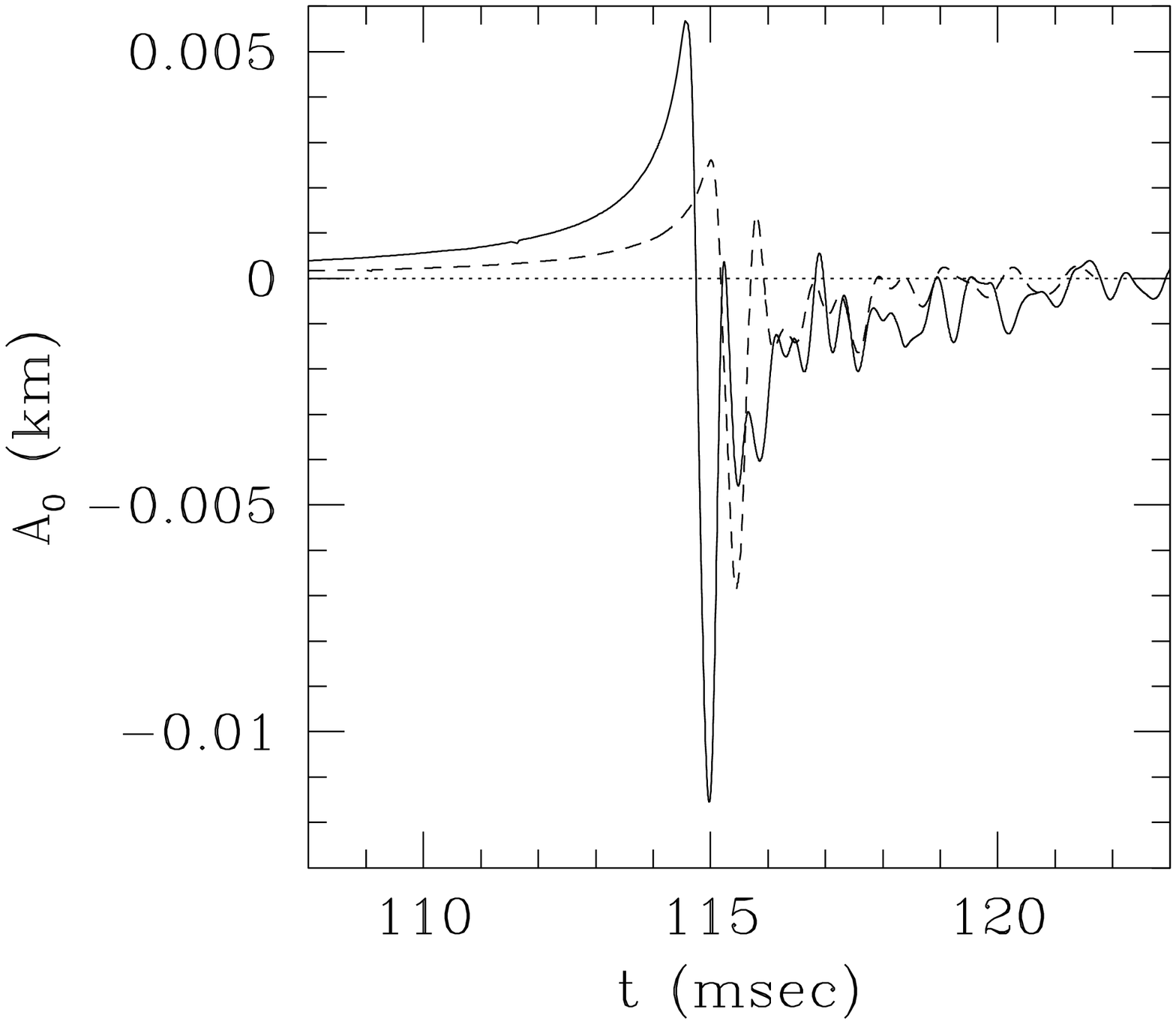}
\epsfxsize=2.8in
\leavevmode
~~~(b)\epsffile{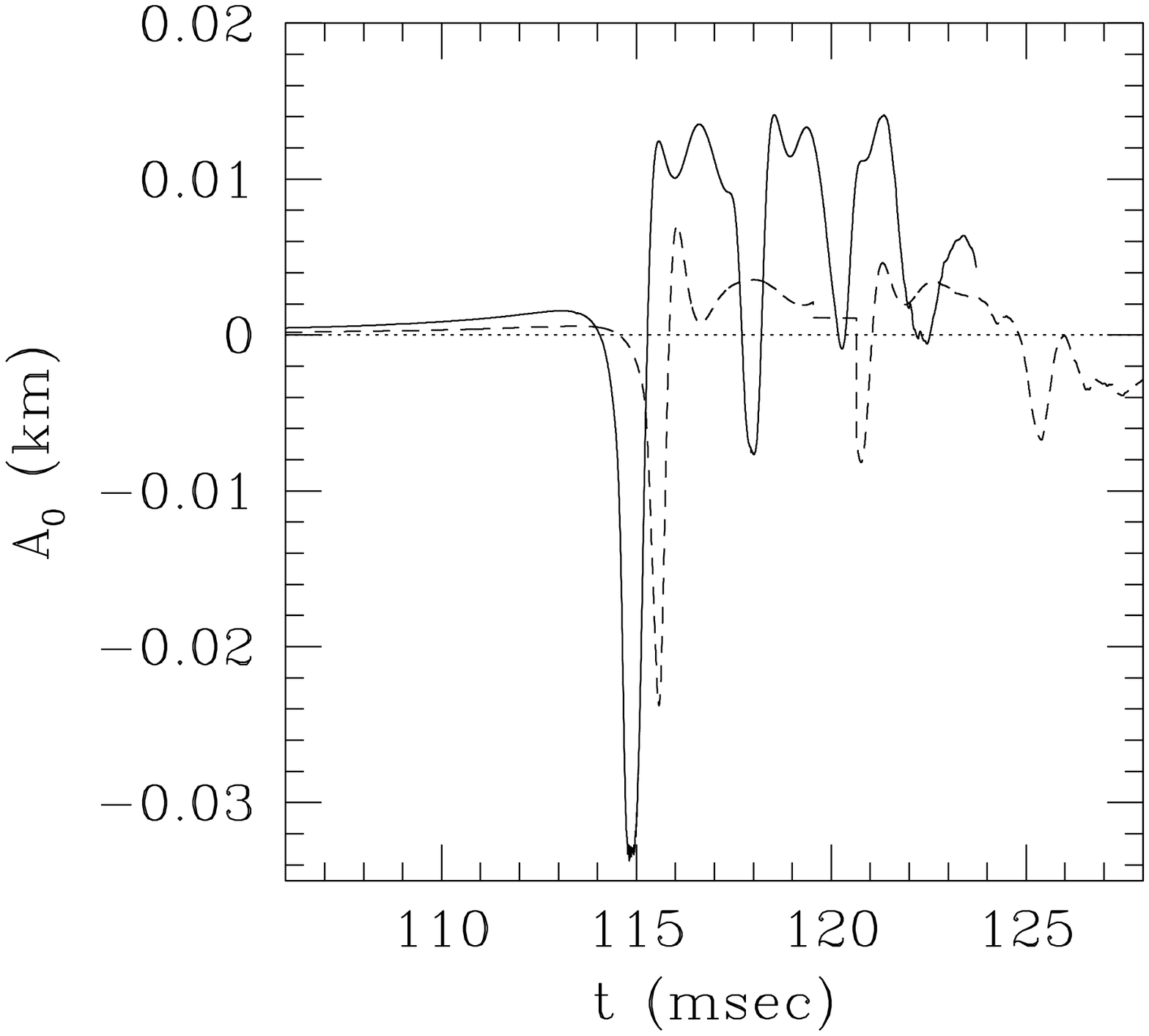}
%\vspace*{-4mm}
\caption{$A_0$ in the axisymmetric simulations
(a) for models M7a1 (solid curves) and M5a1 (dashed curves), and 
(b) for models M7c3 (solid curve) and M5c2 (dashed curve) 
with $\Gamma_1=1.3$ and $\Gamma_2=2.5$. 
\label{FIG16}}
\end{center}
\end{figure}

To compare the amplitude of gravitational waves from the bar-mode
deformation with that from axisymmetric
collapse, we show $A_0$ for models M5a1, M7a1, M5c2, and M7c3 with
$\Gamma_1=1.3$ and $\Gamma_2=2.5$ in the axisymmetric simulations
in Fig. 16. As mentioned in \cite{SS2}, it is difficult 
to extract gravitational waves of small amplitude from the metric in the 
axisymmetric simulations, and hence, only the waveforms by
the quadrupole formula are presented here. Although it provides only 
an approximate waveform, the wave phase can be accurately computed
and the error of the amplitude will be at most $\sim 10\%$. 
Figure 16 indicates that for the initial mass $M \sim 2.5M_{\odot}$, 
the maximum amplitude is at most 0.01 km for the rigidly rotating case 
and 0.02--0.03 km for differentially rotating cases.
The values are by a factor of $\sim 2$ smaller for $M \sim 1.5M_{\odot}$. 
Thus, the amplitude of gravitational waves of the $l=m=2$ modes
from the nonaxisymmetric dynamical instabilities 
is $\sim 10$ times as large as that in the axisymmetric case.
On the other hand, those amplitudes are not as large as 
the maximum amplitude of gravitational waves from 
of coalescing binary neutron stars in close circular orbits \cite{STU}. 
Thus, the nonaxisymmetric deformation in the 
stellar core collapse is not as a stronger emitter as coalescing
binary neutron stars. 

For model M7c2C (M7c3C), the total 
emitted energy and angular momentum are about $9\times 10^{50}$ erg 
($9\times 10^{50}$ erg) and $3\times 10^{47}~{\rm g~cm^2/sec}$
($2 \times 10^{47}~{\rm g~cm^2/sec}$), respectively.
These values are about 0.03\% (0.03\%) of the total mass 
energy ($M_*c^2$) and 0.7\% (0.6\%) of the total angular momentum, 
respectively, and 
are much larger than those in the axisymmetric collapse (e.g., \cite{HD}).
However, they are not as large as those in merger of binary compact
objects in which $\agt 1\%$ of the total mass energy and $\agt 10\%$
of the angular momentum are dissipated by gravitational waves
in the final phase of the merger \cite{STU}.
Thus, in the stellar collapse, the radiation reaction by gravitational waves 
are not likely to play an important role for the dynamics of bounce and 
oscillation of the protoneutron star. For model M5c2C, 
these values are much smaller because of its small mass and 
small compactness achieved. 
Hence, the effect of gravitational wave emission is less important.

\begin{figure}[htb]
\vspace*{-6mm}
\begin{center}
\epsfxsize=2.8in
\leavevmode
(a)\epsffile{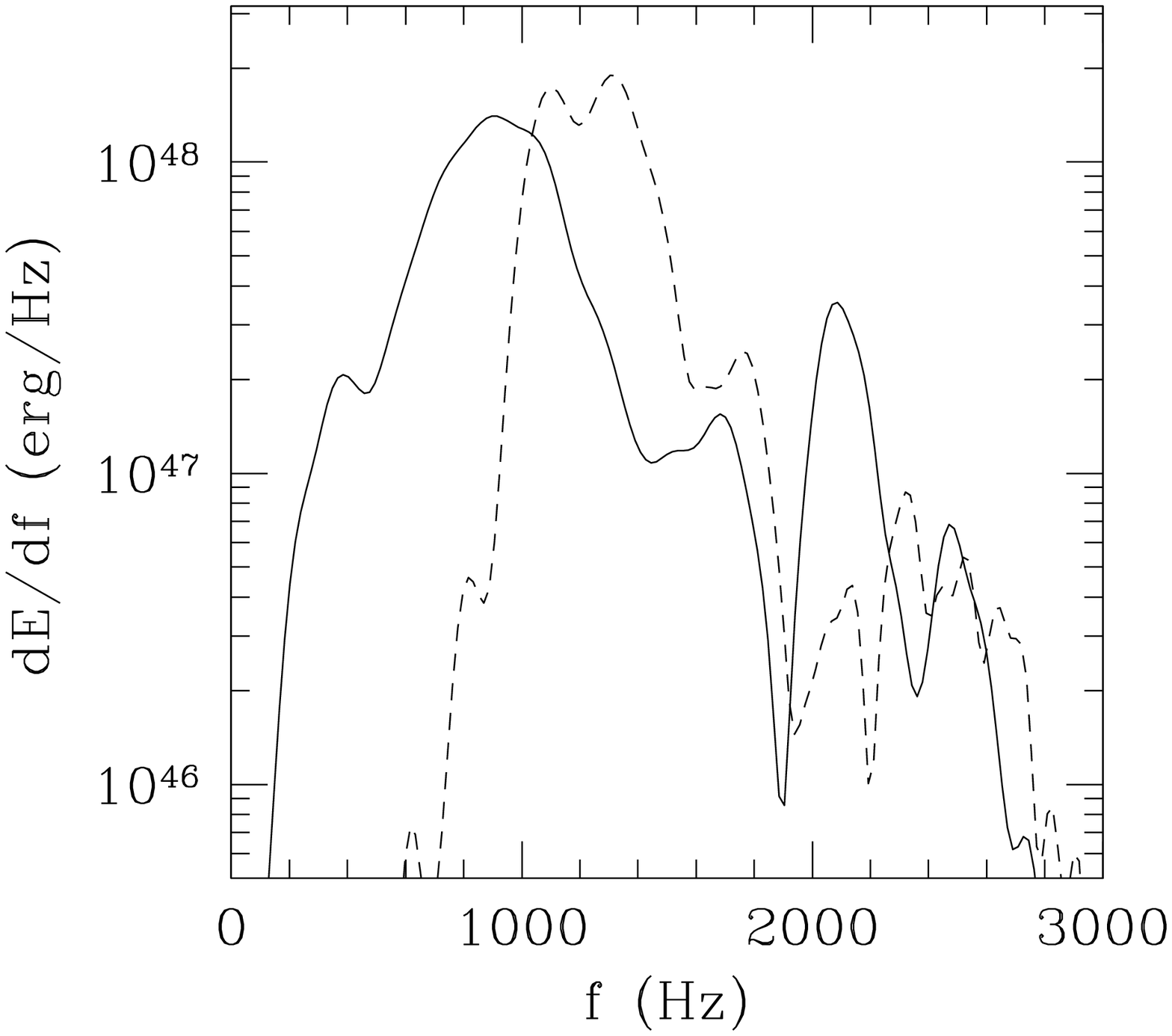}
\epsfxsize=2.8in
\leavevmode
~~~(b)\epsffile{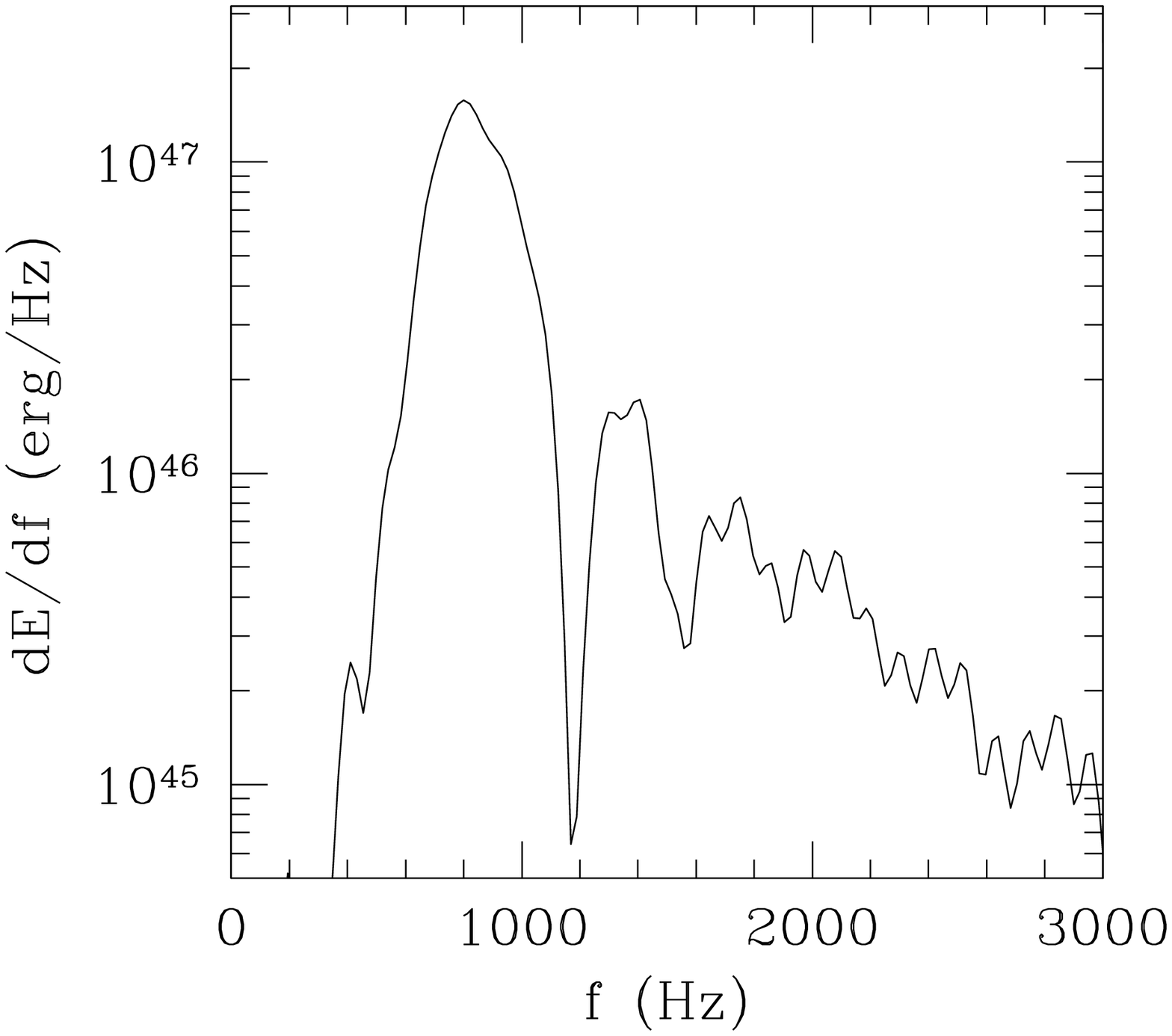}
%\vspace*{-4mm}
\caption{Energy power spectra of $l=m=2$ modes (a) for
models M7c2C (solid curve) and M7c3C (dashed curve), and (b)
for M5c2C.
\label{FIG17}}
\end{center}
\end{figure}

\begin{figure}[htb]
\vspace*{-4mm}
\begin{center}
\epsfxsize=3.in
\leavevmode
(a)\epsffile{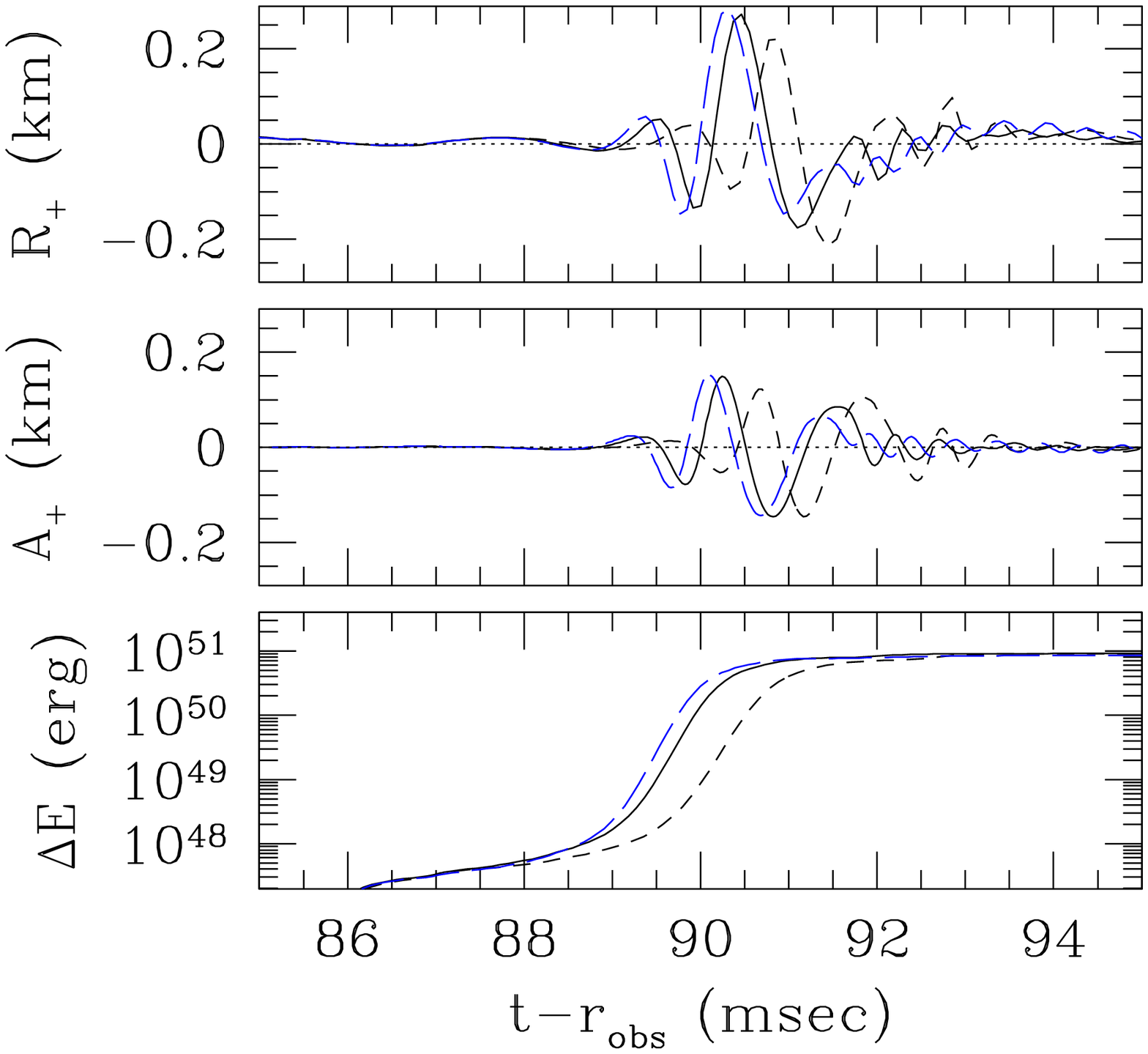}
\epsfxsize=3.in
\leavevmode
~~~(b)\epsffile{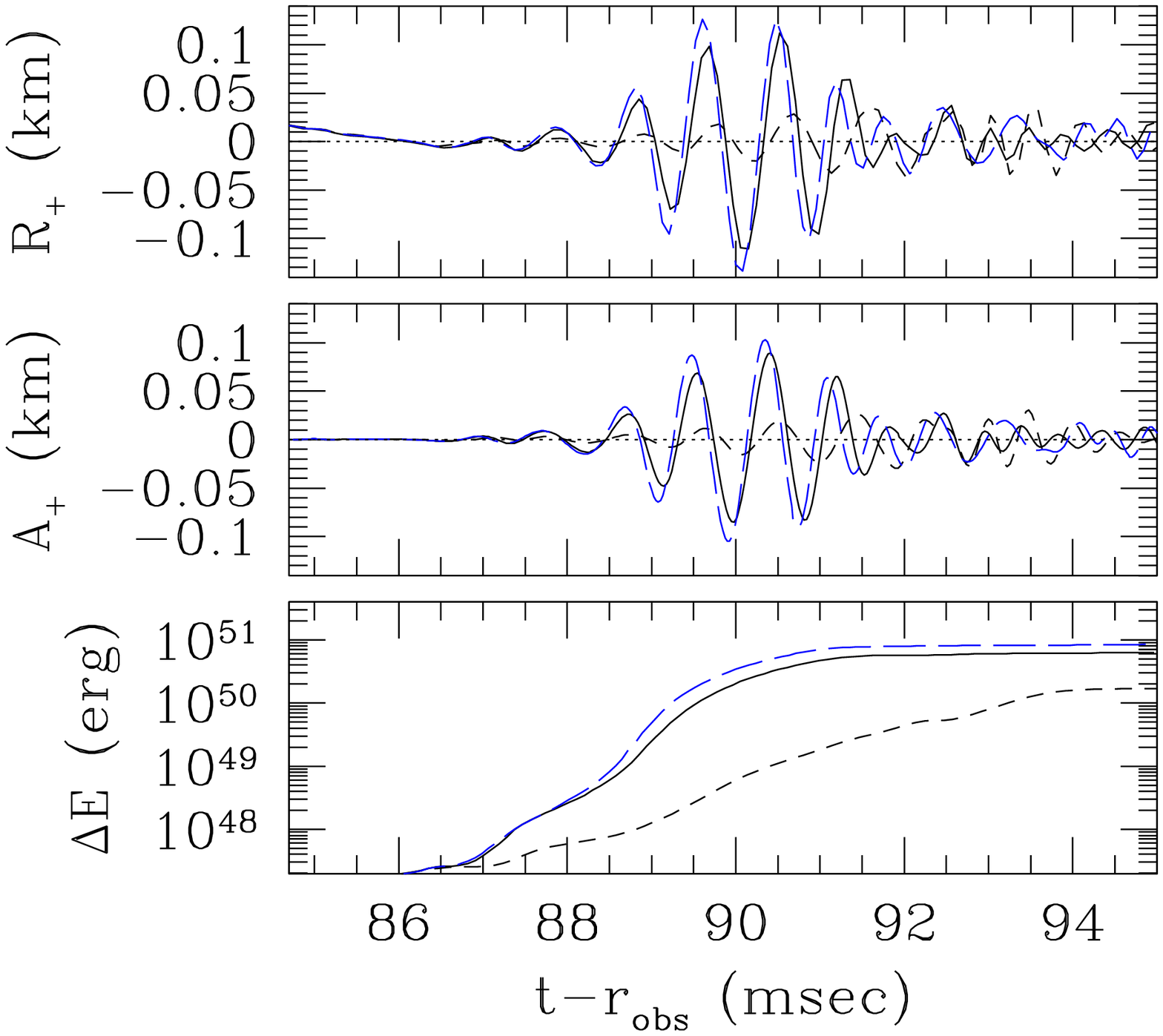}
%\vspace*{-4mm}
\caption{Gravitational waveforms and energy luminosity
(a) for model M7c2C and (b) for model M7c3C with $N=156$, 188, and 220.
The long-dashed, solid, and 
dashed curves denote the results with $N=220$, 188, and 156. 
\label{FIG18}}
\end{center}
\end{figure}

Comparison of gravitational waveforms computed by the 
gauge-invariant wave extraction method and by the quadrupole 
formula shows that the wave phase in the two results 
agree approximately (besides a systematic phase shift).
However, the amplitude disagrees by a factor of $\alt 2$.
As pointed out in \cite{SS2},
in the quadrupole formula, the amplitude is underestimated
by a factor of $M/R \sim 0.1$--0.2 where $R$ here denotes the characteristic
radius of the outcome after the collapse. On the other hand,
the amplitude in the gauge-invariant 
wave extraction method is likely to be overestimated 
because the waveforms are extracted in a local wave zone \cite{SS1}:
In this paper, $L \sim \lambda/2 < \lambda$ where $\lambda$ is the 
wavelength of gravitational waves $\sim 300$ km, and thus, the
amplitude would be overestimated by a factor of 10--20\% \cite{SS1}. 
Hence, the true amplitude would be between two results. However, 
besides the disagreement in the amplitude, two methods 
provide qualitatively the identical results. 
This reconfirms that the quadrupole formula is 
a reasonable method for approximately computing gravitational 
waveforms even in fully general relativistic simulations, 
in the absence of black holes. 

In Fig. 17, we show the energy power spectrum of $m=2$ modes 
(a) for models M7c2C (solid curve) and M7c3C (dashed curve)
and (b) for model M5c2C. 
The Fourier spectrum is computed from gravitational waveforms in terms
of the gauge-invariant wave extraction.
The spectrum for model M7c2C is broader in a low frequency region
with $f < 1$ kHz than those for other models.
This reflects the long oscillation period of this model. 
The peak frequency is about 0.8--1.3 kHz in these models.
These frequencies are determined by the quadrupole f mode frequency of the
deformed star formed after the bounce. Namely, the higher peak frequency 
implies that the outcome is more compact in proportional to $\sqrt{M/R^3}$.
According to a perturbative study for the quadrupole f mode \cite{Kok},
the frequency of neutron stars becomes $\sim 2.5$--4 kHz.
The frequency of the oscillation of unstable protoneutron stars
is much lower than that of neutron stars. The reason is that
the radius of the protoneutron star is larger. Nevertheless, 
the peak frequency is higher than the 
best sensitive frequency (between $\sim 100$ and several 100 Hz)
of kilometer size laser-interferometers such as LIGO \cite{KIP}.
As shown in Eq. (\ref{hamp}), the amplitude of gravitational 
waves is not very high if we assume the distance to the source 
$\agt 10$ Mpc. Thus, gravitational waves from
nonaxisymmetrically deformed protoneutron stars may be promising 
sources for such gravitational wave detectors only when the
stellar collapse happens for $r \ll 10$ Mpc. 
On the other hand, the frequency may be 
in a good range for resonant-mass detectors and/or specially
designed advanced interferometers such as the advanced LIGO \cite{KIP}. 

To summarize this section, we have found that
the amplitude of gravitational waves from dynamically unstable
protoneutron stars against nonaxisymmetric deformation is
$\sim 10$ times as large as that from the axisymmetric collapse.
However, even in the case that the degree of the nonaxisymmetric deformation 
is as large as in model M7c2C, the maximum amplitude is 
$\alt 20$--30\% of that in merger of binary neutron stars (e.g., \cite{STU}).
Since the peak frequency of gravitational waves is fairly high $\sim 1$ kHz, 
gravitational waves from nonaxisymmetric dynamical 
deformation of protoneutron stars may become promising 
sources for the laser-interferometric gravitational wave detectors only 
in the case that the event rate for the nonaxisymmetric deformation 
in the stellar core collapse is large. 

Before closing this section, we demonstrate that 
the convergence with improvement of the grid resolution is 
achieved fairly well for gravitational waveforms. 
In Fig. 18, we show the numerical results for models M7c2C and M7c3C
with $N=156$, 188, and 220. 
For the lower grid resolution, the period of the quasiradial 
oscillation becomes longer. As a result, the 
growth rate of $\eta$ becomes smaller. This causes an 
error in phase of gravitational waves. Also, the lower 
grid resolution results in underestimating the maximum value of $\eta$. 
As a result, the amplitude of gravitational waves 
is underestimated. However, with $N \agt 200$, the 
numerical results appear to converge well. Thus, we conclude that 
with our choice of the grid resolution, a good convergent result is 
obtained.

\section{Summary and discussion}

We have presented the first numerical results of 
three-dimensional hydrodynamic simulations for stellar core 
collapse in full general relativity focusing mainly on 
the criterion for the onset of the bar-mode dynamical instabilities. 
The nonaxisymmetric dynamical instabilities have been widely studied for 
isolated rotating stars in equilibrium to this time not only in 
Newtonian gravity but also in general relativity. However, 
for nonaxisymmetric dynamical instabilities in rotating stellar core 
collapse, very little study has been done even in Newtonian  
gravity \cite{Newton45}. Taking into account such status, we 
performed the simulations for a wide variety of equations of 
state, stellar masses, and velocity profiles to clarify the 
criterion for the onset of the nonaxisymmetric dynamical 
instabilities as well as the outcomes after their onset.

A number of previous works for isolated 
rotating stars in equilibrium have clarified that 
the bar-mode dynamical instabilities 
can set in when the value of $\beta$ exceeds $\sim 0.27$ or
when the degree of differential rotation is sufficiently high. 
Thus, first, we performed axisymmetric simulations of 
rotating stellar collapse to clarify the conditions 
that the value of $\beta$ is amplified beyond $\sim 0.27$ 
and that the degree of differential rotation for the
outcomes of the collapse becomes very large. We have found the 
following conditions are necessary to achieve a state with
$\beta_{\rm max} > 0.27$: 
(A) the initial state of the collapse is highly differentially rotating
with $A \alt 0.1$; 
(B) the progenitor is moderately rapidly rotating
with $0.01 \alt \beta_{\rm init} \alt 0.02$, but 
has to be not very rapidly rotating such as $\beta_{\rm init} \agt 0.02$; 
(C) the progenitor star is massive enough to 
achieve a compact state for which a significant spin-up is achieved.
However, at the same time, the mass should not be very high
to avoid black hole formation. We also found that to achieve a high
degree of differential rotation after the collapse, the depletion
factor of the pressure and the internal energy in an early stage
of collapse in which $\rho \ll \rho_{\rm nuc}$ 
should be large enough to induce a rapid collapse in the central region of
the stellar core and resulting efficient spin-up. 

Next, to clarify the condition for the onset of nonaxisymmetric dynamical 
instabilities, we performed three-dimensional simulations. 
Based on the results of axisymmetric simulations, we picked up 
models which are likely to become unstable during the collapse and bounce.
From the three-dimensional simulations, it is found that 
the nonaxisymmetric dynamical instabilities set in only
for a restricted parameter range as indicated by axisymmetric
simulations. 
Specifically, the following conditions are required to be 
satisfied: (i) the progenitor of the stellar core collapse is
rapidly rotating with $0.01 \alt \beta_{\rm init} \alt 0.02$, 
(ii) the degree of differential rotation for the velocity profile
of the initial condition is very high with $A \alt 0.1$, 
and (iii) the depletion factor of the pressure and internal energy
in the early stage of collapse 
is large enough to induce a rapid collapse in the central region of
the stellar core. Although stellar cores of larger mass have 
more advantage to form a compact protoneutron star, 
the stability property depends weakly on the mass as far as
$M \agt 1.5M_{\odot}$. 

Gravitational waves are computed 
in the case that the bar-mode dynamical instabilities set in. 
For the case that the bar-mode perturbation grows, 
the amplitude of gravitational waves increases exponentially, and as a
result, burst-type waves are emitted. However, since the bar-mode of the core
subsequently damps due to the outward
angular momentum transfer in a short time scale $\sim 2$--3 msec, 
the amplitude of gravitational waves decreases quickly. Thus, 
quasiperiodic gravitational waves, in which
the amplitude can be accumulated effectively, are not emitted efficiently
after the damping of the nonaxisymmetric perturbation. 
The maximum amplitude of gravitational waves at
a distance of 10 Mpc is $\sim 4$--$8 \times 10^{-22}$ with the 
frequency $\sim 1$ kHz for very massive core collapse 
with initial core mass $M \sim 2.5M_{\odot}$. 
The maximum amplitude is approximately proportional to $M^2$
for a given value of $\beta_{\rm init}$. For $M \sim 1.5M_{\odot}$, thus, 
the maximum amplitude is $\sim 1$--$2 \times 10^{-22}$ at 
a distance of 10 Mpc. This amplitude is about 10 times as large as 
that in the axisymmetric collapse, but $\sim 20$--30\% of the maximum
amplitude in the merger of binary neutron stars (e.g., \cite{STU}). 
Thus, the feature of gravitational waves is summarized as follows: 
(i) burst-type (not quasiperiodic) waves are emitted, 
(ii) the frequency is relatively high with $\sim 1$ kHz, and (iii)
the amplitude is about 10 times as large as that from axisymmetric collapse, 
but not as large as that for merge of binary compact objects. 
These facts imply that only when 
nonaxisymmetric dynamical instabilities set in 
for a large fraction of the stellar core collapse, 
gravitational waves induced by nonaxisymmetric dynamical
instabilities of protoneutron stars may become promising sources for 
kilometer size laser-interferometers. 

Besides the dynamical instabilities, 
there is another route for nonaxisymmetric deformation: {\em secular
instabilities}. As found from Figs. 1 and 6, the value of $\beta$ in 
the protoneutron stars formed after the collapse is often larger 
than the critical value for the onset of secular instabilities $\sim 0.14$. 
According to previous works \cite{LS,SK,OTL}, isolated rotating stars of 
$\beta \agt 0.14$ can form an ellipsoidal structure due to 
gravitational radiation, which may become a strong emitter of quasiperiodic 
gravitational waves with the frequency between 10 and several 100 Hz. 
However, these studies were performed for isolated stars. 
In the case of stellar core collapse, the formed protoneutron
stars will be surrounded by massive outer envelopes, and thus,
the bar-mode perturbation excited by the secular instabilities
may be damped quickly due to the angular momentum transfer from
the ellipsoidal protoneutron star to the outer envelope as in the
case of the dynamical instabilities. 
A simulation with massive envelope will be necessary to clarify whether 
the secular instabilities can grow or not. On the other hand, 
in contrast to the dynamical instabilities, the growth 
time scale of the secular instabilities is fairly long $\agt 100$ msec.
In such a long time scale, the surrounding matter may be 
ejected outward or accreted onto the central neutron star
in reality, and hence, the secular instabilities may grow
as in the isolated stars. However, in such a long time scale,
viscous or magnetic dissipation may also play an important role \cite{BSS}
for preventing the growth of the nonaxisymmetric perturbation. 
At present,
it is totally unclear whether the secular instabilities set in or not. 

As reported in this paper, the criterion for the onset 
of nonaxisymmetric dynamical instabilities may depend sensitively
on the equations of state for subnuclear density, since
with the smaller pressure for $\rho < \rho_{\rm nuc}$, 
the collapse is accelerated more for an efficient spin-up of the 
central region. In the present work, we adopted a parametric 
equation of state for simplicity.
To clarify the criterion for the onset of nonaxisymmetric 
dynamical instabilities more strictly, 
however, sophisticated equations of state
should be adopted. In realistic equations of state, 
the increase rate of the pressure as a function of the density
(i.e., an adiabatic index) significantly decreases at density
$\sim 10^{12}{\rm g/cm^3}$ (e.g., \cite{EOS,EOS2}).
This suggests that in a realistic equation 
of state, collapse of the central region is likely to be accelerated
significantly before reaching the nuclear density and, hence, the collapsed
core may be more subject to nonaxisymmetric dynamical instabilities. This
fact suggests that a simulation with more realistic equations of
state is an interesting subject for the future. 

Finally, we note the following issue. 
This paper focuses only on nonaxisymmetric dynamical instabilities
of protoneutron stars in the first 10--20 msec after the bounce.
The formed protoneutron stars subsequently emit neutrinos and
dissipate the thermal energy \cite{Bethe,BL}. 
As a result, they contract gradually in a time scale of $\sim 10$ sec.
Because the angular momentum is conserved approximately, the spin of 
the protoneutron stars will be increased with the contraction and
$\beta$ may be increased beyond $\sim 0.27$. Thus, 
even in the case that they are stable in the first 10--20 msec after the
bounce, they may eventually become unstable after the neutrino cooling.
This issue is not investigated in this paper and remains for the future.

\begin{center}
{\bf Acknowledgments}
\end{center}

We thank Takashi Nakamura for helpful comments. 
Numerical computations were carried out 
on the FACOM VPP5000 machine in the data processing center of 
National Astronomical Observatory of Japan and 
on the NEC SX6 machine in the data processing center of ISAS in JAXA. 
This work is in part supported by Japanese Monbu-Kagakusho Grant
(Nos. 15037204, 15740142, and 16029202). 
YS is supported by JSPS Research Fellowship for Young Scientists
(No. 1611308).

\end{document}